\documentclass{lmcs}
\newcommand{\vlong}[1]{#1}

\usepackage[utf8]{inputenc}
\usepackage{amsmath,amsthm,amssymb}
\usepackage{mathrsfs}

\usepackage{bm}
\usepackage{xparse}
\usepackage{url}
\usepackage{stmaryrd}
\usepackage{textcomp}
\usepackage{proof}
\usepackage{color}
\usepackage{float}
\usepackage{tikz}
\usepackage{wrapfig}
\usetikzlibrary{arrows,positioning,decorations.pathmorphing,arrows.meta,matrix,svg.path}
\usepackage{cmll}
\usepackage{subscript}

\definecolor{mygreen}{rgb}{0,0.6,0.4}
\definecolor{myred}{rgb}{0.8,0.1,0.1}
\definecolor{myblue}{rgb}{0,0.5,0.8}

\usepackage{listings}
\usepackage{doi}
\usepackage{graphicx}
\usepackage{subcaption}
\usepackage{diagbox}
\usepackage{xfrac}
\usepackage{cancel} 
\usepackage{cleveref}
\usepackage[colorinlistoftodos,prependcaption,textsize=scriptsize,textwidth=2cm]{todonotes}
\newcommand{\note}[1]{\todo[color=yellow]{#1}}
\makeatletter
\newcommand{\LeftEqNo}{\let\veqno\@@leqno}
\makeatother

\usepackage[USenglish]{babel}
\usepackage{multicol}
\usepackage{enumitem}
\setlist[itemize]{itemsep=0pt plus 1pt}
\setlist[enumerate]{itemsep=0pt plus 1pt}

\usepackage[strict]{changepage}
\usepackage{array}
\usepackage{ltxcmds}



\newcommand{\automath}[1]{\relax\ifmmode{#1}\else{$#1$}\fi}
\newcommand{\eqtau}{\equiv_\tau}
\newcommand{\nequiv}{\not\equiv}

\theoremstyle{plain}
\makeatletter
\newtheorem*{rep@theorem}{\rep@title}
\newcommand{\newreptheorem}[2]{%
\newenvironment{rep#1}[1]{%
 \def\rep@title{#2 \ref{##1}}%
 \begin{rep@theorem}}%
 {\end{rep@theorem}}}
\makeatother
\newreptheorem{theorem}{Theorem}
\newreptheorem{lemma}{Lemma}
\newreptheorem{proposition}{Proposition}
\newreptheorem{corollary}{Corollary}

\newcommand{\dpaw }{\textrm{\lowercase{d}PA\textsuperscript{$\omega$}}}
\newcommand{\dlpaw}{\textrm{\lowercase{d}LPA\textsuperscript{$\omega$}}}

\newcommand{\nef}{\textsc{nef}}

\newcommand{\cps}{\textsc{CPS}}
\newcommand{\nth}{\texttt{nth}}
\newcommand{\etal}{\emph{et al.}}
\newcommand{\acn}{\automath{\text{AC}_\N}}
\newcommand{\dc}{\automath{\text{DC}}}

\newcommand{\prf}{\mathop{\texttt{prf}}\,}
\newcommand{\wit}{\mathop{\texttt{wit}}\,}
\newcommand{\refl}{\mathop{\texttt{refl}}}
\newcommand{\subst}[2]{\mathop{\texttt{subst}}\, #1\,#2}
\newcommand{\splitop}{\mathop{\texttt{split}}}
\renewcommand{\split}[2]{\splitop\, #1\, \mathop{\texttt{as}}\,(#2)\,\mathop{\texttt{in}}\,}
\newcommand{\case}[3]{\mathop{\texttt{case}} \,#1\, \mathop{\texttt{of}}\, [#2\,|\, #3]}
\newcommand{\exf}{\mathop{\texttt{exfalso}}\,}
\newcommand{\dest}[2]{\mathop{\texttt{dest}} \,#1\, \mathop{\texttt{as}}\, (#2)\, \mathop\texttt{in}\,}
\newcommand{\catch}[1]{\mathop{\texttt{catch}}_{#1}}
\newcommand{\throw}[1]{\mathop{\texttt{throw}}\,{#1}\,}
\newcommand{\letop}{\mathop{\texttt{let}}}
\newcommand{\inop}{\mathop{\texttt{in}}}
\newcommand{\letin}[2]{\letop #1=#2 \inop}
\newcommand{\ind}[4]{\texttt{fix}^{#1}_{#3} [#2\,|\,#4]}
\newcommand{\cofix}[3]{{\texttt{cofix}}^{#1}_{#2}{[#3]}}

\newcommand{\recop}{\mathop{\texttt{rec}}}
\newcommand{\rec}[4]{{\recop}^{#1}_{#2}[#3\,|\,#4]}

\newcommand{\injec}[2]{\iota_{#1}(#2)}

\newcommand{\mut}{\tilde{\mu}}

\newcommand{\coupesep}{|\!|}
\newcommand{\coupe}[2]{
    {\mbox{$\langle$}} {#1} {\coupesep} {#2} \mbox{$\rangle$}}

\newcommand{\typered}{\triangleright}
\newcommand{\reds}{\rightsquigarrow_s}

\newcommand{\dl}{\textrm{\normalfont{d{L}}}}
\newcommand{\dltp}{\textrm{\normalfont{d{L}\textsubscript{$\reset$}}}}
\newcommand{\tp}{{{\texttt{t}\!\texttt{p}}}}
\newcommand{\reset}{{\hat{\tp}}}
\newcommand{\coreset}{{\check{\tp}}}
\renewcommand{\shift}{\mu{\reset}.}
\newcommand{\coshift}{\tmu{\coreset}.}

\newcommand{\tr}[1]{\llbracket #1 \rrbracket}
\newcommand{\trt}[1]{\llbracket #1 \rrbracket_t}

\newcommand{\trp}[1]{\llbracket #1 \rrbracket_p}
\newcommand{\tre}[1]{\llbracket #1 \rrbracket_e}
\newcommand{\trV}[1]{\llbracket #1 \rrbracket_V}

\newcommand{\trc}[1]{\llbracket #1 \rrbracket_c}

\newcommand{\trtp}[1]{\llbracket #1 \rrbracket_{\reset}}

\makeatletter
\newcommand{\ovset}[3][0ex]{%
  \mathrel{\mathop{#3}\limits^{
    \vbox to #1{\kern-2\ex@
    \hbox{#2}\vss}}}}
\makeatother
\newcommand{\oset}[2]{\ovset{\scriptsize #1}{#2}}

\newcommand{\red}{\rightarrow}  
\newcommand{\basearrow}{\longrightarrow}

\newcommand{\bmark}{\beta     }

\newcommand{\bred} {\basearrow_{\bmark}}

\newcommand{\bredn}[1]{\oset{#1}{\basearrow}_{\bmark}}

\makeatletter
\newcommand{\shorteq}{%
  \settowidth{\@tempdima}{-}
  \resizebox{\@tempdima}{\height}{=}%
}
\makeatother

\newcommand{\alphabold}{\bm\kappa}
\newcommand{\cbold}{{{\mathscr{\bm k}}}}
\newcommand{\tmu}{{\tilde\mu}}
\newcommand{\cut}[2]{\coupe{#1}{#2}}
\newcommand{\cmd}[2]{\coupe{#1}{#2}}
\newcommand{\imp}{\rightarrow}
\newcommand{\defeq}{\triangleq}
\newcommand{\lbvtstar}{\automath{\overline{\lambda}_{[lv\tau\star]}}}
\newcommand{\lbv}     {\automath{\overline{\lambda}_{lv}}           }

\newcommand{\pole}{{\bot\!\!\!\bot}}
\newcommand{\orth}{{\pole}}
\newcommand{\Sub}[3]{#1[#3/#2]}

\newcommand{\reduce}{\rightsquigarrow}
\renewcommand{\S}{\mathcal{S}}

\newcommand{\dpt}[1]{\{#1\}}
\newcommand{\eps}{\varepsilon}
\newcommand{\rdpt}[1]{\dpt{\cdot|#1}}
\newcommand{\vide}{\eps}
\newcommand{\negt}{{\bot\!\!\!\bot}}
\newcommand{\dptimp}{\Rrightarrow}
\newcommand{\dptprod}[2]{\Pi{#1}.{#2}}
\newcommand{\dptsum}[2]{\Sigma{#1}.{#2}}
\newcommand{\mustar}{\mu\!\star\!}

\newcommand{\N}{{\mathbb{N}}}

\renewcommand{\P}{\mathcal{P}}

\newcommand{\lmmt}{\automath{\lambda\mu\tmu}}

\newcommand{\fvn}[2]{\|#2\|_{#1}}
\newcommand{\tvn}[2]{|#2|_{#1}}
\newcommand{\fvF}[1]{\fvn{F}{#1}}
\newcommand{\fvf}[1]{\fvn{f}{#1}}
\newcommand{\fvE}[1]{\fvn{E}{#1}}
\newcommand{\fve}[1]{\fvn{e}{#1}}
\newcommand{\tvv}[1]{\tvn{v}{#1}}
\newcommand{\tvV}[1]{\tvn{V}{#1}}
\newcommand{\tvVt}[1]{\tvn{V_t}{#1}}
\newcommand{\fvpi}[1]{\tvn{\pi}{#1}}
\newcommand{\tvt}[1]{\tvn{t}{#1}}
\newcommand{\tvp}[1]{\tvn{p}{#1}}
\newcommand{\real}{\Vdash}

\newcommand{\trn}[2]{\tr{#2}_{#1}}

\newcommand{\trE}[1]{\trn{E}{#1}}

\newcommand{\prfcase}[1] {\paragraph{\normalfont\textbullet~ \textbf{Case}  #1.}}

\newcommand{\cmdt}[2]{\cmd{#1}{#2}_t}

\newcommand{\cmdp}[2]{\cmd{#1}{#2}_p}
\newcommand{\cmde}[2]{\cmd{#1}{#2}_e}
\newcommand{\cmdv}[2]{\cmd{#1}{#2}_v}
\newcommand{\cmdf}[2]{\cmd{#1}{#2}_f}

\newcommand{\cmdpi}[2]{\cmd{#1}{#2}_{\pi}}

\def\limp{\Rightarrow}

\def\<{\langle}
\def\>{\rangle}
\newcommand{\tv}[1]{|#1|}
\newcommand{\fv}[1]{\|#1\|}

\newcommand{\Quote}{\texttt{quote}}

\newcommand{\Fork}{{\pitchfork}}

\newcommand{\dom}{\mathop{\mathrm{dom}}}
\newcommand{\Int}[1]{\llbracket #1 \rrbracket}
\newcommand{\ds}{\displaystyle}
\newcommand{\FV}{{FV}}
\renewcommand{\P}{\mathcal{P}}

\newcommand{\Pow}{\mathcal{P}}

\newcommand{\C}{\mathcal{C}}
\newcommand{\R}{\mathcal{R}}
\newcommand{\V}{\mathcal{V}}

\newcommand{\A}{\mathcal{A}}

\renewcommand{\L}{\mathscr{L}}

\newcommand{\muteq}{\tmu\shorteq}

\newlength{\saut}
\newcommand{\hsep}{\qquad}
\newcommand{\myfig}[1]{\framebox{\vbox{#1}}}
\newcommand{\keyword}[1]{\emph{#1}\index{#1}}

\newcommand{\myquote}[1]{\emph}
\newcommand{\noem}{\vspace{-1em}}
\newcommand{\nomidem}{\vspace{-0.5em}}

\newcommand{\hole}{{[\,\,]}}

\newcommand{\sigdash      }{\vdash^{\sigma}} 
\newcommand{\sigsigdash      }{\vdash^{\sigma\sigma'}} 
\newcommand{\optsigma      }{} 
\newcommand{\sigmaopt      }{} 
\newcommand{\sigsigprimeopt}{} 
\newcommand{\tis}[2]{(#1|#2)}

\newcommand{\compat}[2]{#1\diamond#2}

\makeatletter
\newcommand{\osetb}[3][0ex]{%
  \mathrel{\mathop{#3}\limits^{
    \vbox to#1{\kern-2\ex@
    \hbox{#2}\vss}}}}
\makeatother

\newcommand{\redlv}{\red_{lv}}
\newcommand{\indpt}[2]{{#1\##2}}
\newcommand{\stext}{\vartriangleleft}
\newcommand{\instore}[1]{#1-in-store}
\newcommand{\ct }{\instore{term}}

\newcommand{\ce }{\instore{context}}

\newcommand{\cp }{\instore{proof}}

\newcommand{\stjoin}[2]{\mathsf{join}(#1,#2)}

\newcommand{\str}[2]{\texttt{str}_\infty^{#1}\,#2}
\newcommand{\defrule}{\autorule{\texttt{def}}}
\newcommand{\Ainf}[1]{A_\infty^{#1}}

\newcommand{\nthn}[2]{\nth_{#1}\,#2}
\newcommand{\elimmark}{E}
\newcommand{\intromark}{I}

\newcommand{\autorule}[1]{\relax\ifmmode{\scriptstyle(#1)}\else$(#1)$\fi}

\newcommand{\cutrule}{\autorule{\textsc{Cut} }}
\newcommand{\axrule }{\autorule{\textsc{Ax}  }}
\newcommand{\axnrule}{\autorule{\textsc{Ax}_n}}
\newcommand{\axxrule}{\autorule{\textsc{Ax}_t}}

\newcommand{\axrrule}{\autorule{\textsc{Ax}_r}}
\newcommand{\axlrule}{\autorule{\textsc{Ax}_l}}

\newcommand{\murule }{\autorule{\mu          }}
\newcommand{\mutrule}{\autorule{\mut         }}

\newcommand{\botrule}{\autorule{\bot         }}

\newcommand{\impirule}{\autorule{\imp_\intromark}}
\newcommand{\imperule}{\autorule{\imp_\elimmark }}
\newcommand{\imprrule}{\autorule{\imp_r         }}
\newcommand{\implrule}{\autorule{\imp_l         }}

\newcommand{\falrule}{\autorule{\forall_l         }}
\newcommand{\farrule}{\autorule{\forall_r         }}

\newcommand{\frurule}{\autorule{\forall^1_{\!r}}}
\newcommand{\flurule}{\autorule{\forall^1_{\!l}}}
\newcommand{\frdrule}{\autorule{\forall^2_{\!r}}}
\newcommand{\fldrule}{\autorule{\forall^2_{\!l}}}

\newcommand{\indrule}{\autorule{\texttt{fix}}}
\newcommand{\cofixrule}{\autorule{\texttt{cofix}}}

\newcommand{\exirule}{\autorule{\exists_\intromark}}

\newcommand{\exrrule}{\autorule{\exists_r         }}
\newcommand{\exlrule}{\autorule{\exists_l         }}

\newcommand{\andeurule}{\autorule{\land^1_\elimmark}}
\newcommand{\andedrule}{\autorule{\land^2_\elimmark}}
\newcommand{\andrrule }{\autorule{\land_r          }}
\newcommand{\andlrule }{\autorule{\land_l          }}

\newcommand{\orrrule }{\autorule{\lor_r           }}

\newcommand{\orlrule }{\autorule{\lor_l           }}

\newcommand{\weakrule  }{\autorule{w        }}
\newcommand{\weakdrule }{\autorule{w^d      }}

\newcommand{\prfrule  }{\autorule{\prf    }}

\newcommand{\convrrule}{\autorule{\equiv_r}}
\newcommand{\convlrule}{\autorule{\equiv_l}}
\newcommand{\substrule}{\autorule{\subst{}{}\!\!}}
\newcommand{\reflrule }{\autorule{=_r}}
\newcommand{\witrule  }{\autorule{\wit }}

\newcommand{\lamrule}{\autorule{\lambda}}
\newcommand{\apprule}{\autorule{@      }}

\newcommand{\cboldrule}{\autorule{    \cbold}}

\newcommand{\arule}    {\autorule{a     }}
\newcommand{\crule}    {\autorule{c     }}
\newcommand{\lrule}    {\autorule{l     }}
\newcommand{\dlrule}   {\autorule{l^d   }}
\newcommand{\alpharule}{\autorule{\alpha}}
\newcommand{\alphaboldrule}{\autorule{\alphabold}}
\newcommand{\lft}[1]{\uparrow^{#1}}
\newcommand{\liftVrule}{\autorule{\lft{V}}}
\newcommand{\liftErule}{\autorule{\lft{E}}}
\newcommand{\lifttrule}{\autorule{\lft{t}}}
\newcommand{\lifterule}{\autorule{\lft{e}}}

\newcommand{\eagerrule}{\autorule{\tmu^{[]}}}

\newcommand{\epsrule    }{\autorule{\varepsilon }}
\newcommand{\tautrule   }{\autorule{\tau_t      }}
\newcommand{\tauprule   }{\autorule{\tau_p      }}
\newcommand{\tauerule   }{\autorule{\tau_e      }}

\newcommand{\tauErule   }{\autorule{\tau_E      }}
\newcommand{\taucatrule }{\autorule{\tau\tau'   }}

\newcommand{\resetrule}{\autorule{\reset}}
\newcommand{\shiftrule}{\autorule{\mu\reset}}
\newcommand{\dcutrule  }{\autorule{\textsc{Cut}^d}}
\newcommand{\dtcutrule  }{\autorule{\textsc{Cut}^d_t}}
\newcommand{\dmutrule  }{\autorule{{\mut}^d}}
\newcommand{\dxmutrule  }{\autorule{{\mut}^d_x}}
\newcommand{\eqrule}{\autorule{=_l}}

\newcommand{\splitrule    }{\autorule{ \splitop}        }

\newcommand{\dexlrule}{\autorule{\exists_l^d         }}
\newcommand{\dorlrule }{\autorule{\lor_l  ^d         }}
\newcommand{\dandlrule }{\autorule{\land_l^d          }}

\theoremstyle{plain}
\newtheorem{proposition}[thm]{Proposition}
\crefname{proposition}{proposition}{propositions}

\newtheorem{theorem}[thm]{Theorem}
\crefname{theorem}{theorem}{theorems}

\crefname{corollary}{corollary}{corollarys}

\newtheorem{lemma}[thm]{Lemma}
\crefname{lemma}{lemma}{lemmas}

\theoremstyle{definition}
\newtheorem{definition}[thm]{Definition}
\crefname{definition}{definition}{definitions}

\newtheorem{remark}[thm]{Remark}
\crefname{remark}{remark}{remarks}

\crefname{example}{example}{examples}

\usepackage{booktabs}   
\usepackage{subcaption} 

\begin{document}

\title[A constructive proof of dependent choice in classical arithmetic]{A constructive proof of dependent choice in classical arithmetic via memoization}

\author{Étienne Miquey}
\address{
   {Équipe Gallinette},               
{INRIA, Laboratoire des Sciences du Numérique de Nantes},            
{France}                    
}
\email{etienne.miquey@inria.fr}          


\begin{abstract}
In a recent paper~\cite{Herbelin12}, 
Herbelin developed {\dpaw}, a calculus in which constructive proofs for the axioms of countable and dependent
choices could be derived via the memoization of choice functions.  
However, the property of normalization (and therefore the one of soundness) was only conjectured.
The difficulty for the proof of normalization is due to the simultaneous presence 
of dependent types (for the constructive part of the choice),
of control operators (for classical logic), of coinductive objects (to encode
functions of type $\N\to A$ into streams $(a_0,a_1,\ldots)$) 
and of lazy evaluation with sharing (for memoizing these coinductive objects).

Elaborating on previous works, we introduce in this paper a variant of {\dpaw} presented as a sequent calculus.
On the one hand, we take advantage of a variant of Krivine classical realizability
that we developed to prove the normalization of classical call-by-need~\cite{MiqHer18}.
On the other hand, we benefit from \dltp, a classical sequent calculus with dependent types 
in which type safety is ensured by using delimited continuations together with a syntactic restriction~\cite{Miquey19}.
By combining the techniques developed in these papers,
we manage to define a realizability interpretation \emph{à la} Krivine of our calculus
that allows us to prove normalization and soundness.
This paper goes over the whole process, starting from Herbelin's calculus {\dpaw}
until our introduction of its sequent calculus counterpart {\dlpaw}.
\end{abstract}




\maketitle

\newcommand{\mypar}[1]{\subsection{#1}}
\renewcommand{\todo}[1]{}
\renewcommand{\note}[1]{}

\section{Introduction}
\mypar{The axiom of choice}

The axiom of choice is certainly one of the most intriguing pieces of the foundations of mathematics.
Understanding the axiomatization of a theory
as an intent to give a formal and truthful representation of a given world or structure,
as long as this structure only deals with finite objects 
it is easier to agree on what it ``is'' or ``should be''. 
However, as soon as the theory involves infinite objects, this question quickly turns out
to be much trickier.
In particular, some undeniable properties of finite objects become much more questionable in the case of infinite sets.
The axiom of choice is precisely one of these properties.
Consider for instance the following problem, as presented by Russell~\cite[pp.125-127]{Russell19}:
\begin{quote}
\it
[Imagine a] millionaire who bought a pair
of socks whenever he bought a pair of boots, and
never at any other time, and who had such a passion
for buying both that at last he had $\aleph_0$ pairs 
of boots and $\aleph_0$ pairs of socks.
The problem is: How many boots had he, and how many socks? 
\end{quote}
The cardinal $\aleph_0$ defines exactly the infinite quantity of natural numbers, 
and in particular, since there is a bijection from $\N\times\N$ to $\N$, $\aleph_0$ is not increased by doubling.
Hence, as suggested by Russell:
\begin{quote}
\it
One would naturally suppose that he had twice as many boots and
twice as many socks as he had pairs of each, and that therefore he had $\aleph_0$ of each [...].
\end{quote}
To prove this claim, it is thus necessary and sufficient to give an enumeration of the millionaire's boots and socks.
Yet, if it is easy to enumerate the boots---take each pair one after the other, consider the left boot first then the right one---
this is not possible \emph{a priori} in the case of socks 
without assuming the possibility of making an infinite number of arbitrary choices:
for each pair, one has to decide which sock to pick first.

Formally, the possibility of making these choices is expressed by the axiom of choice,
which was first introduced by Zermelo in the realm of set theory~\cite{Zermelo04}.
Among its different formulations, its relational variant is given by:
$$AC \defeq(\forall x\in A.\exists y \in B. P(x,y)) \imp (\exists f\in B^A.\forall x \in A.P(x,f(x))) $$
which stipulates the existence of a choice function\footnote{If we define the predicate $P(x,y)$ as $y\in x$,
it exactly says that if all the sets $x\in A$ are non-empty, there exists a choice function:
$(\forall x\in A. x\neq \emptyset) \imp \exists f\in (\cup A)^A .\forall x\in A. f(x)\in x$.}.
This axiom was shown to be independent of Zermelo-Fraenkel set theory (ZF)\footnote{Gödel proved that the theory ZF + AC 
is consistent, and Cohen proved the same for the theory ZF + $\neg$AC~\cite{JechAC}.}.

While it might be tempting to consider the axiom of choice as natural,
it leads to very surprising consequences.
The most striking example is certainly the Banach-Tarski paradox~\cite{BanTar1924},
which shows that the unit ball:
$$\mathcal{B}:=\{(x,y,z)\in\R^3: x^2+y^2+z^2=1\}$$
in three dimensions can be disassembled into a finite number of pieces, 
which can then be reassembled (after translating and rotating each of the pieces) 
to form two disjoint copies of the ball $\mathcal{B}$.
Nonetheless, many standard mathematical results require the axiom of choice (\emph{e.g.}, every vector space
has a basis, the product of compact topological spaces is compact, every field has an algebraic closure)
or one of its restricted variants (\emph{e.g.}, the axiom of dependent choice for Bolzano-Weierstrass' theorem,
countable choice to prove that every infinite set is Dedekind-infinite).
Especially, many proofs in analysis relies on the definition of an adequate converging sequence, 
whose existence itself relies on the axiom of dependent choice.

\mypar{Constructive interpretation of AC}
Russell's presentation of his paradox leads us to wonder \emph{how} the millionaire
should perform the choices necessary to obtain the enumeration.
Regarding the axiom of choice as expressed above, this amounts to wonder 
how to compute the choice function whose existence is stated by the axiom.

This point of view is in line with a constructive vision of mathematics
such as the one given by the Brouwer-Heyting-Kolmogorov interpretation~\cite{Troelstra11}.
In that settings, a proof of $A\to B$ is defined as a function that,
given a proof of $A$ constructs a proof of $B$; a proof of $\forall x\in A.B(x)$
is a function that turns any $a\in A$ into a proof of $B(a)$ while
a proof of $\exists x\in A.B(x)$ is a pair $(a,b)$
where $a\in A$ and $b$ is a proof of $B(a)$. 
Therefore, a proof of the premise of the relational axiom of choice (for a given relation $R$)
 has to be a function which, for any $a\in A$, constructs a proof  $(b_a,r_{ab})$ of $\exists y \in B. R(a,y)$.
The function $f:a\mapsto b_a$ then defines a valid choice function, and the axiom is easily satisfied.

While the concepts of ``proof'' and ``function'' above can be interpreted in different ways\footnote{For instance,
Kleene realizability corresponds to the case where ``proofs'' are natural numbers and ``functions'' are taken as being computable functions~\cite{Kleene45}.}
and as such remains meta-notions, the same idea is internalized within Martin-Löf's type theory.
To this end, one of the key features of Martin-L\"of's type theory is the concept
of dependent types, which allows formulas to refer to terms~\cite{MartinLof98}. 
Notably, the existential quantification rule is defined so that 
a proof term of type $\exists x^A\!.B$ is a pair $(t, p)$ where $t$---the \emph{witness}---is
of type $A$, while $p$---the \emph{proof}---is of type $B[t/x]$. 
Dually, the theory enjoys two elimination rules: one with a destructor $\wit$ to extract the witness,
the second one with a destructor $\prf$
to extract the proof.
This allows for a simple and constructive proof of the 
full axiom of choice~\cite{MartinLof98}:
$$
\begin{array}{rll}
 AC_A & := & \lambda H.(\lambda x.\wit(H x),\lambda x.\prf(H x)) \\
 &:& (\forall x^A\!.\exists y^B\!. P(x,y)) \imp \exists f^{A\imp B}\!.\forall x^A\!. P(x,f(x))
\end{array}
$$
This term is nothing more that an implementation of Brouwer-Heyting-Kolomogoroff (BHK) interpretation 
of the axiom of choice:
given a proof $H$ of $\forall x^A\!.\exists y^B\!. P(x,y)$, 
it constructs a choice function which simply maps any $x$ to the witness of $Hx$, while the proof that this function
is sound w.r.t. $P$ returns the corresponding certificate.

\mypar{Incompatibility with classical logic}

Unsurprisingly, this proof does not scale to classical logic. 
Imagine indeed that we could dispose, in a type theoretic (or BHK interpretation, realizability) fashion, 
of a classical framework including a proof term $t$ for the axiom of choice:
$$\vdash t : \forall x\in A.\exists y \in B. P(x,y) \imp \exists f\in B^A.\forall x \in A.P(x,f(x))$$
Consider now any undecidable predicate $U(x)$ over a domain $X$. 
Taking advantage of the exceluded middle, we can strengthen the formula $U(x) \lor \neg U(x)$ (which is true for any $x\in X$)
into:
$$\forall x\in X.\exists y\in\{0,1\}.(U(x)\land y=1) \lor (\neg U(x) \land y = 0)$$
which is provable as well in a classical framework and thus should have a proof term $u$. 
Now, this has the shape of the hypothesis of the axiom of choice, so that by application of $t$ to $u$, we should obtain a term:
$$\vdash t\,u : \exists f\in \{0,1\}^X.\forall x \in X.(U(x)\land f(x)=1) \lor (\neg U(x) \land f(x) = 0)$$
In particular, the term $\wit(t\,u)$ would be a function which, for any $x\in X$, outputs $1$ if $U(x)$ is true, and $0$ otherwise.
This is absurd, since $U$ is undecidable.
This informal explanation gives us a metamathematical argument on the impossibility of having a proof system 
which is classical as a logic, entails the axiom of choice and where proofs fully compute. 
Since the existence of consistent classical theories with the axiom of choice (like set theory) has been proven,
the incompatibility is to be found within the constructive character of proofs.


\subsection{The computational content of classical logic}
\label{s:inconsistency} 
In 1990, Griffin discovered that the control operator \texttt{call/cc} (for \emph{call with current continuation})
of the Scheme programming language could be typed by Peirce's law $((A\to B)\to A)\to A)$.
As Peirce's law is known to imply, in an intuitionistic framework, all the other forms of classical reasoning 
(excluded middle, \emph{reductio ad absurdum}, double negation elimination, etc.),
this discovery opened the way for a direct computational interpretation of classical proofs,
using control operators and their ability to \emph{backtrack}. 
Several calculi were born from this idea, such as Parigot's $\lambda\mu$-calculus~\cite{Parigot97}, 
Barbanera and Berardi's symmetric $\lambda$-calculus~\cite{BarBer96}, Krivine's $\lambda_c$-calculus~\cite{Krivine09} 
or Curien and Herbelin's $\bar{\lambda}\mu\tilde\mu$-calculus~\cite{CurHer00}.

Notably, in order to address the incompatibility of intuitionistic realizability with classical logic,
Krivine used the $\lambda_c$-calculus to develop
in the middle of the 90s the theory of classical realizability~\cite{Krivine09}.
Krvine realizability is a complete reformulation\footnote{As observed in several articles~\cite{OliStr08,Miquel10},
classical realizability can in fact be seen as a reformulation of Kleene's realizability 
through Friedman's $A$-translation~\cite{Friedman78}.} of
the very principles of realizability to make them compatible with classical reasoning. Although it was
initially introduced to interpret the proofs of classical second-order arithmetic (PA2), the theory of classical
realizability can be scaled to more expressive theories such as Zermelo-Fraenkel set theory~\cite{Krivine01} or the
calculus of constructions with universes~\cite{Miquel07}.
Krivine realizability has shown in the past twenty years to be a very powerful framework,
both as a way to build new models of set theory~\cite{Krivine12,Krivine14,Krivine18}
and as a tool to analyze programs~\cite{JabTab10,GuiMiq16,Geoffroy18}; 
we shall use it several times in this paper to prove the correctness of different calculi.

Last but not least, Griffin's discovery led to an important 
paradigmatic shift from the point of view of logic.
Instead of trying to get an axiom by means of logical translations (\emph{e.g.} Gödel's negative translation for classical reasoning), 
and then transfer this translation to program along the Curry-Howard 
correspondence (\emph{e.g.} continuation-passing style for negative translation),
one can rather try to directly add an operator whose computational behavior is adequate with the expected axiom. 
Several works over the past twenty years have emphasized the fact
that adding new programming primitives (and in particular side effects) 
to a calculus may bring new reasoning principles~\cite{Krivine03,JabSozTab12,Miquel11b,Herbelin10,JaberEtAl16}.
The present work is totally in line with this philosophy:
we will show how to use a form of memoization 
to give a computational content to the axioms of dependent and countable choices in a framework that
is compatible with classical logic.
That is, we will take advantage of several computational features (namely: control operators, dependent types, streams, 
lazy evaluation and shared memory) to build proof terms for these axioms.

\mypar{Realizing AC$_\N$ and DC in presence of classical logic}
As we explained earlier, the full axiom of choice and computational classical logic 
are \emph{a priori} incompatible. 
In particular, extending Martin-Löf's type theory with control operators
leads to an inconsistent theory.
This was observed by Herbelin~\cite{Herbelin05}:
starting from a sound minimal language with dependent types, he showed
that the further addition of control operators allows to derive a proof of false.
The main idea behind Herbelin's paradox, which we will detail in \Cref{s:Herbelin},
is that control operators allows to define a proof $p$ of type
$\exists x^\N\!.x=1$ 
that will first give a wrong witness for $x$ (say 0) together with a 
certificate which, if it is evaluated, will backtrack to furnish the appropriate witness (\emph{i.e.} $1$)
and the usual equality proof $\refl$ for the equality $1=1$.
Yet, extracting the witness from this proof with $\wit p$ returns $0$, 
while extracting the certificate with $\prf p$  reduces to the usual proof of equality $\refl$ (after backtracking). 
The paradox arises from the fact that $\prf p$ is of type $\wit p=1$, 
while $\wit p$ is convertible to $0$.

The bottom line of this example is that the same proof $p$
is behaving differently in different contexts thanks to control operators, 
causing inconsistencies between the witness and its certificate.
In 2012, Herbelin proposed a way of scaling up Martin-L\"of's proof to classical logic
while preventing the previous paradox to occur.
The first idea is to restrict dependent types to the fragment 
of \emph{negative-elimination-free} proofs (\nef) 
which, intuitively, only contains constructive proofs behaving as values\footnote{For instance,
in the previous example $p$ is not {\nef}, and the restriction thus precludes us from writing $\prf p$ or $\wit p$.}.
The second idea is to share the computation of the choice function, in order
to avoid inconsistencies coming from classical proofs behaving differently in different places.
To this purpose, Herbelin considered the cases of choice functions whose domain is the set of natural numbers
(\emph{i.e.}, the axioms of countable and dependent choices) and proposed to memoize their computations.
While the presence of control operators induces the possibility for a proof of type 
$\forall x\in\N.\exists y\in B.R(x,y)$ to furnish different $y$ for a same $n\in\N$ depending on the evaluation context,
the memoization ensures that it will be computed at most once.
Technically, the key is to encode functions of 
type $\mathbb{N} \to B$ by streams $(b_0,b_1,\dots)$
that are lazily evaluated, that is to say infinite sequences represented coinductively
and whose components are evaluated only if necessary.

This allows us to internalize into a formal system
 the realizability approach~\cite{BerBezCoq98,EscOli14} as a direct 
proofs-as-programs interpretation. 
The resulting calculus, called \dpaw, thus contains
dependent types (for the constructive part of choices),
control operators (for classical logic),
coinductive objects (to define streams)
and a lazy evaluation strategy with shared memory for these objects.

\mypar{Normalization of {\textbf{dPA}$^{\omega}$}}
In~\cite{Herbelin12}, the property of normalization (on which relies the one of consistency)
was only conjectured, and the proof sketch that was given turned out to be difficult to formalize properly.
Our first attempt to prove the normalization of {\dpaw} was to derive 
a continuation-passing style translation (CPS)\footnote{Indeed, removing the control part of {\dpaw} 
(which CPS translations do through an encoding) would leave us with an intuitionistic 
system whose normalization would be easier to prove (in particular, 
it would be a subsystem of the Calculus of Inductive Constructions).},
but translations appeared to be hard to obtain for \dpaw~as such. 
In addition to the difficulties caused by control operators and co-fixpoints, 
{\dpaw} reduction system is defined in a natural deduction fashion, with contextual rules 
involving meta-contexts of arbitrary depth. 
This kind of rules are particularly difficult to faithfully translate through a CPS.

Rather than directly proving the normalization of {\dpaw}, we choose to first 
give an alternative presentation of the system under the shape of a sequent calculus, which we call {\dlpaw}.
Indeed, sequent calculus presentations of a calculus usually provides good intermediate steps 
for {\cps} translations~\cite{Munch13PhD,MunSch15,AriDowMauJon16} since they enforce a decomposition of the reduction system
into finer-grained rules.
To this aim, we first handled separately the difficulties peculiar to the definition of such a calculus:
on the one hand, we proved with Herbelin the normalization of a calculus with control operators and lazy evaluation~\cite{MiqHer18};
on the other hand, we defined a classical sequent calculus with dependent types~\cite{Miquey19}.
By combining the techniques developed in these frameworks, we finally managed to define {\dlpaw}, 
which we present here and prove to be sound and normalizing.

In each case, we were guided by a methodology first described by Danvy,
who highlighted what he called the \emph{unity of semantic artifacts}~\cite{Danvy09}:
for a given notion of operational semantics, there is a calculus, an abstract machine, 
and a continuation-passing style translation that correspond exactly with one another. 
Therefore, from any one of these semantic artifacts, the others may be systematically derived. 
We extend this observation by showing how it can be used to ease the definition of 
a realizability interpretation \emph{à la} Krivine, which we actually do in each case to prove the
normalization and consistency of the considered calculi.

\mypar{Outline of the paper}
In this paper, we describe the complete path from Herbelin's conjecture~\cite{Herbelin12}
to the normalization proof of {\dlpaw} in~\cite{Miquey18a} (of which this paper is 
an extended version). This article will thus cover and summarize a significant amount of different papers
authored by Herbelin and myself during the whole process~\cite{Herbelin12,AriEtAl12,Miquey19,MiqHer18,Miquey18a}.

In details, we will start in \Cref{s:dpaw} by presenting the main ingredients of Herbelin's {\dpaw} calculus,
and the intuitions underlying the proof terms for the axioms of countable and dependent choices.
In \Cref{s:lmmt}, we will recap the main ideas of the Curien-Herbelin's \lmmt-calculus~\cite{CurHer00}, 
the sequent calculus on which we will base all the calculi developed afterwards, and in particular \dlpaw.
We shall also take advantage of this section to illustrate Danvy's 
\begin{wrapfigure}{r}{0.4\textwidth}
\scalebox{0.8}{
 \tikzstyle{base}   = [rounded corners=0.1cm]
\tikzstyle{format} = [draw,minimum width=2cm, thin,rounded corners=0.1cm]
\tikzstyle{medium} = [ellipse, draw, thin, minimum height=2.5em]
\begin{tikzpicture}[node distance=4cm, auto,>=latex', thick]
\draw (0,0) node[format,fill=black!10] (lmmt) {
  \begin{tabular}{l} 
    \lmmt-calculus\\[-0.3em]
    \small + sequent calculus \\[-0.4em]
  \end{tabular}
};
\draw(lmmt)+(-2,-2) node[format, fill=black!10] (lbv) {
  \begin{tabular}{l}
    \lbvtstar-calculus\\[-0.3em]
    \small + sequent calculus \\[-0.4em]
    \small + sharing \& lazyness\qquad
  \end{tabular}
};
\draw(lmmt)+(2,-4) node[format, fill=black!10] (dl) {
  \begin{tabular}{l}
    \dltp\\[-0.3em]
    \small + sequent calculus \\[-0.4em]
    \small + dependent types\\[-0.4em]
  \end{tabular}
};
\draw(lmmt)+(0,-6) node[format, fill=black!10] (dlpaw) {
  \begin{tabular}{l}
    $\dlpaw$\\[-0.3em]
    \small + sequent calculus \\[-0.4em]
    \small + dependent types\\[-0.4em]
    \small + co-fixpoints    \\[-0.4em]
    \small + sharing \& lazyness\qquad
  \end{tabular}
};        										  
\path[->,dashed]  (lmmt) edge (lbv);
\path[->,dashed]  (lmmt) edge (dl);
\path[->,dashed]  (lbv) edge (dlpaw);
\path[->,dashed]  (dl) edge (dlpaw);
\draw (lmmt)+(0.9,0.7) node {\small \em Sec. \ref{s:lmmt}};
\draw (lbv)+(-1.2,0.9) node {\small \em Sec. \ref{s:lbvtstar}};
\draw (dl)+(0.9,0.85) node {\small \em Sec. \ref{s:dl}};
\draw (dlpaw)+(-1.3,1.2) node {\small \em Sec. \ref{s:dlpaw}};
\end{tikzpicture}
}\label{fig:outline}
\end{wrapfigure}
methodology on the call-by-name \lmmt-calculus
in order to obtain a continuation-passing style translation and a Krivine realizability interpretation.
We then present in \Cref{s:lbvtstar} the \lbvtstar-calculus, a lazy classical sequent calculus, which was originally
defined by Ariola {\etal} using Danvy's method of semantic artifacts~\cite{AriEtAl12}.
After summarizing the successive steps leading to the definition of the \lbvtstar-calculus, we will recall 
the type system and realizability interpretation we developed together with Herbelin 
in order to prove its normalization~\cite{MiqHer18}.
In \Cref{s:dl}, we will present {\dltp}, a classical sequent calculus with dependent types~\cite{Miquey19}.
In particular, we will explain how the seek of a dependently typed continuation-passing style
unveiled the need for using delimited continuations and the restriction to the {\nef} fragment.
Finally, we shall gather these two systems to define {\dlpaw} in \Cref{s:dlpaw},
whose normalization is proved by means of a realizability interpretation combining
the interpretations for {\dltp} and the \lbvtstar-calculus.

Along the way, we will use the unity of semantic artifacts for each of the calculi aforementioned.
While we would like to avoid boring the reader stiff with recalling 
the full process each time, we still intend to outline it in each case. 
Indeed, we believe that the core ideas used in our realizability interpretation for {\dlpaw}
directly stem from the subcases of {\lbvtstar}-calculus and {\dltp}, and that they are easier to 
understand in these specific contexts which only contains the core ingredients. 
We hope that this will help the reader to understand 
the different technicalities induced by the expressiveness of {\dlpaw} 
 before its introduction in \Cref{s:dlpaw}.
%
%
%
%
%
%


\section{A proof of dependent choice compatible with classical logic}
We shall begin by presenting \dpaw, the proof system that was introduced by Herbelin
as a mean to give a computational content to the axiom of choice in a classical setting~\cite{Herbelin12}.
As explained in the introduction, 
the calculus is a fine adaptation of Martin-Löf proof which circumvents the different difficulties
caused by classical logic. 
Rather than restating {\dpaw} in full details, for which we refer the reader to~\cite{Herbelin12},
let us describe informally the rationale guiding its definition and the properties that it verifies.
We shall then dwell on the missing piece in the puzzle that led us to this work, namely the normalization property,
and outline our approach to prove it.

\subsection{A constructive proof of dependent choices compatible with classical logic}
The dependent sum type of Martin-L\"of's type theory provides a strong 
 existential elimination, 
 which allows us to prove the full axiom of choice.
 The proof is simple and constructive:
$$
\begin{array}{rll}
 AC_A & := & \lambda H.(\lambda x.\wit(H x),\lambda x.\prf(H x)) \\
 &:& \forall x^A.\exists y^B. P(x,y) \imp \exists f^{A\imp B}.\forall x^A. P(x,f(x))
\end{array}
$$

To scale up this proof to classical logic, the first idea in Herbelin's work~\cite{Herbelin12} 
is to restrict the dependent sum type to a fragment of his  
system which is called \emph{negative-elimination-free} (\nef).
This fragment contains slightly more proofs than just values, but is still computationally
compatible with classical logic. We shall see in \Cref{s:nef} how this restriction naturally appears
when trying to define a dependently-typed continuation-passing style translation.

The second idea is to represent a countable universal quantification as an 
infinite conjunction. 
This allows us to internalize into a formal system the realizability approach of
Berardi, Bezem and Coquand or Escardi and Oliva~\cite{BerBezCoq98,EscOli14} 
as a direct proofs-as-programs interpretation.
Informally, let us imagine that given a proof $H:\forall x^\N\!.\exists y^B \!.P(x,y)$,
we could create the sequence $H_\infty=(H 0,H 1,\ldots,H n,\ldots)$
and select its $n^{\textrm{th}}$-element with some function \nth.
Then one might wish that:
$$\lambda H.(\lambda n.\wit(\nth~n~H_\infty),\lambda n.\prf(\nth~n~H_\infty))$$
could stand for a proof for $AC_{\N}$.

However, even if we were effectively able to build such a term, $H_\infty$ might still
contain some classical proofs. Therefore, two copies of $H\,n$ might end up behaving 
differently according to the contexts in which they are executed, and then return
two different witnesses. 
This problem could be fixed by using a shared version of $H_\infty$, say: 
$$\lambda H.\letin{a}{H_\infty}~(\lambda n.\wit(\nth~n~a),\lambda n.\prf(\nth~n~a))\,.$$

It only remains to formalize the intuition of $H_\infty$. This is done by means of a coinductive fixpoint operator.
We write $\cofix{t}{bx}{p}$ for the co-fixpoint operator binding the variables $b$ and $x$, where $p$ is a proof and $t$ a term. 
Intuitively, such an operator is intended to reduce according to the rule:
$$\cofix{t}{bx}{p} ~\to~ p[t/x][\lambda y.\cofix{y}{bx}{p}/b]$$
This is to be compared with the usual inductive fixpoint operator which we write 
$\ind{t}{p_0}{bx}{p_S}$ (which binds the variables $b$ and $x$) and which reduces as follows:
$$\ind{0}{p_0}{bx}{p_S} ~\to~ p_0
\qquad\qquad\qquad
\ind{S(t)}{p_0}{bx}{p_S}~\to~p_S[t/x][\ind{t}{p_0}{bx}{p_S}/b]$$
The presence of coinductive fixpoints allows us to consider the term $\cofix{0}{bn}{(H n, b(S(n)))}$, 
which implements a stream eventually producing the (informal) infinite sequence $H_\infty$.
Indeed, this proof term reduces as follows:
$$\begin{array}{rcl}
   \cofix{0}{bn}{(H n, b(S(n)))} & \to &(H\,0,\cofix{1}{bn}{(H n, b(S(n)))})\\
   &\to &(H\,0,(H\,1,\cofix{2}{bn}{(H n, b(S(n))))})\\
   & \to & \dots 
  \end{array}$$
This allows for the following definition of a proof term for the axiom of countable choice:
$$
\begin{array}{rll}
 AC_{\N} & := & \lambda H.\letop a=\cofix{0}{bn}{(H n, b(S(n)))}
		 \inop\, (\lambda n.\wit(\nth~n~a),\lambda n.\prf(\nth~n~a))\, \\
 &:& \forall x^\N.\exists y^B. P(x,y) \imp \exists f^{\N\imp B}.\forall x^A. P(x,f(x))
\end{array}
$$
Whereas the construction $\letin{a}{\dots}\dots$ suggests a call-by-value discipline,
we cannot afford to pre-evaluate each component of the stream.
In turn, this imposes a \emph{lazy} call-by-value evaluation discipline
for coinductive objects.
However, this still might be responsible for some non-terminating reductions,
all the more as classical proofs may backtrack.

If we analyze what this construction does at the level of types\footnote{We delay the formal introduction of a type system 
and the given of the typing derivation for $AC_\N$ to \Cref{s:macros}.}, 
in first approximation it turns a proof ($H$) of the formula $\forall x^\N.A(x)$ (with $A(x)=\exists y.P(x,y)$ in that case) into
a proof (the stream $H_\infty$) of the (informal) infinite conjunction $A(0)\land A(1)\land A(2)\land \dots$.
Formally, a proof $\cofix{t}{bx}{p}$ is an inhabitant of a coinductive formula, written $\nu^t_{X x}A$ 
(where $t$ is a term and which binds the variables $X$ and $x$). The typing rule is given by:
$$\infer[\cofixrule]{\Gamma\vdash \cofix{t}{bx}{p}: \nu^t_{Xx} A }{
	\Gamma\vdash t:T & \Gamma, x:T,b:\forall y^T\!. X y\vdash p:A  }$$
with the side condition that $X$ can only occurs in positive position in $A$.  
Coinductive formulas are defined with a reduction rules which is very similar to the rule for the co-fixpoint:
$$
\nu^t_{Xx} A ~~~\typered~~~ A[t/x][\nu^t_{Xy} A/Xy]
$$
In particular, the term $\cofix{0}{bn}{(H n, b(S(n))}$ is thus an inhabitant of a coinductively defined (infinite) conjunction, 
written $\nu^0_{X n}(A(n)\wedge X(S(n)))$.
This formula indeed reduces accordingly to the reduction of the stream:
$$\begin{array}{rcl}
\nu^0_{X n}(A(n)\wedge X(S(n))) & \typered & A(0)\wedge [\nu^1_{X n}(A(n)\wedge X(S(n))) ]\\
& \typered & A(0)\wedge (A(1) \wedge [\nu^2_{X n}(A(n)\wedge X(S(n)))])\\
&\typered& \dots
\end{array}$$

More generally, at the level of formulas, the key is to identify the formula $A(x)$ and a suitable law $g:\N\to T$ 
to turn a proof of $\forall x^T\!.A(x)$ into the conjunction $A(g(0))\land A(g(1))\land A(g(2))\land \dots$.
In the case of the axiom of countable choice, this law is simply this identity.
As for the axiom of dependent choice, the law $g$ we are looking for is precisely the choice function.
We can thus use the same trick to define a proof term for the axiom of dependent choice:
$$DC ~~\defeq ~~ \forall x^T.\exists y^T. P(x,y)  \to  \forall x_0^T\! .\exists  f\in T^\N.( f(0) = x_0  \land \forall n^\N\!. P( f (n), f (S (n))))$$
The stream we actually construct corresponds again to a coinductive formula, defined here by
$\nu^{{x_0}}_{X n}[\exists y^\N\!.(P(x,y)\wedge X(y))]$ and which
ultimately unfolds into\footnote{As we shall explain afterwards, $\dest{p}{x,a}q$ is a non-dependent eliminator for the 
existential type, it destructs a proof $p$ of $\exists x.A(x)$ as $(x,a)$ in $q$ and reduces as follows:
$\dest{(t,p)}{x,a}q \to q[t/x,p/a]$.}:
$$\begin{array}{r@{~~}c@{~~}l}
\nu^{x_0}_{X n}[\exists y.(P(x,y)\wedge X(y))]&\typered& \exists x_1^\N\!.(P(x_0,x_1)\wedge \nu^{x_1}_{X n}[\exists y.(P(x,y)\wedge X(y))])  \\[0.2em]
& \typered & \exists x_1^\N\!.(P(x_0,x_1)\wedge \exists x_2^\N\!.(P(x_1,x_2) \land \nu^{x_2}_{X n}[\exists y.(P(x,y)\wedge X(y))])))\\
& \typered & \dots
\end{array}$$
Given a proof $H:\forall x.\exists y.P(x,y)$ and a term $x_0$ , we can define a stream corresponding to this coinductive 
formula by $\bm{str}\,x_0:=\cofix{x_0}{bn}{\dest{H\,n}{(y,c)}(y,(c,(b\,y))}$. This term reduces as expected:
$$(x_0,\bm{str}\,x_0)~\red~(x_0,(x_1,(p_1,\bm{str}\,x_1)))~ \red~(x_0, (x_1,(p_1,(x_2,(p_2,\bm{str}\,x_2)))))~\red~\dots $$
where $p_i:P(x_{i-1},x_i)$. 
From there, it is almost direct to extract the choice function $f$ (which maps any $n\in\N$ to $x_n$) and the corresponding certificate that 
$( f(0) = x_0 \land \forall n\in\N. P( f (n), f (S(n))))$. In practice, it essentially amounts to define the adequate $\nth$ function.
We will give a complete definition of the proof term for the axiom of dependent choice in \Cref{thm:dc}.

\subsection{An overview of \dpaw}\label{s:dpaw}
Formally, the calculus {\dpaw} is a proof system for the language of classical arithmetic in finites types (abbreviated PA$^\omega$), 
where the `d' stands for ``dependent''. 
It adopts a stratified presentation of dependent types, by syntactically distinguishing \emph{terms}---
that represent mathematical objects---from \emph{proof terms}---that represent mathematical proofs. 
In other words, we syntactically separate the categories corresponding to witnesses and proofs in dependent
sum types.
Finite types and formulas are thus separated as well, corresponding to the following syntax:
$$\begin{array}{>{}l@{~~\quad}r@{~~}c@{~~}l}
\text{\bf Types}        &T,U & ::= & \N \mid T\to U \\
\text{\bf Formulas    } &A,B  &::= &\top\mid \bot \mid t = u\mid A\land B \mid A\lor B \mid \exists x^T.A \mid \forall x^T. A \mid \dptprod{a:A}{B} \mid \nu^t_{x,f}A
\end{array}$$
Terms, denoted by $t,u,...$ are meant to represent arithmetical objects, their syntax thus includes:
\begin{itemize}
\item a term $0$ and a successor $S$;
\item an operator $\rec{t}{xy}{t_0}{t_S}$ for recursion, which binds the variables $x$ and $y$: where $t$ is the term on which the recursion is performed,
$t_0$ is the term for the case $t=0$ ans $t_S$ is the term for case $t=S(t')$;
\item $\lambda$-abstraction $\lambda x.t$ to define functions and terms application $t\,u$;
\item a $\wit$ constructor to extract the witness of a dependent sum.
\end{itemize}
As for proofs, denoted by $p,q,...$, they contain:
\begin{itemize}
 \item a proof term $\refl$ which is the proof of atomic equalities $t=t$;
 \item $\subst{p}{q}$ which eliminates an equality proof $p:t=u$ to get a proof of $B[u]$ from a proof $q:B[t]$;
 \item pairs $(p,q)$ to prove logical conjunctions and destructors of pairs $\split{p}{a_1,a_2}{q}$ (which binds the variables $a_1$ and $a_2$ in $q$);
 \item injections $\injec{i}{p}$ for the logical disjunction and pattern-matching $\case{p}{a_1.p_1}{a_2.p_2}$ which binds the variables $a_1$ in $p_1$ and $a_2$ in $p_2$;
 \item pairs $(t,p)$ where $t$ is a term and $p$ a proof for the dependent sum type;
 \item $\prf p$ which allows us to extract the certificate of a dependent pair;
 \item non-dependent destructors $\dest{p}{x,a}q$ which binds the variables $x$ and $a$ in $q$;
 \item abstractions over terms $\lambda x.p$ for the universal quantification and applications $p\,t$;
 \item (possibly) dependent abstractions over proofs $\lambda a.p$ and applications $p\,q$;
 \item a construction ${\letin{a}{p}q}$, which binds the variable $a$ in $q$ and which allows for sharing;
 \item operators $\ind{t}{p_0}{ax}{p_S}$ and $\cofix{t}{bx}{p}$ that we already described for inductive and coinductive reasoning;
 \item control operators ${\catch{\alpha} p}$ (which binds the variable $\alpha$ in $p$) and ${\throw{\alpha}\,p}$ (where $\alpha$ is a variable and $p$ a proof)
 \item $\exf p$ where $p$ is intended to be a proof of false.
\end{itemize}
This results in the following syntax:
$$\begin{array}{l@{\qquad}rcl}
\text{\bf Terms}  & t,u & ::= & x \mid 0 \mid S(t) \mid \rec{t}{xy}{t_0}{t_S}\mid \lambda x.t \mid t~u \mid \wit p \\[0.1em]
\text{\bf Proofs} & p,q & ::= & a\mid \refl \mid \subst{p}{q} \mid  \injec{i}{p} \mid  {\case{p}{a_1.p_1}{a_2.p_2}}\\[0.1em]
		  &     & \mid& (p,q) \mid {\split{p}{a_1,a_2}q}\\[0.1em]
                  &	& \mid& (t,p) \mid \prf p \mid  {\dest{p}{x,a}q} \mid \lambda x.p\mid p\,t \\[0.1em]
                  &	& \mid& \lambda a.p \mid p\,q  \mid {\letin{a}{p}q}  \mid \ind{t}{p_0}{ax}{p_S} \mid \cofix{t}{bx}{p}  \\[0.1em]
                  &	& \mid& {\exf{p}}  \mid {\catch{\alpha}p} \mid {\throw{\alpha}p}
\end{array}$$ 

The problem of degeneracy caused by the conjoint presence of classical proofs and dependent types is solved by 
enforcing a {compartmentalization} between them.
Dependent types are restricted to the set of \emph{negative-elimination-free} proofs (\nef),
which are a
generalization of values preventing backtracking evaluations from occuring by 
excluding expressions of the form $p\,q$, $p\,t$, ${\exf{p}}$, ${\catch{\alpha}p}$ or ${\throw{\alpha}p}$
which are outside the body of a $\lambda x$ or $\lambda a$. 
Syntactically, they are defined by:
$$\begin{array}{l@{\quad~}r@{~}c@{~}l}
\text{\bf Values} & V_1,V_2& ::= & a \mid \injec{i}{V} \mid  (V_1,V_2) \mid (t,V) \mid \lambda x.p \mid  \lambda a.p\mid \refl\\
\text{\textbf{\scriptsize{NEF}}} & N_1,N_2 & ::= & a  \mid \refl \mid \subst{N_1}{N_2} \mid  \injec{i}{N} \mid  {\case{p}{a_1.N_1}{a_2.N_2}} \\
		      &	& \mid& (N_1,N_2) \mid {\split{N_1}{a_1,a_2}N_2}\\
                      &	& \mid& (t,N) \mid \prf N \mid  {\dest{N_1}{x,a}N_2} \mid \lambda x.p \\
                      &	& \mid& \lambda a.p \mid {\letin{a}{N_1}N_2} \mid \ind{t}{N_0}{ax}{N_S} \mid \cofix{t}{bx}{N}\\
\end{array}$$ 
This allows us to restrict typing rules involving dependencies, notably the rules for $\prf$ or $\letin{a}{\dots}\dots$:
$$
\infer[\prfrule]{\Gamma \vdash \prf p : A(\wit p) }{\Gamma \vdash p: \exists x^T. A(x)  & p \in\text{\nef}}
\qquad
\infer[\cutrule]{\Gamma\vdash \letin{a}{p}{q}:B[p/a]}{\Gamma \vdash p:A & \Gamma,a:A \vdash q:B & a\notin\FV(B)~\text{if }p\notin\nef}
$$

About reductions, let us simply highlight the fact that they globally follow a call-by-value discipline, for instance in this sample:
$$\begin{array}{lcl}
(\lambda a.p)\,q 	& \red & \letin{a}{q}p \\
\letin{a}{(p_1,p_2)}p   & \red & \letin{a_1}{p_1}\letin{a_2}{p_2}p[(a_1,a_2)/a] \\
\letin{a}{V}p      	& \red & p[V/a] \\
  \end{array}
  $$
except for co-fixpoints which are lazily evaluated:
$$\begin{array}{lcl}
F[\letin{a}{\cofix{t}{bx}q}p]	& \red & \letin{a}{\cofix{t}{bx}q}F[p] \\
\letin{a}{\cofix{t}{bx}q}D[a]	& \red & \letin{a}{q[\lambda y.\cofix{y}{bx}q/b][t/x]}D[a] \\
\end{array}
  $$
In the previous rules, the first one expresses the fact that evaluation of co-fixpoints under contexts $F[\,]$ are momentarily delayed.
The second rules precisely corresponds to a context where the co-fixpoint is linked to a variable $a$ whose value is needed, 
a step of unfolding is then performed. 
The full type system, as well as the complete set of  reduction rules, are given in~\cite{Herbelin12}, and will be restated with a different presentation
in~\Cref{s:dlpaw}. 
In the same paper, some important properties of the calculus are given. 
In particular, {\dpaw} verifies the property of subject reduction, and provided it is normalizing, there is no proof of false. 

\begin{theorem}[Subject reduction]
 If ~$\Gamma \vdash p:A$ and $p\red q$, then $\Gamma \vdash q:A$.
\end{theorem}
\begin{proof}[Proof (sketch)]
By induction on the derivation of $p \to q$, see~\cite{Herbelin12}. 
\end{proof}

\begin{theorem}[Conservativity]
 Provided {\dpaw} is normalizing, if $A$ is $\imp$-$\nu$-$\wit$-$\forall$-free, and $\vdash_{\dpaw} p: A$,
 there is a value $V$ such that $\vdash_{\text{HA}^\omega} V:A$.
\end{theorem}
\begin{proof}[Proof (sketch)]
 Considering a closed proof $p$ of $A$, $p$ can be reduced. 
 By analysis of the different possible cases, it can be found a closed value of type $A$.
 Then using the fact that $A$ is a $\imp$-$\nu$-$\wit$-$\forall$-free formula, $V$ does
 not contain any subexpression of the form $\lambda x.p$ or $\lambda a.p$, 
 by extension it does not contain either any occurrence of ${\exf{p}}$, ${\catch{\alpha}p}$ or ${\throw{\alpha}p}$
 and is thus a proof of $A$ already in $\text{HA}^\omega$.
\end{proof}

\begin{theorem}[Consistency]
 Provided {\dpaw} is normalizing, it is consistent, that is: $\nvdash_{dPA^\omega}p: \bot$.
\end{theorem}
\begin{proof}
The formula $\bot$ is a particular case of $\imp$-$\nu$-$\wit$-$\forall$-free formula,
thus the existence of a proof of false in {\dpaw} would imply the existence of a contradiction already in $HA^\omega$, which is absurd.
\end{proof}

The last two results rely on the property of normalization. 
Unfortunately, the proof sketch that is given in~\cite{Herbelin12} to support the claim
that {\dpaw} normalizes turns out to be hard to formalize properly. 
Since, moreover, {\dpaw} contains both control operators (allowing for backtrack)
and co-fixpoints (allowing infinite objects, like streams), which can be combined and interleaved,
we should be very suspicious \emph{a priori} about this property.
Anyhow, the proof sketch from~\cite{Herbelin12} relies on metamathematical arguments while
we are rather interested in a fine analysis of the interactions between the different computational features of \dpaw.
Indeed, our goal is not limited to adding an axiom to the theory or to finding a way to realize it.
In turns, we are looking for an adapted calculus with an appropriate operational semantics:
such an approach mixing syntax and semantics let us hope to develop a better calculus (from a programming point of view)
and to get a better understanding of the proof arguments (from the point of view of logic).
The proposal of Herbelin is already in line with this philosophy: as we saw, the proofs terms for {\acn} and {\dc} 
in {\dpaw} are build using usual programming primitives rather than one monolithic extra-instruction.

\subsection{Toward a proof of normalization for \dpaw}
This paper recounts a long process devoted to the search for a proof of normalization for {\dpaw}
by means of a realizability interpretation or by a continuation-passing-style translation.
Aside from the very result of normalization, such an approach is of interest in itself in that, 
as advocated earlier, it requires and provides us with a fine understanding of the computational features of \dpaw.
In particular, the difficulty in obtaining a continuation-passing style translation directly for {\dpaw}
highlights several problems.
The first one is the simultaneous presence of control operators and of lazily evaluated streams, thus 
allowing to define programs that may backtrack and require the evaluation of of potentially infinit object.
program that 
The second is the presence of dependent types. Indeed, dependent types (and in particular the dependent sum) are
known to be incompatible with a continuation-passing style translation~\cite{BarUus02}. 
Last, the reduction system is defined in a natural-deduction style with contextual rules (as in the rule to reduce proofs of the shape $\letin{a}{\cofix{t}{bx}{p}}D[a]$)
where the contexts involved can be of arbitrary depth. 
This kind of rules are, in general and especially in this case, very difficult to translate faithfully through a continuation-passing style translation.

All in all, there are several difficulties in getting a direct proof of normalization for \dpaw.
Hence, we shall study them separately before combining the solutions to each subproblem in order
to attack the main problem.
Roughly, our strategy consists of two steps:
\begin{enumerate}
 \item reduce {\dpaw} to an equivalent presentation in a sequent calculus fashion,
 \item use the methodology of semantic artifacts to define a CPS or a realizability interpretation.
\end{enumerate}
Defining a sequent calculus presentation of a calculus is indeed known to be
a good intermediate step for the definition of continuation-passing style
translations~\cite{Munch13PhD,MunSch15,AriDowMauJon16}. 
In our case it forces to us to rephrase the contextual rules into 
abstract-machine like reduction rules which only depend of the top-level command,
thus solving the last difficulty evoked above.
This presentation should of course verify at least the property of subject reduction
and its reduction system should mimic the one of {\dpaw}. 
Schematically, this corresponds to the following roadmap where question marks indicate what is to be done:
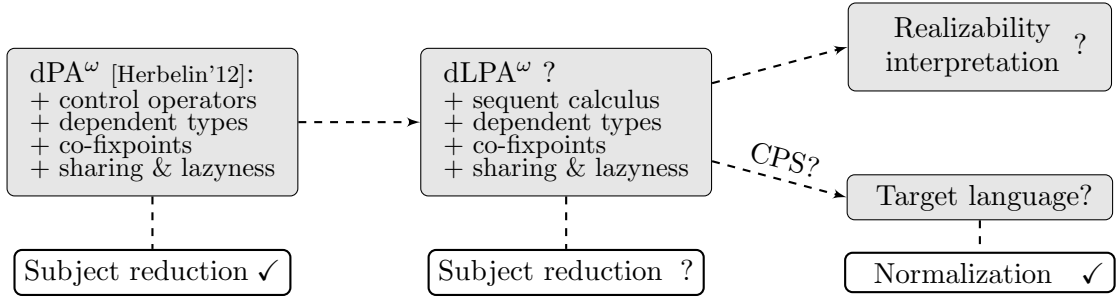
\begin{center}
 \tikzstyle{base}   = [rounded corners=0.1cm]
\tikzstyle{format} = [draw,minimum width=3.5cm, thin,rounded corners=0.1cm]
\tikzstyle{medium} = [ellipse, draw, thin, minimum height=2.5em]
\begin{tikzpicture}[node distance=4cm, auto,>=latex', thick]
     \draw (0,0) node[format,fill=black!10] (ch) {\begin{tabular}{l}
										  $\dpaw ~{\footnotesize \text{[Herbelin'12]}}$:\\[-0.3em]
										  \small + control operators\\[-0.4em]
										  \small + dependent types\\[-0.4em]
										  \small + co-fixpoints    \\[-0.4em]
										  \small + sharing \& lazyness
										  \end{tabular}};
     \draw(ch)+(5.5,0) node[format, fill=black!10] (cs) {\begin{tabular}{l}
										  $\dlpaw$ ?\\[-0.3em]
										  \small + sequent calculus \\[-0.4em]
										  \small + dependent types\\[-0.4em]
										  \small + co-fixpoints    \\[-0.4em]
										  \small + sharing \& lazyness\qquad
										  \end{tabular}};        										  
      \draw (cs)+(5.5,1) node[format, fill=black!10] (real) {\begin{tabular}{c}Realizability\\ interpretation\end{tabular}?};
      \draw (cs)+(5.5,-1) node[format, fill=black!10] (lc) {~~Target language?~~};
	\draw (ch)+(0,-2) node[base,draw,fill=black!00] (rs) { Subject reduction\;\checkmark};
	\draw (cs)+(0,-2) node[base,draw,fill=black!00] (rss) {Subject reduction\; ?};
	\draw (lc)+(0,-1.) node[base,draw,fill=black!00] (nm) {~~Normalization~~\;\checkmark};
	\path[dashed]  (ch) edge (rs);
	\path[dashed]  (cs) edge (rss);
	\path[dashed]  (lc) edge (nm);
	\path[dashed,->] (cs) edge node[sloped, anchor=center,above] {CPS?}(lc.west);
	\path[dashed,->] (ch) edge node[base,fill=red!00] {}(cs);
	\path[dashed,->] (cs) edge node[base,fill=red!00] {}(real.west);
\end{tikzpicture}

\end{center}
To be fair, this approach is idealistic. 
In particular, we will not formally define an embedding for the first arrow,
since we are not interested in {\dpaw} for itself, but rather in 
the computational content of the proofs for countable and dependent choice.
Hence, we will content ourself with a sequent calculus presentation of {\dpaw} which
allows for similar proof terms, which we call {\dlpaw},
without bothering to prove that the reduction systems are strictly equivalent (see \Cref{s:macros}).
{As for the second arrow, as we shall explain in the next sections,
the search for a continuation-passing style translation or a realizability interpretation
can coincide for a large part. We shall thus apply the methodology of semantic artifacts
and in the end, choose the easiest possibility (in that case, defining a realizability interpretation).}

From this roadmap actually arises two different subproblems that are
already of interest in themselves. Forgetting about the general context of \dpaw, 
we shall first wonder whether these easier questions have an answer:
\begin{enumerate}
 \item Can we prove the normalization of a call-by-need calculus with control operators? 
 Can we define a Krivine realizability interpretation of such a calculus?
 \item Is it possible to define a (classical) sequent calculus with a form of dependent types? 
 If so, would it be compatible with a typed continuation-passing style translation?
\end{enumerate}
We shall treat the first question in \Cref{s:lbvtstar} and the second one in \Cref{s:dl},
before combining the ideas of both systems to define \dlpaw~in \Cref{s:dlpaw}.
As explained in \Cref{fig:outline}, in each case we will define
variants of Curien-Herbelin's \lmmt-calculus, so that we shall begin with presenting the latter.
We will take advantage of this introduction to illustrate our global methodology (that is, refining the system
into an abstract-machine like calculus in order to get a continuation-passing style translation 
or a realizability interpretation that allows us t
o prove normalization and soundness).

\section{The \lmmt-calculus and Danvy's semantics artifacts}
\label{s:lmmt}
\subsection{Continuation-passing style translation and Danvy's methodology}
The terminology of \emph{continuation-passing style} (CPS) was first introduced 
in 1975 by Sussman and Steele in a technical report about the Scheme 
programming language~\cite{SusSte75}. 
In this report, after giving the usual recursive definition of the factorial,
they explained how the same computation could be driven differently:
\begin{quote}
 \em ``It is always possible, if we are willing to specify explicitly 
 what to do with the answer, to perform any calculation in this way: 
 rather than reducing to its value, it reduces to an application of a 
 continuation to its value. That is, in this continuation-passing programming
 style, a function always ``returns'' its result by ``sending'' it 
 to another function. This is the key idea.''
\end{quote}
Interestingly, by making explicit the order in which reduction steps 
are computed, continuation-passing style translations indirectly specify
an operational semantics for the translated calculus.
In particular, different evaluation strategies for a calculus correspond 
to different continuation-passing style translations. 
This was for instance studied by Plotkin for the call-by-name and call-by-value 
strategies within the $\lambda$-calculus \cite{Plotkin75}.

Continuations and their computational benefits have been deeply studied 
since then, and there exists a wide literature on continuation-passing 
style translations.
Among other things, these translations have been used to ease the definitions 
of compilers~\cite{Appel92,FelSab99}, one of their interests being that they 
make explicit the flow of control.
As such, continuation-passing style translations \emph{de facto} provide us 
with an operational semantics for control operators,
as observed in~\cite{FelEtAl87} for the $\mathcal{C}$ operator.
Continuation-passing style translations therefore bring 
an indirect computational interpretation of classical logic. 
This observation can be strengthened on a purely logical aspect 
by considering the logical translations they induce at the level
of types: 
the translation of types through a CPS mostly amounts to a negative translation
allowing to embed classical logic into intuitionistic
logic~\cite{Griffin90,Murthy90}.

In addition to the operational semantics, continuation-passing style translations allow to benefit from properties already proved 
for the target calculus. 
For instance, we will see how to define a translation $p\mapsto \tr{p}$ from the simply-typed 
call-by-name \lmmt-calculus to the simply-typed $\lambda$-calculus along which the properties of normalization and soundness can be transfered.
In details, these translations will preserve reduction, 
in that a reduction step in the source language gives rise to a step (or more) in the target language:
\begin{equation}
\label{eq:cps:1}
 c \oset{1}{\longrightarrow} c' \qquad\Rightarrow\qquad \tr{c}\bredn{+} \tr{c'}
\end{equation}
We say that a translation is \emph{typed} when it comes with  a translation $A\mapsto \tr{A}$ 
from types of the source language to types of the target language,
such that a typed proof in the source language is translated into a typed proof of the target language: 
\begin{equation}
\label{eq:cps:2}
 \Gamma \vdash p : A \mid \Delta \qquad\Rightarrow\qquad \tr{\Gamma}, \tr{\Delta}\vdash \tr{p}:\tr{A}
\end{equation}
Last, we are interested in translations mapping the type $\bot$ into a type $\tr{\bot}$ which is not inhabited:
\begin{equation}
\label{eq:cps:3}
 \nvdash p : \tr{\bot}
\end{equation}
Assuming that the previous properties hold, one automatically gets:
\begin{proposition}[Benefits of the translation]
\label{thm:benefits_cps}
If the target language of the translation is sound and normalizing, and if besides 
the equations (\ref{eq:cps:1}), (\ref{eq:cps:2}) and (\ref{eq:cps:3}) hold, then:
\begin{enumerate}
 \item If $\tr{p}$ normalizes, then $p$ normalizes
 \item If~ $p$ is typed, then $p$ normalizes
 \item The source language is sound, \emph{i.e.} there is no proof ~$\vdash p:\bot$
\end{enumerate} 
\end{proposition}
\begin{proof}
 \begin{enumerate}
  \item By contrapositive, if $p$ does not normalizes, then according to equation (\ref{eq:cps:1}) neither does $\tr{p}$.
  \item If $p$ is typed, then $\tr{p}$ is also typed by(\ref{eq:cps:2}), and thus normalizes. Using the first item, $p$ normalizes.
  \item By \emph{reductio ad absurdum}, direct consequence of (\ref{eq:cps:3}).\qedhere
 \end{enumerate}
\end{proof}

Continuation-passing style translations are thus a powerful tool both on the computational and the logical facets 
of the proofs-as-programs correspondence.
We shall illustrate afterwards its use to prove normalization and soundness of the call-by-name {\lmmt}-calculus.
Rather than giving directly the appropriate definitions, 
we would like to insist on a convenient methodology to obtain 
CPS translations as well as realizability interpretations (which are deeply connected).
This methodology is directly inspired from Danvy \emph{et al.} 
method to derive hygienic semantics artifacts for a call-by-need calculus \cite{DanEtAl10}.
Reframed in our setting, it essentially consists in the successive definitions of:
\begin{enumerate}
 \item[(1)\,] an operational semantics,
 \item[(2)\,] a small-step calculus or abstract machine,
 \item[(3)\,] a continuation-passing style translation,
 \item[(3')] a realizability model.
\end{enumerate}
The first step is nothing more than the usual definition of a reduction system. 
The second step consists in refining the reduction system to obtain small-step reduction rules (as opposed to big-step ones),
that are finer-grained reduction steps. These steps should be as atomic as possible, and in particular, 
they should correspond to an abstract machine in which the sole analysis of the term (or the context)
should determine the reduction to perform.
Such a machine is called in \emph{context-free form}~\cite{DanEtAl10}. 
If so, the definition of a CPS translation is almost straightforward, as well as the realizability interpretation.
Let us now illustrate this methodology on the call-by-name {\lmmt}-calculi.

\subsection{A short primer to the \lmmt-calculus}
\label{ss:mumutilde}
We recall here the spirit of the \lmmt-calculus, for further details and references please refer to the original article~\cite{CurHer00}.
The key notion of the \lmmt-calculus is the notion of \emph{command}.
A command $\cmd{p}{e}$ can be understood as a state of an abstract machine, representing the evaluation of a \emph{proof} $p$ (the program) 
against a co-proof $e$ (the stack) that we call \emph{context}.
The syntax and reduction rules (parameterized over a subset of proofs ${\mathcal{V}}$ and a subset of evaluation contexts~${\mathcal E}$) are given in
Figure~\ref{fig:lmmt}, where $\mut a.c$ can be read as a context $\letin{a}{[~]}{c}$.
The $\mu$ operator comes from Parigot's $\lambda\mu$-calculus~\cite{Parigot97}, $\mu\alpha$ binds a context to a context variable $\alpha$
in the same way that $\mut a$ binds a proof to some proof variable $a$.

The \lmmt-calculus can be seen as a proof-as-program correspondence between sequent calculus and abstract machines. 
Right introduction rules correspond to typing rules for proofs, while left introduction are seen as typing rules for evaluation contexts.
In contrast with Gentzen's original presentation of sequent calculus,
the type system of the \lmmt-calculus explicitly identifies at any time which formula is being worked on.
In a nutshell, this presentation distinguishes between three kinds of sequents:
\begin{enumerate}
 \item sequents of the form $\Gamma \vdash p: A \mid \Delta$ for typing proofs, where the focus is put on the (right) formula $A$;
 \item sequents of the form $\Gamma \mid e:A \vdash \Delta$ for typing contexts, where the focus is put on the (left) formula $A$;
 \item sequents of the form $c : (\Gamma \vdash \Delta)$ for typing commands, where no focus is set.
\end{enumerate}
In a right (resp. left) sequent $\Gamma\vdash p:A\mid \Delta$, 
the singled out formula\footnote{This formula is often referred to as the formula in the \emph{stoup}, a terminology due to Girard.}
$A$ reads as the conclusion \emph{``where the proof shall continue''} (resp. hypothesis \emph{``where it happened before''}).    

\begin{figure}[t]
\framebox{\vbox{
\input{figures/lmmt}
}}
\caption{The $\lambda\mu\tilde\mu$-calculus}
\label{fig:lmmt}
\end{figure}

For example, the left introduction rule of implication can be seen as a typing rule 
for pushing an element $q$ on a stack $e$ leading to the new stack $q\cdot e$:
$$\infer[\imp_l]{\Gamma \mid  {q\cdot e} : A\imp B\vdash \Delta}
    {\Gamma\vdash {q}:{A} \mid \Delta & \Gamma \mid {e}:B\vdash \Delta}$$

As for the reduction rules, we can see that there is a critical pair if ${\mathcal V}$ and ${\mathcal E}$ are not restricted enough:
$$
  \Sub{c}{\alpha}{\tmu x.c'}   \quad\longleftarrow\quad 
  \cut{\mu\alpha.c}{\tmu x.c'} \quad\longrightarrow\quad
  \Sub{c'}{x}{\mu\alpha.c}.\\
$$
The difference between call-by-name and call-by-value can be characterized by how 
this critical pair\footnote{Observe that this critical pair can be also interpreted in terms of non-determinism. 
Indeed, we can define a fork instruction by 
$\Fork \defeq \lambda ab.\mu\alpha.\cut{\mu\_\cut{a}{\alpha}}{\tmu\_.\cut{b}{\alpha}}$, which verifies indeed that
$\cut{\Fork}{p_0\cdot p_1\cdot e}\rightarrow \cut{p_0}{e}$
  and
  $\cut{\Fork}{p_0\cdot p_1\cdot e}\rightarrow \cut{p_1}{e}$.}
is solved,
by defining ${\mathcal V}$ and ${\mathcal E}$ such that the two rules do not overlap.
Defining the subcategories of values $V\subset p$ and co-values $E\subset e$ by:
$$ \text{(Values)}\qquad V::= a \mid \lambda a.p \qquad \qquad\qquad\qquad \text{(Co-values)}\qquad  E::= \alpha \mid q\cdot e$$
the call-by-name evaluation strategy amounts to the case where ${\mathcal V}\defeq$~\emph{Proofs} and ${\mathcal E}\defeq$~\emph{Co-values}, while
  call-by-value corresponds to ${\mathcal V}\defeq$~\emph{Values} and ${\mathcal E}\defeq$~\emph{Contexts}.
  Both strategies can also be characterized through different CPS translations~\cite[Section 8]{CurHer00}.

\begin{remark}[Application]
\label{rmk:application}
The reader unfamiliar with the \lmmt-calculus might be puzzled by the absence of a syntactic construction for the application of proof terms.
Intuitively, the usual application $p\, q$ of the $\lambda$-calculus is recovered
by considering the command it reduces to in an abstract machine, that is, given an evaluation context $e$, the application of the proof $p$ 
to the stack of the shape $q\cdot e$\footnote{To pursue the analogy with the $\lambda$-calculus,
the rest of the stack $e$ can be viewed as a context $C_e[\,]$ surrounding the application $p\, q$, 
the command $\cmd{p}{q\cdot e}$ thus being identified with the term $C_e[p\,q]$. Similarly,
the whole stack can be seen as the context $C_{q\cdot e}[\,] = C_e[[\,]q]$, whence the terminology.}:
$$\cmd{p\,q}{e} ~~\to~~ \cmd{p}{q\cdot e}$$
The usual application can thus be obtained by solving the previous equation, that is through the following shorthand:
$$p\,q \defeq \mu\alpha.\cmd{p}{q\cdot\alpha}$$
\end{remark}

Finally, it is worth noting that the $\mu$ binder is a \emph{control operator},
since it allows for catching evaluation contexts and backtracking further in the execution.
This is the key ingredient that makes the \lmmt-calculus a proof system for classical logic.
To illustrate this, let us draw the analogy with the \texttt{call/cc}
operator of Krivine's $\lambda_c$-calculus~\cite{Krivine09}.
Let us define the following proof terms:
$$ \mathtt{call/cc}\defeq \lambda a.\mu\alpha.\cut{a}{{\bm k}_\alpha\cdot \alpha} \qquad \qquad
{\bm k}_e \defeq \lambda a'.\mu\beta.\cut{a'}{e}$$
The proof ${\bm k}_e$ can be understood as a proof term where the context $e$ has been encapsulated.
As expected, \texttt{call/cc} is a proof for Peirce's law (see Figure~\ref{fig:peirce}), which
is known to imply other forms of classical reasoning (\emph{e.g.},
the law of excluded middle, the double negation elimination).

Let us observe the behavior of \texttt{call/cc} (in call-by-name evaluation strategy, as in Krivine $\lambda_c$-calculus):
in front of a context of the shape $q\cdot e$ with $e$ of type $A$,
it will catch the context $e$ thanks to the $\mu\alpha$ binder and reduce as follows:
$$\cut{\lambda a.\mu\alpha.\cut{a}{{\bm k}_\alpha\cdot \alpha}}{q\cdot e}
\quad\!\to\quad\!\!
\cut{q}{\tmu a.\cut{\mu\alpha.\cut{a}{{\bm k}_\alpha\cdot \alpha}}{ e}}
\quad\!\!\to\quad\!\!
\cut{\mu\alpha.\cut{q}{{\bm k}_\alpha\cdot \alpha}}{ e} 
\quad\!\!\to\quad\!\!
\cut{q}{{\bm k}_e\cdot e}$$
We notice that the proof term ${\bm k}_e = \lambda a'.\mu\beta.\cut{a'}{e}$ on top of the stack 
(which, if $e$ was of type $A$, is of type $A\to B$, see Figure~\ref{fig:peirce}) contains a second binder $\mu\beta$.
In front of a stack $q'\cdot e'$, this binder will now catch the context $e'$ and replace it by the former context $e$:
$$
\cut{\lambda a'.\mu\beta.\cut{a'}{e}}{q'\cdot e'}
\quad\!\to\quad\!
\cut{q'}{\tmu a'.\cut{\mu\beta.\cut{a'}{e}}{e'}}
\quad\!\to\quad\!
\cut{\mu\beta.\cut{q'}{e}}{e'}
\quad\!\to\quad\!
\cut{q'}{e}
$$
This computational behavior corresponds exactly to the usual reduction rule for \texttt{call/cc} in 
the Krivine machine~\cite{Krivine09}:
$$\begin{array}{r@{~}c@{~}lcr@{~}c@{~}l}
    \mathtt{call/cc}&\star&t\cdot\pi &\succ & t&\star&{\bm k}_{\pi}\cdot\pi \\
    {\bm k}_{\pi}&\star&t\cdot\pi' &\succ& t&\star&\pi \\
  \end{array}$$

\begin{figure}[t]
 \framebox{\vbox{
 \scalebox{0.9}{
$$
\infer[\!\imprrule]{\vdash \lambda a.\mu\alpha.\cut{a}{\lambda a'.\mu\beta.\cut{a'}{\alpha}\cdot \alpha}:((A\to B)\to A)\to A\mid}{
  \infer[\!\murule]{a:(A\to B)\to A \vdash \mu\alpha.\cut{a}{\lambda a'.\mu\beta.\cut{a'}{\alpha}\cdot \alpha}: A\mid}{
    \infer[\!\cutrule]{\cut{a}{\lambda a'.\mu\beta.\cut{a'}{\alpha}\cdot \alpha} :(a:(A\to B)\to A \vdash \alpha:A)}{
       \scalebox{0.85}{\infer[\!\!\axrrule]{a:(A\to B)\to A \vdash a : (A\to B)\to A \mid \bullet}{}}
      &
      \hspace{-0.8cm}\infer[\!\!\implrule]{\bullet\mid \lambda a'.\mu\beta.\cut{a'}{\alpha}\cdot \alpha :(A\to B)\to A \vdash \alpha:A}{
	\infer[\!\!\imprrule]{\bullet\vdash \lambda a'.\mu\beta.\cut{a'}{\alpha}:A\to B\mid \alpha:A}{
	  \infer[\!\murule]{\bullet,a':A \vdash \mu\beta.\cut{a'}{\alpha}:B \mid \alpha:A}{
 	    \infer[\!\cutrule]{ \cut{a'}{\alpha}: (\bullet,a':A \vdash \alpha:A,\beta:B)}{
	      \infer[\!\!\axrrule]{\bullet,a':A \vdash a':A\mid \bullet}{}
	      &
	      \infer[\!\!\axlrule]{\bullet \mid \alpha : A \vdash \alpha:A,\bullet}{}
	    }
	  }
	}
	&
	\hspace{-0.9cm}\scalebox{0.85}{\infer[\!\!\axlrule]{\mid \alpha : A \vdash \alpha:A}{}}
      }
    }
  }
}
$$
}
\vspace{-0.5em}
\begin{flushleft}
 \em \footnotesize (where $\bullet$ is used to shorten useless parts of typing contexts.)
\end{flushleft}

}}
\caption{Proof term for Peirce's law}
\label{fig:peirce}
\end{figure}

\subsection{Small-step abstract machine}
Following Danvy's methodology, the next step towards the definition of continuation-passing 
style translation consists in refining the reduction rules to obtain context-free reduction rules.
We recall here the (big-step) reduction rules of the call-by-name {\lmmt}-calculus,
where the $\mut$ operator gets the priority over the $\mu$ operator:
$$
\begin{array}{c@{\qquad \rightarrow \qquad}c}
  \cmd{p}{\mut a.c}		& c[p/a] \\
  \cmd{\mu \alpha.c}{E}		& c[E/\alpha] \\ 
  \cmd{\lambda a.p}{q\cdot e} 	& \cmd{q}{\mut a.\cmd{p}{e}}
\end{array}    
$$
As such, these rules define an abstract machine which is not in context-free since
to reduce a command one need to analyze simultaneously the term and the context.

\label{s:cbn:small_step}
To alleviate this ambiguity, we refine the reduction system into
small-step rules always specifying which part of the command is being analyzed.
If we examine the big-step rules, the only case where the knowledge of only one side suffices is
when the context is of the form $\mut a.c$, which has the absolute priority.
Therefore, we can start our analysis of a command by looking at its left-hand side.
If it is a $\tmu a.c$, we reduce it, otherwise, we can look at the right-hand side.
Now, if the term is of the shape $\mu \alpha.c$, it should be reduced, otherwise, 
we can analyze the left-hand side again. The only case left is when the context is a stack $q\cdot e$ 
and the term is a function $\lambda a.p$, in which case the command reduces.

The former case suggests two things: first, that the reduction should proceed by alternating 
examination of the left-hand and the right-hand side of commands. 
Second, that there is a descent in the syntax from the most general level (context $e$) to the most
specific one (values\footnote{Observe that values usually include variables, but here we rather consider
them in the category $p$. This is due to the fact that the operator $\tmu$ catches proofs at level $p$ 
and variables are hence intended to be substituted by proofs at this level.
Through the CPS, we will see that we actually need values to be considered at level $p$
as they are indeed substituted by proofs translated at this level.} $V$), passing by $p$ and $E$ in the middle:
$$
\begin{array}{l@{\hspace{1.2cm}}ccl@{\qquad\quad}}
    \text{\bf Terms   }  & p & ::= & \mu\alpha.c \mid a \mid V	\\
    \text{\bf Values  }  & V & ::= & \lambda a.p \\
\end{array}
\qquad\qquad
\begin{array}{l@{\hspace{1.2cm}}ccl@{\qquad\quad}}
    \text{\bf Contexts } & e & ::= & \tmu a.c \mid E \\
    \text{\bf Co-values} & E & ::= & \alpha\,  \mid p\cdot e	\\
  \end{array}
$$
So as to stick to this intuition, we denote commands with the level of syntax we are examining ($c_e,c_t,c_E,c_V$),
and define a new set of reduction rules which are of two kinds: computational steps, which reflect the former reduction steps,
and administrative steps, which organize the descent in the syntax. 
For each level in the syntax, we define one rule for each possible construction.
For instance, at level $e$, there is one rule if the context is of the shape $\tmu a.c$,
and one rule if it is of shape $E$.
This results in the following set of small-step reduction rules:
$$
\begin{array}{c@{\qquad \rightsquigarrow \qquad}c}
  \cmd{p}{\mut a.c}_e		& c_e[p/a] \\
  \cmd{p}{E}_e			& \cmd{p}{E}_p \\[0.3em]
  \cmd{\mu \alpha.c}{E}_p	& c_e[E/\alpha] \\ 
  \cmd{V}{E}_p			& \cmd{V}{E}_E \\[0.3em]
  \cmd{V}{q\cdot e}_E	        & \cmd{V}{q\cdot e}_V\\[0.3em]
  \cmd{\lambda a.p}{q\cdot e}_V	& \cmd{q}{\mut a.\cmd{p}{e}}_e
\end{array}    
\eqno\text{\begin{tabular}{c}\begin{tikzpicture}[color=black!40,scale=0.7]
 \node (0,0) (start) {};
 \path[->] (start) edge (0,-4.5);
 \path[-]  (-0.15,-0.75) edge (0.15,-0.75);
 \draw  (0.4,-0.75) node (e) {$e$};
 \path[-]  (-0.15,-2) edge (0.15,-2);
 \draw  (0.4,-2) node (p) {$p$};
 \path[-]  (-0.15,-3.2) edge (0.15,-3.2);
 \draw  (0.4,-3.2) node (E) {$E$};
 \path[-]  (-0.15,-4.1) edge (0.15,-4.1);
 \draw  (0.4,-4.1) node (V) {$V$};
\end{tikzpicture}\end{tabular}}
$$
where the last two rules could be compressed in one rule:
$$  \cmd{\lambda a.p}{q\cdot e}_E \rightsquigarrow \cmd{q}{\mut a.\cmd{p}{e}}_e$$
Note that there is no rule for variables and co-variables, since they block the reduction.
It is obvious that theses rules are indeed a decomposition of the previous ones,
in the sense that if $c,c'$ are two commands such that $c \oset{$1$}\red c'$, 
then there exists $n>1$ such that $c\oset{$n$}\rightsquigarrow c'$.

The previous subdivision of the syntax and reductions also suggests a fine-grained type system, 
where sequents are annotated with the adequate syntactic categories (see \Cref{fig:cbn_types}).
While this does not bring any benefit when building typing derivations (when collapsed at level $e$ and $p$,
this type system is exactly the original one), it has the advantage of splitting the
rules in more atomic ones which are closer from the reduction system.
Hence it will be easier to prove that the CPS translation is typed using these rules as induction bricks.

\begin{figure}[t]
\myfig{ 
 $$
 \begin{array}{c}
\infer[\!\scriptstyle (V)]{\Gamma \vdash_p V: A\mid \Delta}{\Gamma \vdash_V V: A\mid \Delta}
\qquad\qquad
\infer[\!\axrrule]{\Gamma \vdash_p a: A\mid \Delta}
      {(a:A)\in\Gamma }
\qquad\qquad
\infer[\!\murule]{\Gamma \vdash_p \mu\alpha.c: A\mid\Delta}
      {c : (\Gamma \vdash_c \Delta, \alpha: A )} 
\\[0.8em]
\infer[\!\imprrule]{\Gamma \vdash_V \lambda a.p: A \imp B \mid \Delta }
      {\Gamma, a:A \vdash_p p:B \mid \Delta}      
\qquad\qquad
\infer[\!\!\scriptstyle (E)]{\Gamma \mid E: A\vdash_e \Delta}{\Gamma \mid E: A\vdash_E \Delta}
\qquad\qquad
\infer[\!\!\mutrule]{\Gamma\mid \tmu a.c : A \vdash_e \Delta}
      {c : (\Gamma, a:A \vdash_c \Delta)}
\\[0.8em]

\infer[\!\!\axlrule]{\Gamma \mid \alpha:A\vdash_E \Delta}
      {(\alpha:A )\in\Delta}
\qquad\qquad\qquad
\infer[\!\!\implrule]{\Gamma \mid p \cdot e:A\to B \vdash_E \Delta}
      {\Gamma \vdash_p p:A \mid \Delta & \Gamma\mid e: B \vdash_e \Delta}
\end{array}      
$$
}
\caption{Refined call-by-name type system}
\label{fig:cbn_types}
\end{figure}

\subsection{Continuation-passing style translation}
\label{s:lmmt:cbn_cps} 
\subsubsection{Translation of terms}
Once we have an abstract-machine in context-free form at hands, the corresponding continuation-passing style translation
is straightforward. It suffices to start from the higher level in the descent (here $e$) and to define a translation
for each level which, for each element of the syntax, simply describe the corresponding small-step rule.
In the current case, this leads to the following definition:
$$
\text{\begin{tabular}{c}\begin{tikzpicture}[color=black!40,scale=0.7]
 \node (0,0) (start) {};
 \path (start) edge (0,-2.2);
 \path[->,dashed] (0,-2.2) edge (0,-3);
 \path[-]  (-0.15,-0.75) edge (0.15,-0.75);
 \draw  (0.4,-0.75) node (e) {$e$};
 \path[-]  (-0.15,-2.4) edge (0.15,-2.4);
 \draw  (0.4,-2.4) node (p) {$p$};
 \end{tikzpicture}\end{tabular}}\qquad
\begin{array}{l@{~\defeq~}l}
  \tr{\mut a.c}_e	\,p	& (\lambda a. \tr{c}_c)\,p \\
  \tr{E}_e		\,p	& p\, \tr{E}_E \\[0.3em]
  \tr{\mu \alpha.c}_p	\,E	& (\lambda \alpha.\tr{c}_c)\,E\\ 
  \tr{a}_p			& a \\
  \end{array}
  \qquad\qquad\quad
  \begin{array}{l@{~\defeq~}l}
  \tr{V}_p		\,E	& E \,\trV{V} \\[0.3em]
  \tr{q\cdot e}_E	\,V	& V\,\tr{q}_p\,\tr{e}_e \\
  \tr{\alpha}_E			& \alpha \\[0.3em]
  \tr{\lambda a.p}_V	\,q\,e	& (\lambda a.e\,\tr{p}_p)\,q \\[0.2em]
  \end{array}\qquad
  \text{\begin{tabular}{c}\begin{tikzpicture}[color=black!40,scale=0.7]
 \path[dashed] (0,-1.8) edge (0,-2.5);
 \path[->] (0,-2.5) edge (0,-4.75);
 \path[-]  (-0.15,-2.1) edge (0.15,-2.1);
 \draw  (0.4,-2) node (p) {$p$};
 \path[-]  (-0.15,-3.2) edge (0.15,-3.2);
 \draw  (0.4,-3.2) node (E) {$E$};
 \path[-]  (-0.15,-4.4) edge (0.15,-4.4);
 \draw  (0.4,-4.3) node (V) {$V$};
\end{tikzpicture}\end{tabular}}
$$
where administrative reductions peculiar to the translation (like continuation-passing) are compressed, 
and where $\trc{\cut{p}{e}} \defeq \tre{e}\,\trp{p}$.
The expanded version is simply:
$$
\begin{array}{l@{~\defeq~}l}
  \tr{\mut a.c}_e		& \lambda a. \tr{c}_c \\
  \tr{E}_e			& \lambda p.p\, \tr{E}_E \\
  \tr{\mu \alpha.c}_p		& \lambda \alpha.\tr{c}_c\\ 
  \tr{a}_p			& a \\
  \end{array}
  \qquad\qquad\quad
  \begin{array}{l@{~\defeq~}l}
  \tr{V}_p			& \lambda E.E \,\trV{V} \\
  \tr{q\cdot e}_E		& \lambda V. V\,\tr{q}_p\,\tr{e}_e \\
  \tr{\alpha}_E			& \alpha \\
  \tr{\lambda a.p}_V		& \lambda q e.(\lambda a.e\,\tr{p}_p)\,q \\
  \end{array}
$$
This induces a translation of commands at each level of the translation:
 $$ \begin{array}{c@{\qquad\qquad}c}
\trc{\cut{p}{e}}^e \defeq \tre{e}\,\trp{p} & \trc{\cut{V}{E}}^E \defeq \trE{E}\,\trV{V}                \\
\trc{\cut{p}{E}}^p \defeq \trp{p}\,\trE{E} & \trc{\cut{V}{q\cdot e}}^V \defeq \trV{V}\,\trp{q}\,\tre{e}\\
\end{array}
$$
which is easy to prove correct with respect to computation,
since the translation is defined from the reduction rules. 
We first prove that substitution is sound through the translation, and then prove that the whole
translation preserves the reduction.
\begin{lemma}
 For any variable $a$ (co-variable $\alpha$) and any proof $q$ (co-value $E$), the following holds for any command $c$:
 $$
        \trc{c[q/a]} ~=~ \trc{c} [\trp{q}/a]
\qquad\qquad \qquad \trc{c[E/\alpha]} ~=~ \trc{c} [\trE{E}/\alpha]
 $$
 The same holds for substitution within proofs and contexts.
\end{lemma}
\begin{proof}
 Easy induction on the syntax of commands, proofs and contexts, the key cases corresponding to (co-)variables:
$$\tre{\alpha}[\trE{E}/\alpha] ~=~ (\lambda p.p\,\alpha)[\trE{E}/\alpha] ~=~ \lambda p.p\,\trE{E} ~=~ \tre{E} ~=~  \tre{\alpha[E/\alpha]}\noem$$
\end{proof}
\begin{proposition}\label{prop:cps_reduc_cbn} 
For all levels $\iota,o$ of $e,p,E$, and any commands $c,c'$,
 if $c_\iota\oset{$1$}\rightsquigarrow c'_o$, then $\trc{c}^\iota \oset{$+~~$}{\bred} \trc{c'}^o$.
\end{proposition}
\begin{proof}
The proof is an easy induction on the reduction $\rightsquigarrow$.
Administrative reductions are trivial, the cases for $\mu$ and $\tmu$ correspond to the previous lemma, which leaves us with the case for $\lambda$:
$$\trc{\cmd{\lambda a.p}{q\cdot e}}^V 
= (\lambda q e.(\lambda a.e\,\tr{p}_p)\,q)\, \trp{q}\,\tre{e}
\bredn{2} (\lambda a.\tre{e}\,\tr{p}_p)\,\trp{q} =  \trc{\cmd{q}{\mut a.\cmd{p}{e}}}^e\noem$$
\end{proof}

\subsubsection{Translation of types}
The computational translation naturally induces a translation on types, which follows the
same descent in the syntax:
 $$\begin{array}{c|c|c}
  \begin{array}{l@{~\defeq~}l}
  \tre{A}		& \trp{A} \to \bot \\
  \trp{A}		& \trE{A} \to \bot \\
  \end{array}
  &
  \begin{array}{r@{~\defeq~}l}
  \trE{A}		& \trV{A} \to \bot \\
  \trV{A\to B}		& \trp{A} \to \tre{B} \to \bot \\
  \end{array}
  &
  \begin{array}{l@{~\defeq~}l}
  \trV{X}		& X\\
  \end{array}
  \end{array}
  \eqno\begin{array}{r}\text{($X$ variable)}\end{array}$$
 and where we take $\bot$ as return type for continuations. 
 This extends naturally to typing contexts, where the translation of $\Gamma$ is defined at level $p$ 
 while $\Delta$ is translated at level $E$:
 $$\trp{\Gamma,a:A} \defeq \trp{\Gamma},a:\trp{A} 
 \hsep
 \trE{\Delta,\alpha:A} \defeq \trE{\Delta},\alpha:\trE{A} $$
  As we did not include any constant of atomic types, the choice for the translation of atomic types
 is somehow arbitrary, and corresponds to the idea that a constant ${\bm c}$ would be translated into $\lambda k.k\,{\bm c}$.
 We could also have translated atomic types at level $p$, with constants translated as themselves.
 In any case, the translation of proofs, contexts and commands is well-typed:
 \begin{proposition}\label{prop:cps_typed_cbn} 
  For any contexts $\Gamma$ and $\Delta$, we have
  \begin{enumerate}
   \item if ~$\Gamma\vdash p:A\mid\Delta$  ~then $\trp{\Gamma},\trE{\Delta}\vdash \trp{p}:\trp{A}$
   \item if ~$\Gamma\mid e:A\vdash \Delta$ ~then $\trp{\Gamma},\trE{\Delta}\vdash \tre{e}:\tre{A}$
   \item if ~~~$c:\Gamma\vdash \Delta$~~~~~\;then $\trp{\Gamma},\trE{\Delta}\vdash \trc{c}:\bot$
  \end{enumerate}
 \end{proposition}
\begin{proof}
 The proof is done by induction over the typing derivation. 
 We can refine the statement by using the type system presented in \Cref{fig:cbn_types},
 and proving two additional statements:  if ~$\Gamma\vdash_V V:A\mid\Delta$ then $\trp{\Gamma},\trE{\Delta}\vdash \trV{V}:\trp{A}$
 (and similarly for $E$).
 We only give two cases, other cases are easier or very similar.
 \prfcase{$c$}
 If $c = \cut{p}{e}$ is a command typed under the hypotheses $\Gamma,\Delta$:
 $$\infer[\cutrule]{\cut{p}{e}:\Gamma\vdash_c\Delta}{\Gamma\vdash_p p:A\mid \Delta & \Gamma \mid e:A\vdash_e\Delta}$$
 then by induction hypotheses for $e$ and $p$, we have that 
 $\trp{\Gamma},\trE{\Delta}\vdash \tre{e}:\trp{A}\to\bot$ and that $\trp{\Gamma},\trE{\Delta}\vdash \trp{p}:\trp{A}$,
 thus we deduce that $\trp{\Gamma},\trE{\Delta}\vdash \tre{e}\,\trp{p}:\bot$.
 \prfcase{$V$}
 If $\lambda a.p$ has type $A\imp B$:
 $$\infer[\!\imprrule]{\Gamma \vdash_V \lambda a.p: A \imp B \mid \Delta }{\Gamma, a:A \vdash_p p:B \mid \Delta} $$
 then by induction hypothesis, we get that $\trp{\Gamma},\trE{\Delta}, a:\trp{A} \vdash \trp{p}:\trp{B}$.
 By definition, we have  $\tr{\lambda a.p}_V = \lambda q e.(\lambda a.e\,\tr{p}_p)\,q$, which we can type:
  $$
 \infer[\impirule]{\trp{\Gamma},\trE{\Delta} \vdash \lambda q e.(\lambda a.e\,\tr{p}_p)\,q:\trp{A}\to\tre{B}\to\bot}{
    \infer[\imperule]{\trp{\Gamma},\trE{\Delta},q:\trp{A},e:\tre{B} \vdash (\lambda a.e\,\tr{p}_p)\,q:\bot}{
      \infer[\impirule]{\trp{\Gamma},\trE{\Delta},e:\tre{B} \vdash \lambda a.e\,\tr{p}_p:\trp{A}\to\bot}{
	\infer[\imperule]{\trp{\Gamma},\trE{\Delta},e:\tre{B},a:\trp{A} \vdash e\,\tr{p}_p:\bot}{
	  \infer[\axrule]{e:\tre{B}\vdash e:\trp{B}\to\bot}{}
	  &
	  \trp{\Gamma},\trE{\Delta},a:\trp{A} \vdash \tr{p}_p:\trp{B}
	}
      }
      &
      \hspace{-.5cm}
      \infer[\axrule]{q:\trp{A} \vdash q:\trp{A}}{}
    }
  }\noem
  $$
 \end{proof}

Up to this point, we already proved enough to obtain the normalization of the {\lmmt}-calculus for the operational semantics considered:
\begin{theorem}[Normalization]
Typed commands of the simply typed call-by-name {\lmmt}-calculus are normalizing.
\end{theorem}
\begin{proof}
 By applying the generic result for translations (\Cref{thm:benefits_cps}) since the required conditions are satisfied:
 the simply-typed $\lambda$-calculus is normalizing and \Cref{prop:cps_reduc_cbn,prop:cps_typed_cbn} 
 correspond exactly to Equations (5.1) and (5.2).
\end{proof}

It only remains to prove that there is no term of the type $\trp{\bot}$ to
 ensure the soundness of the {\lmmt}-calculus.
 \begin{proposition}
 There is no term $t$ in the simply typed $\lambda$-calculus such that $\vdash t:\trp{\bot}$.
\end{proposition}
\begin{proof}
 By definition, $\trp{\bot} = (\bot \to \bot) \to \bot$. Since $\lambda x.x$ is of type $\bot\to\bot$,
 if there was such a term $t$, then we would obtain $\vdash t\,\lambda x.x:\bot$, which is absurd.
\end{proof}
\begin{theorem}
There is no proof $p$ (in the simply typed call-by-name {\lmmt}-calculus) such that $~\vdash p:\bot\mid~ $. 
\end{theorem}
\begin{proof}
 Simple application of \Cref{thm:benefits_cps}.
\end{proof}

\subsection{Krivine classical realizability}
\label{s:lmmt:cbn_real}
We shall present in this section a realizability interpretation \emph{à la} Krivine
for the call-by-name {\lmmt}-calculus.
In a nutshell\footnote{For a more detailed introduction to the topic, we refer
the reader to Krivine introductory paper~\cite{Krivine09} or
Rieg's Ph.D. thesis~\cite{RiegPhD}.}, 
Krivine realizability associates to each type $A$ a set $|A|$ 
of terms whose execution is guided by the structure of $A$.
These terms are the ones usually called \emph{realizers} in Krivine's classical realizability.
Their definition is in fact indirect, 
that is from a set $\fv{A}$ of execution contexts that are intended to challenge the truth of $A$.
Intuitively, the set $\fv{A}$---which we shall call the \emph{falsity value} of $A$—can be understood as
the set of all possible counter-arguments to the formula $A$. In this framework, a program
realizes the formula $A$---\emph{i.e.} belongs to the truth value $|A|$---if and only if it is able to defeat
all the attempts to refute $A$ using a context in $\fv{A}$. 
Realizability interpretations are thus parameterized by a set of ``correct'' computations, called a \emph{pole}. 
The choice of this set is central when studying the models induced by classical realizability,
but in what follows we will mainly pay attention to the particular pole of terminating 
computations\footnote{As such, the proof of normalization that this interpretation provides is also very close 
to a proof by reducibility (see for instance the proof of normalization 
for system $D$ presented in~\cite[3.2]{Krivine93}).}.
The central piece of the interpretation, called the \emph{adequacy lemma}, consists in proving that typed terms 
belong to the corresponding sets of realizers, and are thus normalizing.

\subsubsection{Extension to second-order}
As Krivine classical realizability is naturally suited for a second-order setting, 
we shall first extend the type system to second-order logic. As we will see, the adequacy
of the typing rules for universal quantification almost comes for free.
However, we could also have sticked to the simple-typed setting, 
whose interpretation would have required to explicitly interpret each atomic type by a falsity value.
We give the usual typing rules \emph{à la} Curry for first- and second-order universal quantifications in the framework of the 
{\lmmt}-calculus. Note that in the call-by-name setting, these rules are not restricted and
defined at the highest levels of the hierarchy ($e$ for context, $p$ for proofs).

$$
\begin{array}{c@{\qquad\qquad}c}
\infer[\flurule]{\Gamma \mid e:\forall x.A\vdash \Delta}{\Gamma \mid e:A[n/x] \vdash \Delta}
&
\infer[\frurule]{\Gamma \vdash p:\forall x.A\mid\Delta}{\Gamma \vdash p:A\mid\Delta & x\notin FV(\Gamma,\Delta)}
\\[1em]
\infer[\fldrule]{\Gamma \mid e:\forall X.A\vdash \Delta}{\Gamma \mid e:A[B/X]\vdash \Delta}
&
\infer[\frdrule]{\Gamma \vdash p:\forall X.A\mid\Delta}{\Gamma \vdash p:A\mid\Delta & X\notin FV(\Gamma,\Delta)}
\end{array}
$$

\subsubsection{Realizability interpretation}
We shall now present the realizability interpretation for the 
call-by-name \lmmt-calculus\footnote{As shown in \Cref{ss:mumutilde},
the call-by-name evaluation strategy allows to fully embed the $\lambda_c$-calculus. 
It is no surprise that the respective realizability interpretations for these calculi are very close.
The major difference lies in the presence of the $\tmu$ operator which has no equivalent in the $\lambda_c$-calculus,
and which will force us to add a level in the interpretation.}.
Rather than directly stating the definition of the interpretation,
we wish to attract the reader attention to the fact that this definition 
is again a consequence of the small-steps operational semantics.
Indeed, we are intuitively looking for sets of proofs (truth values) and set of contexts (falsity values) which are
``well-behaved'' against their respective opponents. 
That is, given a formula $A$, we are looking for players for $A$ which compute ``correctly'' in front of any
contexts opposed to $A$. 
If we take a closer look at the definition of the context-free abstract machine (cf. \Cref{s:cbn:small_step}),
we see that the four levels $e$, $p$, $E$, $V$ are precisely defined as sets of objects computing ``correctly''
in front of any object in the previous category: for instance, proofs in $p$ are defined 
together with their reductions in front of any context in  $E$. 
This was already reflected in the continuation-passing style translation.
This suggests a four-level definition of the realizability interpretation, which we compact in three levels as
the lowest level $V$ can easily be inlined at level $p$ (this was already the case in the small-step
operational semantics and we could have done it also for the CPS).

As is usual in Krivine realizability, the interpretation uses the standard model $\N$ 
for the interpretation of first-order expressions and is parameterized by a pole $\pole$:
\begin{definition}[Pole]
A \emph{pole} is any subset $\pole$ of commands which is closed by anti-reduction, that is
for all commands $c, c'$, if $c \in \pole$ and $c\rightarrow c'$, then $c\in\pole$.
\end{definition}

We try to stick as much as possible to the notations and definitions of Krivine realizability~\cite{Krivine09}.
In particular, we define $\Pi$ (the base set for falsity values) as the set of all co-values: $\Pi \defeq E$.
In order to interpret second-order variables that occur in a given formula~$A$, 
it is convenient to enrich the language of PA2 with a new
predicate symbol $\dot{F}$ of arity~$k$ for every \emph{falsity value function~$F$} of arity~$k$, 
that is, for every function $F:\N^{k}\to\Pow(\Pi)$ that associates a falsity value
$F(n_1,\ldots,n_k)\subseteq\Pi$ to every $k$-tuple
$(n_1,\ldots,n_k)\in\N^k$. 
A formula of the language enriched with the predicate
symbols~$\dot{F}$ is then called a \keyword{formula with parameters}.
Formally, this corresponds to the formulas\ defined by:  
$$A,B~::=~X(e_1,\ldots,e_k)\mid A\imp B \mid \forall x. A\mid \forall X .A\mid \dot{F}(e_1,\ldots,e_k)\eqno X\in\V_2,F\in\Pow(\Pi)^{\N^k}$$
where $e_1,\ldots,e_k$ are first-order expressions which we will interpret in the standard model $\N$.

The interpretation of formulas with parameters is defined by induction on the structure of formulas:
 $$\begin{array}{rcl}
 \fvE{\dot{F}(e_1,\ldots,e_k)} &\defeq&  F(\Int{e_1},\ldots,\Int{e_k}) \\
 \fvE{A\imp B}                 &\defeq& \{p\cdot e:~~ p \in \tvp{A} \land e\in\fve{B}\}\\
 \fvE{\forall x. A}            &\defeq&  \ds \bigcup_{n\in\N}\|A[n/x]\|_E \\
 \fvE{\forall X.A}             &\defeq& \ds \bigcup_{\!\!\!F:\N^k\to\Pow(\Pi)\!\!\!}\fvE{A[\dot{F}/X]} \\
     \tvp{A} 	               &\defeq& \fvE{A}^\pole ~=~ \{p:~~ \forall e\in\fvE{A},\cut{p}{e}\in\pole\}\\
    \fve{A} 	               &\defeq& \;\tvp{A}^\pole \;~=~ \{e:~~ \forall e\in\fvE{A},\cut{p}{e}\in\pole\}\\
   \end{array}$$
This definition exactly matches Krivine's interpretation for the $\lambda_c$-calculus, considering that the ``extra'' level of 
interpretation $\fve{A}$ is hidden in the latter, since all stacks are co-values.
The expected monotonicity properties are satisfied:
\begin{proposition}[Monotonicity]For any formula $A$, the following hold:
\begin{multicols}{2}
 \begin{enumerate}
  \item $\fvE{A} \subseteq \fve{A}$
  \item ${\tvp{A}^{\pole\pole}} =\tvp{A}$
  \item $\tvp{\forall x.A}  = \bigcap_{n\in\N} \tvp{A[n/x]}$
  \item $\tvp{\forall X.A}  = \bigcap_{F:\N^k\to\Pow(\Pi)} \tvp{A[\dot F/X]}$
  \item $\fve{\forall x.A}  \supseteq \bigcup_{n\in\N} \fve{A[n/x]}$
  \item $\fve{\forall X.A}  \supseteq \bigcup_{F:\N^k\to\Pow(\Pi)} \fve{A[\dot F/X]}$
 \end{enumerate}
\end{multicols}
\label{prop:cbn_lmmt_mon}
\end{proposition}
\begin{proof}
These properties actually hold for arbitrary sets $A$ and orthogonality relation $\bot$.
 Facts 1 and 2 are simply the usual properties of bi-orthogonal sets: $A\subseteq A^{\bot\bot}$ and $A^{\bot\bot\bot}=A^\bot$.
 Facts 3 and 4 are the usual equality $(\bigcup_{A\in\A} A)^\bot = \bigcap_{A\in\A} A^\bot$.
 Facts 5 and 6 are the inclusion $(\bigcap_{A\in\A} A)^\bot \supseteq \bigcup_{A\in\A} A^{\bot}$.
\end{proof}

In order to state the central lemma, we need to introduce a few more technical concepts.
A \emph{valuation} is defined as a function $\rho$ which associates a natural number 
 $\rho(x)\in\N$ to every first-order variable~$x$ and
a falsity value function $\rho(X):\N^k\to\Pow(\Pi)$ to every second-order variable~$X$ of arity~$k$.
A \emph{substitution}, written $\sigma$, is a function mapping variables to closed proofs (written $\sigma,a:=p$) 
and co-variables to co-values (written $\sigma,\alpha:=E$).
We denote by $A[\rho]$ (resp. $p[\sigma],e[\sigma,...$) the closed formula (resp. proofs, context,...) where
all variables are substituted by their values through $\rho$.

Given two closed left and right contexts $\Gamma, \Delta$, we say that a substitution $\sigma$ realizes $\Gamma\cup\Delta$, 
which we write $\sigma \Vdash \Gamma$, if for any $(a:A) \in\Gamma$, $\sigma(a) \in \tvp{A}$
and if for any $\alpha:A \in\Delta$, $\sigma(\alpha) \in \fvE{A}$.
We are now equipped to prove the adequacy of the typing rules for the (call-by-name) {\lmmt}-calculus
with respect to the realizability interpretation we defined.
\begin{proposition}[Adequacy]
Let $\Gamma, \Delta$ be typing contexts, $\rho$ be any valuation and $\sigma$ be a substitution 
such that $\sigma \Vdash (\Gamma\cup \Delta)[\rho]$, then 
\begin{enumerate}
 \item if ~$\Gamma \vdash p:A\mid \Delta$, then $p[\sigma] \in \tvp{A[\rho]}$
 \item if ~$\Gamma \mid e:A\vdash \Delta$, then $e[\sigma] \in \fve{A[\rho]}$
 \item if ~$c: \Gamma \vdash \Delta$, then $c[\sigma] \in \pole$ 
\end{enumerate}
\end{proposition}
\begin{proof}
 By mutual induction over the typing derivation. 
\prfcase{\cutrule} We are in the following situation:
  $$ \infer[\cutrule]{{\cmd{p}{e}}:\Gamma\vdash \Delta}{
       \Gamma \vdash {p}:{A}  \mid \Delta & \Gamma \mid  {e}:{A} \vdash \Delta
 } $$
  By induction, we have $p[\sigma]\in\tvp{A[\rho]}$ and $e[\sigma]\in\fve{A[\rho]}$, thus $\cut{p[\sigma]}{e[\sigma]}\in\pole$.
 
\prfcase{\axrrule}
 We are in the following situation: $$\infer[\axrrule]{\Gamma \vdash a:A \mid \Delta}{(a:A)\in\Gamma}$$
 Since $\sigma\Vdash \Gamma[\rho]$, we deduce that $\sigma(a)\in \tvp{A}\subset \tv{A[\rho]}$.
 
\prfcase{\axlrule}
 We are in the following situation: $$\infer[\axlrule]{\Gamma \mid \alpha:A \vdash \Delta}{(\alpha:A)\in\Delta}$$
 Since $\sigma\Vdash \Delta[\rho]$, we deduce that $\sigma(\alpha)\in \fv{A[\rho]}$.

\prfcase{\murule}
 We are in the following situation:
  $$\infer[\murule]{\Gamma \vdash {\mu\alpha.c} : {A}  \mid  \Delta }{{c}:(\Gamma\vdash \Delta, {\alpha}:{A})}$$
 Let $E$ be any context in $\fvE{A[\rho]}$, then $(\sigma,\alpha:=E)\Vdash (\Gamma\cup(\Delta,\alpha:A))[\rho]$.
 By induction, we can deduce that $c[\sigma,\alpha:=E]~=~(c[\sigma])[E/\alpha] \in \pole$.
 By definition, 
 $$\cut{(\mu \alpha.c)[\sigma]}{E}~=~\cut{\mu \alpha.c[\sigma]}{E} \rightarrow {c[\sigma][E/\alpha]}\in\pole$$
 thus we can conclude by anti-reduction.
 
\prfcase{\mutrule}
  We are in the following situation:
  $$\infer[\mutrule]{\Gamma \mid  {\mut a.c} : {A} \vdash \Delta}    {{c}:(\Gamma,{a}:{A} \Vdash \Delta)}$$
  Let $p$ be a proof in $\tvp{A[\rho]}$, by assumption we have $(\sigma,a:=p)\Vdash ((\Gamma,a:A)\cup\Delta)[\rho]$.
  As a consequence, we deduce from the induction hypothesis that $c[\sigma,a:=p] = (c[\sigma])[p/a] \in \pole$.
  By definition, we have:
  $$\cut{p}{(\mut a.c)[\sigma]}~=~\cut{p}{\mut a.c[\sigma]}\rightarrow {(c[\sigma])[p/a]}\in\pole$$
  so that we can conclude by anti-reduction.

\prfcase{\imprrule}We are in the following situation:
 $$\infer[\imprrule]{\Gamma \vdash {\lambda a.p} : {A\imp B} \mid \Delta }{\Gamma,{a}:{A} \vdash {p}:{B} \mid  \Delta}$$
 Let $q\cdot e$ be a stack in $\fvE{(A\imp B)[\rho]}$, that is to say that $q\in\tvp{A[\rho]}$ and $e\in\fve{B[\rho]}$.
 By definition, since $q\in\tvp{A[\rho]}$, we have  $(\sigma,a:=q)\Vdash ((\Gamma,a:A)\cup\Delta)[\rho]$. 
 By induction hypothesis, this implies in particular that $p[\sigma,a:=q]\in\tvp{B[\rho]}$
 and thus $\cut{p[\sigma,a:=q]}{e}\in\pole$.
 We can now use the closure by anti-reduction to get the expected result:
 $$\cut{\lambda a.p[\sigma]}{q\cdot e} \rightarrow \cut{q}{\mut a.\cut{p[\sigma]}{e}} \rightarrow \cut{p[\sigma,a:=q]}{e} \in \pole $$

\prfcase{\implrule}We are in the following situation:
   $$\infer[\imp_E]{\Gamma \mid  {q\cdot e} : {A\imp B}\vdash \Delta}{\Gamma\vdash {q}:{A} \mid \Delta & \Gamma \mid {e}:{B}\vdash \Delta}$$
   By induction hypothesis, we obtain that $q[\sigma]\in\tvp{A[\rho]}$ and $e[\sigma]\in\fve{B[\rho]}$.
   By definition, we thus have that $(q\cdot e)[\sigma]\in\fvE{A\imp B}\subseteq \fve{A\imp B}$.

\prfcase{\frurule}
We are in the following situation:
$$\infer[\frurule]{\Gamma \vdash p:\forall x.A\mid\Delta}{\Gamma \vdash p:A\mid\Delta & x\notin \FV(\Gamma,\Delta)}$$
By induction hypothesis, since $x\notin\FV(\Gamma,\Delta)$, we have 
$(\Gamma\cup\Delta)[\rho,x\leftarrow n] = (\Gamma\cup\Delta)[\rho]$  for any $n\in\N$ 
and thus $\sigma \Vdash (\Gamma\cup\Delta)[\rho,x\leftarrow n]$.
We obtain by induction hypothesis
that $p[\sigma]\in\tvp{A[\rho,x\leftarrow n]}$ for any $n\in\N$, \emph{i.e.} that 
$p[\sigma]\in\bigcap_{n\in\N}\tvp{A[\rho,x\leftarrow n]} = \tvp{\forall x.A[\rho]}$.
The case {\frdrule} is identical to this one.

\prfcase{\flurule}
We have that
$$\infer[\flurule]{\Gamma \mid e:\forall x.A\vdash \Delta}{\Gamma \mid e:A[n/x] \vdash \Delta}$$
thus by induction hypothesis we get that $e[\sigma]\in\fve{(A[n/x])[\rho]}$.
Therefore we have in particular that $e[\sigma]\in\bigcup_{n\in\N}\fve{(A[n/x])[\rho]} \subseteq \fve{\forall x.A[\rho]}$ (\Cref{prop:cbn_lmmt_mon}).
The case {\frdrule} is identical to this one.
\end{proof}

Once the adequacy is proved, normalization and soundness almost come for free.
The normalization is a direct corollary of the following observation, whose proof is the same as for \Cref{prop:norm_pole}:
\begin{proposition}
 The set $\pole_\Downarrow \defeq \{c: c~ \text{normalizes}\}$ of normalizing commands defines a valid pole.
\end{proposition}
\begin{proof}
We only have to check that $\pole_\Downarrow$ is closed under antireduction, which is indeed the case:
if $c\tau \rightarrow c'\tau'$ and $c'\tau'$ normalizes, then $c\tau$ normalizes too.
\end{proof}

\begin{theorem}[Normalization]
 For any contexts $\Gamma,\Delta$ and any command $c$, if $c:\Gamma\vdash \Delta$, then $c$ normalizes.
\end{theorem}
\begin{proof}
 By adequacy, any typed command $c$ belongs to the pole $\pole_\Downarrow$ modulo the closure under a 
 substitution $\sigma$ realizing the typing contexts.
 It suffices to observe that to obtain a closed term, any free variable $a$ of type $A$ in $c$ can be substituted
 by an inert constant ${\bm a}$ which will realize its type (since it forms a normalizing command in front of any $E$ in $\fvE{A}$). 
 Thus $c[{\bm a}/a,{\bm b}/b,\dots]$ normalizes and so does $c$.
\end{proof}

Similarly, the soundness is an easy consequence of adequacy, since the existence of a proof $p$ of type $\bot=\forall X.X$
would imply that $p\in\tvp{\bot}$ for any pole $\pole$. For any consistent pole (say the empty pole), this is absurd.  
\begin{theorem}[Soundness]
There is no proof $p$ (in the second-order call-by-name {\lmmt}-calculus) such that $~\vdash p:\bot\mid~ $. 
\end{theorem}


\section{Classical call-by-need}
\label{s:lbvtstar}

Let us start this section with a story, borrowed from \cite{DanEtAl10}, 
illustratingdemand-driven computation and memoization of intermediate results,
two key features of the call-by-need evaluation strategy 
that distinguish it from the call-by-name and call-by-value evaluation strategies:

\begin{quote}
\it A famous functional programmer once was asked to give an overview talk. 
He began with : ``This talk is about lazy functional programming and call by need.'' and
paused. Then, quizzically looking at the audience, he quipped: ``Are there any
questions?'' There were some, and so he continued: ``Now listen very carefully, I
shall say this only once.''
\end{quote}

The \emph{call-by-name} evaluation strategy indeed passes arguments to functions without evaluating
them, postponing their evaluation to each place where the argument is
needed, re-evaluating the argument several times if needed.
 It has in common with the call-by-value evaluation strategy that all places
where a same argument is used share the same value.  
Nevertheless, it observationally behaves like the 
call-by-name evaluation strategy, in the sense that a given computation
eventually evaluates to a value if and only if it evaluates to the same
value (up to inner reduction) along the call-by-name evaluation\footnote{In 
particular, in a setting with non-terminating 
computations, it is not observationally equivalent to the call-by-value
evaluation: if the evaluation of a useless argument loops in
the call-by-value evaluation, the whole computation loops (\emph{e.g.} in $(\lambda\_.I)\,\Omega$), which is not
the case of call-by-name and call-by-need evaluations.}.
The call-by-name, call-by-value and call-by-need evaluation strategies
can be turned into equational theories. For call-by-name and
call-by-value, this was done by Plotkin~\cite{Plotkin75} through
continuation-passing style semantics characterizing these
theories. 
For the call-by-need evaluation strategy, a continuation-passing style semantics
was proposed in the 90s by Okasaki, Lee and Tarditi~\cite{OkaLeeTar94}.
However, this semantics does not ensure normalization of simply-typed call-by-need
evaluation, as shown in~\cite{AriEtAl12}, thus failing to
ensure a property which holds in the simply-typed call-by-name
and call-by-value cases.

The semantics of calculi with control can also be
reconstructed from an analysis of the duality between programs and
their evaluation contexts, and the duality between the {\tt let}
construct (which binds programs) and a control operator such as
Parigot's $\mu$ (which binds evaluation contexts). 
As explained in \Cref{s:lmmt}, such an analysis can be done 
in the context of the \lmmt-calculus~\cite{CurHer00,HerbelinHdR}.
To attack the problem, Ariola \emph{et al.}~\cite{AriHerSau11}
first proposed the $\lbv$-calculus, 
a call-by-need variant of Curien-Herbelin's {\lmmt}-calculus~\cite{CurHer00}.
Thanks to Danvy's methodology of semantics artifacts,
they then refined the reduction system until to a get a calculus
with context-free reduction rules, from which they derived an untyped continuation-passing 
style translation~\cite{AriEtAl12}.
This calculus, named the $\lbvtstar$-calculus,
relies on the use of an explicit environment to store substitutions.
By pushing one step further Danvy's methodology, 
we finally showed with Herbelin how to obtain a realizability interpretation \emph{à la}
Krivine for this framework~\cite{MiqHer18}.
The main idea, in contrast to usual models of Krivine realizability~\cite{Krivine09},
is that realizers are defined as pairs of a term and a substitution.
The adequacy of the interpretation directly provided us with a proof of normalization,
and we shall follow the same methodology in~\Cref{s:dlpaw} to prove the normalization of~\dlpaw.

We shall now recap the different aforementionned steps leading to the correct definition of 
a classical call-by-need sequent calculus.

\subsection{The $\lbv$-calculus: call-by-need with control}
Recall from \Cref{s:lmmt} that the 
reduction rules of the \lmmt-calculus are given by:
$$\begin{array}{l@{\qquad\rightarrow\qquad}ll}
\cut{p}{\tmu a.c }           & \Sub{c}{a}{p} & p\in {\mathcal V}\\
\cut{\mu\alpha.c}{e}         & \Sub{c}{\alpha}{e} & e \in {\mathcal E}\\
\cut{\lambda a.p}{q\cdot e}  & \cut{q}{\tmu a.\cut{p}{e}}\quad
\end{array}$$
where the set of terms ${\mathcal V}$ and the set of evaluation contexts~${\mathcal E}$
parameterize the evaluation strategy:
the call-by-name evaluation strategy amounts to the case where ${\mathcal E}\defeq$~\emph{Co-values} while call-by-value 
dually corresponds to ${\mathcal V}\defeq$~\emph{Values}.
For the call-by-need case, 
intuitively, we would like to set ${\mathcal V}\defeq$~\emph{Values} (we only substitute evaluated terms
of which we share the value) and ${\mathcal E}\defeq$~\emph{Co-values} 
(a term is only reduced if it is in front of a co-value). However, such a definition is clearly not enough 
since any command of the shape $\cmd{\mu\alpha.c}{\tmu a.c'}$ would be blocked.
We thus need to understand how the computation is driven forward, 
that is to say when we need to reduce terms.

Observe that applicative contexts\footnote{Note that we need
to restrict the shape of applicative contexts: 
the general form $q\cdot e$ is not necessarily a valid application, since for example
in $\cmd{\mu\alpha.c}{q \cdot\tmu a\cmd{b}{\alpha}}$, the context $p \cdot\tmu a\cmd{b}{\alpha}$ forces
the execution of $c$ even though its value is not needed. Applicative contexts are thus considered of the restricted
shape $q\cdot E$.}
$q\cdot E$ eagerly demand a value. Such contexts are called \emph{forcing contexts}, and denoted by $F$. 
When a variable $a$ is in front of a forcing context, that is in $\cut{a}{F}$, the variable $a$ 
is said to be \emph{needed} or \emph{demanded}.
This allows us to identify meta-contexts $C$ 
which are nesting of commands of the form $\cut{p}{e}$ for
which neither $p$ is in ${\mathcal V}$ (meaning it is some
$\mu\alpha.c$) nor $e$ in ${\mathcal E}$ (meaning it is an instance of some $\tmu a.c$ which is not a forcing context).
These contexts, defined by the following grammar:
    $$\begin{array}{l@{\hspace{0.8cm}}lll}
     & C[~] & ::= & [~] \mid  \cut{\mu\alpha.c}{\tmu a.C[~]}\\
    \end{array}\leqno\text{\bf Meta-contexts}$$
are such that in a $\tmu$-binding of the form $\tmu a.C[\cut{a}{F}]$,
$a$ is needed and a value is thus expected. 
These contexts, called \emph{demanding contexts}, are evaluation contexts whose
evaluation is blocked on the evaluation of $a$, therefore requiring the
evaluation of what is bound to $a$.
In this case, we say that the bound variable $a$ has been \emph{forced}.

All this suggests another refinement of the syntax, introducing a division between \emph{weak} co-values (resp. \emph{weak} values), 
also called \emph{catchable} contexts (since they are the one caught by a $\mu\alpha$ binder),
and \emph{strong} co-values (resp. \emph{strong} values), which are precisely the forcing contexts.
Formally, the syntax (to which we add constants $\cbold$ and co-constants $\alphabold$) is defined 
by\footnote{In the definition, we implicitly assume $\tmu a.c$ to only cover the cases which are not of the form $\tmu a.C[\cut{a}{F}]$.}:
$$\begin{array}{c|c}
\begin{array}{l@{\hspace{0.4cm}}l@{~}l@{~}l}
\text{\bf Strong values} & v   & ::= & \lambda a.p \mid \cbold  \\
\text{\bf Weak values}   & V   & ::= & v \mid  a		\\
\text{\bf Proofs}        & p   & ::= & V \mid  \mu\alpha.c  	\\
\end{array}
&
\begin{array}{l@{\hspace{0.4cm}}l@{~}l@{~}l}
\text{\bf Forcing   contexts} & F & ::= & p \cdot E \mid  \alphabold\\
\text{\bf Catchable contexts} & E & ::= & F \mid  \alpha \mid  \tmu a.C[\cut{a}{F}]\\
\text{\bf Contexts} 	      & e & ::= & E \mid  \tmu a.c\\
\end{array}
\end{array}
$$

We can finally define ${\mathcal V}\defeq$~\emph{Weak values} and ${\mathcal E}\defeq$~\emph{Catchable contexts}.
The so-defined call-by-need calculus is close to the calculus called $\lbv$ in Ariola {\em et al}
~\cite{AriEtAl12}\footnote{The difference
      lies in the fact that we add constants to preserve the duality.
      }.

The $\lbv$ reduction, written as $\redlv$, 
denotes thus the compatible reflexive transitive closure of the rules:
$$\begin{array}{c@{\qquad\redlv\qquad}c}
\cut{V}{\tmu a.c }	        & \Sub{c}{a}{V} \\
\cut{\mu\alpha.c}{E}	  	& \Sub{c}{\alpha}{E} \\
\cut{\lambda a.p}{q\cdot E}	& \cut{q}{\tmu a.\cut{p}{E}} \\
\end{array}$$

Observe that the next reduction is not necessarily at
the top of the command, but may be buried under several bound
computations $\mu\alpha.c$.
For instance, the command 
$\cut{\mu \alpha.c}{\tmu a_1.\cut{a_1}{\tmu a_2.\cut{a_2}{F}}}$,
where $a_1$ is not needed, reduces to 
$\cut{\mu \alpha.c}{\tmu a_1.\cut{a_1}{F}}$, which now demands $a_1$.

The $\lbv$-calculus can easily be equipped with a type system 
made of the usual rules of the classical sequent calculus~\cite{CurHer00}, 
and adopting the convention that constants $\cbold$ and co-constants $\alphabold$ come with a signature $\S$
which assigns them a type. 
We delay the introduction of such a type system for our next object of study, the \lbvtstar-calculus.


\subsection{The $\lbvtstar$-calculus}
Since the call-by-need evaluation strategy imposes
to share the evaluation of arguments across all the places where thay are need, 
abstract machines implementing it require a form of global memory~\cite{Sestoft97,Cregut07,Lang07,AccBarMaz14}.
In order to obtain such an abstract machine from the \lbv-calculus, Ariola \etal~ first define the 
\lbvtstar-calculus, which is reformulation of the former using explicit environments.
We call \emph{stores} these environments, which we denote by $\tau$.
Stores consists of a list of bindings of the shape $[a:=p]$, where $a$ is a term variable and $p$ a term,
and of bindings of the shape $[\alpha:=e]$ where $\alpha$ is a context variable and $e$ a context. 
For instance, in the closure $c\tau[a:=p]\tau'$, the variable $a$ is bound to $p$ in $c$ and $\tau'$.
Besides, the term $p$ might be an unevaluated term (\emph{i.e.} lazily stored),
so that if $a$ is eagerly demanded at some point during the execution of this closure, $p$ will be reduced in order to obtain a value.
In the case where $p$ indeed produces a value $V$, the store will be updated with the binding $[a:=V]$. 
However, a binding of this shape (with a value) is fixed for the rest of the execution.
As such, our so-called stores somewhat behave like lazy explicit substitutions or mutable environments~\footnote{To draw the comparison
between our structures and the usual notions of stores and environments, two things should be observed.
First, the usual notion of store refers to a structure of list that is fully mutable, in the sense that the cells can be updated at any time and thus values might be replaced. 
Second, the usual notion of environment designates a structure in which variables are bounded to closures made of a term and an environment. 
In particular, terms and environments are duplicated, \emph{i.e.} sharing is not allowed. Such a structure resemble to a tree whose nodes
are decorated by terms, as opposed to a machinery allowing sharing (like ours) whose the underlying structure is broadly a directed acyclic graphs.}.

The lazy evaluation of terms allows us to reduce a command $\cut{\mu\alpha .c}{\tmu a.c'}$ to the command $c'$ together with the binding $[a:=\mu\alpha.c]$.
In this case, the term $\mu\alpha.c$ is left unevaluated (``frozen'') in the store, until possibly reaching a command in which the variable $a$ is needed.
When evaluation reaches a command of the form $\cut{a}{F}\tau[a:=\mu\alpha.c]\tau'$, the binding is opened and the term is evaluated in front of the context $\tmu[a].\cut{a}{F}\tau'$:
 $$\cut{a}{F}\tau[a:=\mu\alpha.c]\tau'\red\cut{\mu\alpha.c}{\tmu[a].\cut{a}{F}\tau'}\tau$$
The reader can think of the previous rule as the ``defrosting'' operation of the frozen term $\mu\alpha.c$:
this term is evaluated in the prefix of the store $\tau$ which predates it, in front of the context $\tmu[a].\cut{a}{F}\tau'$
where the $\tmu[a]$ binder is waiting for an (unfrozen) value. This context keeps trace of the suffix of the store $\tau'$ that was after the binding for $a$.
This way, if a value $V$ is indeed furnished for the binder $\tmu[a]$, the original command $\cut{a}{F}$ is evaluated in the updated full store:
 $$\cut{V}{\tmu[a].\cut{a}{F}\tau'}\tau \red \cut{V}{F}\tau[a:=V]\tau'$$
The brackets are used 
to express the fact that the variable $a$ is forced at top-level (unlike contexts of the shape $\tmu a.C[\cut{a}{F}]$ in the $\lbv$-calculus).
The reduction system resembles the one of an abstract machine. 
Especially, it allows us to keep the standard redex at the top of a command and avoids searching through the
meta-context for work to be done.

Note that our approach slightly differ from \cite{AriEtAl12}
in that we split values into two categories: strong values ($v$) and
weak values ($V$). The strong values correspond to values strictly
speaking. The weak values include the variables which force the
evaluation of terms to which they refer into shared strong
value. Their evaluation may require capturing a continuation.
The syntax of the language is given by:
$$
\begin{array}{c|c}
\begin{array}{l@{\hspace{0.4cm}}l@{~}l@{~}l}
\text{\bf Strong values} & v & ::= & \lambda a.p \mid \cbold \\ 
\text{\bf Weak values}   & V & ::= & v \mid  a               \\ 
\text{\bf Proofs}         & p & ::= & V \mid  \mu\alpha.c     \\ 
\end{array}&
\begin{array}{l@{\hspace{0.4cm}}l@{~}l@{~}l}
\text{\bf Forcing contexts} & F & ::= & \alphabold \mid  p \cdot E\\
\text{\bf Catchable contexts} & E & ::= & F \mid  \alpha \mid  \tmu[a].\cut{a}{F}\tau\\
\text{\bf Evaluation contexts} & e & ::= & E \mid  \tmu a.c\\
\end{array}           \\
\multicolumn{2}{c}{
\begin{array}{l@{\hspace{0.4cm}}l@{~}l@{~}l}
\\
\text{\bf Closures} & l & ::= & c\tau \\ 
\text{\bf Commands} & c & ::= & \cut{p}{e} \\ 
\text{\bf Stores} & \tau & ::= & \varepsilon \mid  \tau[a:=p] \mid   \tau[\alpha:=E]\\
\end{array}}
\end{array}$$
The reduction, written $\rightarrow$, is the compatible reflexive transitive closure of the rules
  given in Figure~\ref{fig:reduction-rules}.
 {
\begin{figure}[t]
  \framebox{\vbox{
$$\begin{array}{c@{\qquad\rightarrow\qquad}c}
\cut{p}{\tmu a.c }\tau               & c\tau[a:=p] \\
\cut{\mu\alpha.c}{E}\tau             & c\tau[\alpha:=E] \\
\cut{V}{\alpha}\tau[\alpha:=E]\tau'  & \cut{V}{E}\tau[\alpha:=E]\tau' \\
\cut{a}{F}\tau[a:=p]\tau'            & \cut{p}{\tmu[a].\cut{a}{F}\tau'}\tau \\
\cut{V}{\tmu[a].\cut{a}{F}\tau'}\tau & \cut{V}{F}\tau[a:=V]\tau' \\
\cut{\lambda a.p}{u\cdot E}\tau      & \cut{u}{\tmu a.\cut{p}{E}}\tau
\end{array}\leqno
\begin{array}{l}	
 (\textsc{Let}          )\\
 (\textsc{Catch}        )\\
 (\textsc{Lookup}_\alpha)\\
 (\textsc{Lookup}_a     )\\
 (\textsc{Restore}      )\\
 (\textsc{Beta}         )\\
\end{array}$$
}}
\caption{Reduction rules of the $\lbvtstar$-calculus}

\label{fig:reduction-rules}
\end{figure}
}

The different syntactic categories can again be understood as the different
levels of alternation in a context-free abstract machine:
the priority is first given to contexts at level $e$ (lazy storage of terms),
then to terms at level $p$ (evaluation of $\mu\alpha$ into values),
then back to contexts at level $E$ and so on until level~$v$.
These different categories are thus directly reflected in the definition of the 
context-free abstract machine (that we will present in \Cref{s:context-free})
and of the realizability interpretation
but also in the type system we define.
We indeed consider nine kinds of (one-sided\footnote{To this end, observe that we write $A^\negt$ for a type $A$
that would have been in the context $\Delta$ in two-sided sequents. 
While this is only used to compact notations, this will become crucial when using dependent types in the next sections.})
sequents, one for typing each of the nine
syntactic categories. We write them with an annotation on the $\vdash$
sign, using one of the letters $v$, $V$, $p$, $F$, $E$, $e$, $l$, $c$,
$\tau$.
Sequents typing values and
terms are asserting a type, with the type written on the right;
sequents typing contexts are expecting a type $A$ with the type written
$A^\negt$; sequents typing commands and closures are black boxes neither 
asserting nor expecting a type; sequents typing substitutions are 
instantiating a typing context.
In other words, we have the following nine kinds of sequents:
\begin{center}
\begin{tabular}{l@{\qquad\quad}l@{\qquad\quad}l}
\begin{tabular}{l}
$\Gamma \vdash_l l$\\
$\Gamma \vdash_c c$\\
$\Gamma \vdash_\tau \tau:\Gamma' $\\
\end{tabular}
&
\begin{tabular}{l}
$\Gamma \vdash_p p:A $\\
$\Gamma \vdash_V V:A $\\
$\Gamma \vdash_v v:A $\\
\end{tabular}
&
\begin{tabular}{l}
$\Gamma \vdash_e e:A^\negt $\\
$\Gamma \vdash_E E:A^\negt $\\
$\Gamma \vdash_F F:A^\negt $\\
\end{tabular}
\end{tabular}
\end{center}
where types and typing contexts are defined by:
$$
 A,B  ::=  X \mid  A \imp B \qquad\qquad\qquad
 \Gamma ::=  \varepsilon \mid  \Gamma, a:A\mid \Gamma,\alpha:A^\negt
$$
\label{sec:typing}
\begin{figure}[t]
  \framebox
{\vbox{
\input{figures/typing_rules_lbvtstar}
}}
\caption{Typing rules of the $\lbvtstar$-calculus}
\label{fig:typing-rules}
\end{figure}
The typing rules are given on Figure \ref{fig:typing-rules} where we assume 
that a variable $a$ (resp. co-variable $\alpha$) only occurs once in a context $\Gamma$
(we implicitly assume the possibility of renaming variables by $\alpha$-conversion).
We also adopt the convention that constants $\cbold$ and co-constants $\alphabold$ come with a signature $\S$
which assigns them a type.
Regarding the type system introduced earlier for the \lmmt-calculus, the main novelties
are the rules \lrule, {\tautrule} and {\tauErule} to handle stores, and the rule {\eagerrule} for the new binder $\tmu[a].c\tau$.

This type system enjoys the property of subject reduction:
\begin{theorem}[Subject reduction]
\label{thm:subject}
 If $\Gamma \vdash_l c\tau$ and $c\tau\rightarrow c'\tau'$ then $\Gamma\vdash_l c'\tau'$.
\end{theorem}
\begin{proof}
 By induction on typing derivations\vlong{, see \cite{these} for the complete proof}.
\end{proof}

\subsection{Abstract machine in context-free form}
\label{s:context-free}

Reduction rules of the \lbvtstar-calculus can again be refined
in order to finally obtain small-step reduction rules of a context-free abstract machine~\cite{AriEtAl12}.
This essentially consists in annotating again commands with the level of syntax we are examining ($c_e,c_p,\dots$),
and in defining a new set of reduction rules which separates computational steps (corresponding to big-step reductions),
and administrative steps, which organize the descent in the syntax.
In order, a command first put the focus on the context at level $e$, 
then on the term at level $p$, and so on following the hierarchy $e,p,E,V,F,v$.
This results again in an abstract machine in context-free form, 
since each step only analyzes one component of the command, 
the ``active'' term or context, and is parametric in the other ``passive'' component. 
In essence, for each phase of the machine, either the term or the context is fully in control and independent,
regardless of what the other half happens to be.

We recall the resulting abstract machine from \cite{AriEtAl12} in \Cref{fig:cbn:small_step}.
These rules directly lead to the definition of the CPS in~\cite{AriEtAl12} that we shall type in the next sections.
Furthermore, the realizability interpretation \emph{à la} Krivine (that we are about to present in the coming section)
is deeply based upon this set of rules.
Indeed, remember that a realizer is precisely a term which is going to behave well 
in front of any opponent in the opposed falsity value. We shall thus take advantage of 
the context-free rules where, at each level, the reduction step is defined independently of the passive component.

\begin{figure}[t]
\newcommand{\trait}[2]{
\path[-]  (-0.15,#2) edge (0.15,#2);
\draw  (0.4,#2) node  {$#1$};
}
  \framebox{\vbox{
  \vspace{-0.8em}
  \setlength{\saut}{0.5em}
$$\begin{array}{r@{\qquad\rightarrow\qquad}l}
\cut{p}{\tmu a.c }_e\tau		& c_e\tau[a:=p] \\
\cut{p}{E}_e\tau 	     		& \cut{p}{E}_p\tau \\[\saut]

\cut{\mu\alpha.c}{E}_p\tau             	& c_e\tau[\alpha:=E] \\
\cut{V}{E}_p\tau			& \cut{V}{E}_E\tau \\[\saut]

\cut{V}{\alpha}_E\tau[\alpha:=E]\tau'  	& \cut{V}{E}_E\tau[\alpha:=E]\tau' \\
\cut{V}{\tmu[a].\cut{a}{F}\tau'}_E\tau 	& \cut{V}{F}_V\tau[a:=V]\tau' \\
\cut{V}{F}_E\tau			& \cut{V}{F}_V\tau \\[\saut]

\cut{a}{F}_V\tau[a:=p]\tau'            	& \cut{p}{\tmu[a].\cut{a}{F}\tau'}\tau \\
\cut{v}{E}_V\tau			& \cut{v}{F}_V\tau \\[\saut]

\cut{v}{u\cdot E}_F\tau      		& \cut{v}{e\cdot {E}}_v\tau \\[\saut]
\cut{\lambda a.p}{q\cdot E}_v\tau      	& \cut{q}{\tmu a.\cut{p}{E}}_e\tau
\end{array}\vspace{-1em}\leqno
\text{\begin{tabular}{c}
\\[-0.5em]
\begin{tikzpicture}[color=black!40,scale=0.7]
 \path[->] (0,-0.3) edge (0,-9.2);
 \trait{e}{-1}
 \trait{p}{-2.6}
 \trait{E}{-4.4}
 \trait{V}{-6.4}
 \trait{F}{-7.7}
 \trait{v}{-8.8}
 \end{tikzpicture}\end{tabular}}\vspace{-1.5em}
$$
}}
\caption{Context-free abstract machine for the $\lbvtstar$-calculus}
\label{fig:cbn:small_step} 
\end{figure}

\subsection{Realizability interpretation}
\label{s:lbvtstar_real}
We shall now see how to adapt the realizability interpretation for the \lmmt-calculus
to this setting, and in particular to handle stores.
As for the \lmmt-calculus, we are guided by the small-step abstract-machine.
First of all, given a formula $A$ we now define its interpretation at each 
typing level $o$ (of $e,t,E,V,F,v$) 
a set $|A|_o$ (resp. $\|A\|_o$) of proof terms (resp. contexts) in the corresponding syntactic category.
Second, we generalize the usual notion of closed term to the notion of closed \emph{\cp}.
Intuitively, this is due to the fact that we are no longer interested in closed terms and substitutions to close opened terms, 
but rather in terms that are closed when considered in the current store. 
This is based on the simple observation that a store is nothing more than a shared substitution whose 
content might evolve along the execution.
Last, we adapt the notion of \emph{pole} $\pole$ to be sets of closures, which ask to be closed by anti-evaluation and store extension.
In particular, the set of normalizing closures defines a valid pole.

We begin with a bunch of definitions related to stores. These notions will be re-used later 
when defining the interpretation for \dlpaw.
\begin{definition}[Closed store]
 We extend the notion of free variable to stores:
 $$\begin{array}{c@{~~\defeq~~}l}
 FV(\varepsilon) 	& \emptyset \\
 FV(\tau[a:=p]) 	& FV(\tau)\cup\{y\in FV(t):y\notin \dom(\tau)\} \\
 FV(\tau[\alpha:=E])	& FV(\tau)\cup\{\beta \in FV(E):\beta \notin \dom(\tau)\}
 \end{array}$$
so that we can define a \emph{closed store} to be a store $\tau$ such that $FV(\tau) = \emptyset$.
\end{definition}

\begin{definition}[Compatible stores]\label{def:compat}
 We say that two stores $\tau$ and $\tau'$ are \emph{independent} and note $\indpt{\tau}{\tau'}$ when
 ${\dom(\tau)\cap\dom(\tau')=\emptyset}$.
 We say that they are \emph{compatible} and note $\compat{\tau}{\tau'}$ 
 whenever for all variables $a$ (resp. co-variables $\alpha$) present in both stores: ${x\in \dom(\tau)\cap\dom(\tau')}$;
 the corresponding terms (resp. contexts) in $\tau$ and $\tau'$ 
 coincide: formally $\tau = \tau_0[a:=p]\tau_1$ and $\tau' = \tau'_0[a:=p]\tau'_1$. 
 Finally, we say that $\tau'$ is an \emph{extension} of $\tau$ and note $\tau\stext \tau'$ whenever 
 $\dom(\tau)\subseteq\dom(\tau')$ and $\compat{\tau}{\tau'}$.
 \end{definition}

\begin{definition}[Compatible union]  
  We denote by $\overline{\tau\tau'}$ the compatible union $\stjoin{\tau}{\tau'}$ of closed stores $\tau$ and $\tau'$, defined by:\vspace{-0.4em}
 $$\begin{array}{r@{~~\defeq~~}l}
 \stjoin{\tau_0[a:=p]\tau_1}{\tau'_0[a:=p]\tau'_1} & \tau_0\tau'_0[a:=p]\stjoin{\tau_1}{\tau'_1} \\
  \stjoin{\tau}{\tau'} & \tau\tau'                                                             \\
  \stjoin{\varepsilon}{\tau} & \tau \\
 \stjoin{\tau}{\varepsilon} &\tau\\
 \end{array}\eqno\begin{array}{r}(\text{if } \indpt{\tau_0}{\tau_0'})\\(\text{if }\indpt{\tau}{\tau'})\\\\\\\end{array} 
 $$
 \end{definition}
The following lemma (which follows easily from the previous definition) states the main property  
we will use about union of compatible stores.
\begin{lemma}
\label{lm:st_union}
If $\tau$ and $\tau'$ are two compatible stores, then $\tau\stext\overline{\tau\tau'}$ and $\tau'\stext\overline{\tau\tau'}$.
Besides, if $\tau$ is of the form $\tau_0[a:=p]\tau_1$,
then $\overline{\tau\tau'}$ is of the form $\overline{\tau_0}[a:=p]\overline{\tau_1}$ with $\tau_0 \stext \overline{\tau_0}$
and $\tau_1\stext\overline{\tau_1}$.
\end{lemma}

As we explained, we will not consider closed terms in the usual sense.
Indeed, while it is frequent in the proofs of normalization (\emph{e.g.} by realizability or reducibility) of a calculus to consider
only closed terms and to perform substitutions to maintain the closure of terms, this only makes sense if it corresponds to the 
computational behavior of the calculus. For instance, to prove the normalization of $\lambda a.p$ in
typed call-by-name $\lambda\mu\tmu$-calculus, one would consider a substitution $\rho$ that is suitable
with respect to the typing context $\Gamma$, then a context $u\cdot e$ of type $A\to B$, and evaluates :
$$\cut{\lambda a.p_\rho}{q\cdot e} ~~\rightarrow~~\cut{p_\rho[q/a]}{e}$$
Then we would observe that $p_\rho[q/a] = p_{\rho[a:=q]}$ and deduce that $\rho[a:=q]$ is suitable for realizing $\Gamma,a:A$, 
which would allow us to conclude by induction.

However, in the $\lbvtstar$-calculus we do not perform global substitution when reducing a command, but rather add a new binding $[a:=p]$ in the store:
$$\cut{\lambda a.p}{q\cdot E}\tau ~~\rightarrow~~\cut{p}{E}\tau[a:=q]$$
Therefore, the natural notion of closed term invokes the closure under a store, 
which might evolve during the rest of the execution (this is to contrast with a substitution).

\begin{definition}[Terms-in-store]
We call \emph{closed \cp} (resp. \emph{closed \ce}, \emph{closed closures}) 
the combination of a proof $p$ (resp. context $e$, command $c$) with a closed store $\tau$ such that
$FV(t)\subseteq \dom(\tau)$. 
We use the notation $\tis{p}{\tau}$ to denote such a pair. 
\end{definition}
We should note that in particular, if $p$ is a closed term, then $\tis{p}{\tau}$ is a {\cp} for any closed store $\tau$. 
The notion of {closed \cp} is thus a generalization of the notion of closed terms, and we will (ab)use of this 
terminology in the sequel. We denote the sets of closed closures by $\C_0$, and will identify $\tis{c}{\tau}$ and the closure $c\tau$ when $c$ is closed in $\tau$.
Observe that if $c\tau$ is a closure in $\C_0$ and $\tau'$ is a store extending $\tau$, then $c\tau'$ is also in $\C_0$.
We are now ready to define the notion of pole, and verify that the set of normalizing closures is indeed a valid pole.
\begin{definition}[Pole]
 A subset $\pole\subseteq \C_0$ is said to be \emph{saturated} or \emph{closed by anti-reduction} 
 whenever for all $\tis{c}{\tau},\tis{c'}{\tau'}\in\C_0$,  if $c'\tau' \in \pole$ and $c\tau\rightarrow c'\tau'$ then $c\tau\in\pole$.
 It is said to be \emph{closed by store extension} if whenever $c\tau\in\pole$, for any store $\tau'$ extending $\tau$: $\tau\stext\tau'$, $c\tau'\in\pole$.
 A \emph{pole} is defined as any subset of $\C_0$ that is closed by anti-reduction and store extension.
\end{definition}

\begin{proposition}
\label{prop:norm_pole}
 The set $\pole_{\Downarrow}=\{c\tau\in\C_0:~c\tau\text{ normalizes }\}$ is a pole.
\end{proposition}
\begin{proof}
 As we only considered closures in $\C_0$, both conditions (closure by anti-reduction and store extension) are clearly satisfied:
 \begin{itemize}
  \item if $c\tau \rightarrow c'\tau'$ and $c'\tau'$ normalizes, then $c\tau$ normalizes too;
  \item if $c$ is closed in $\tau$ and $c\tau$ normalizes, if $\tau\stext \tau'$ then $c\tau'$ will reduce as $c\tau$ does
  (since $c$ is closed under $\tau$, it can only use terms in $\tau'$ that already were in $\tau$) and thus will normalize.\qedhere
 \end{itemize}
\end{proof}

\begin{definition}[Orthogonality]
Given a pole $\pole$, we say that a {\cp} $\tis{p}{\tau}$ is {\em orthogonal} to a {\ce} $\tis{e}{\tau'}$
and write $\tis{p}{\tau}\orth\tis{e}{\tau'}$
if $\tau$ and $\tau'$ are compatible and $\cut{p}{e}\overline{\tau\tau'}\in\pole$.
\end{definition}
\begin{remark}
 The reader familiar with Krivine's forcing machine~\cite{Krivine11} might recognize his definition of orthogonality 
 between terms of the shape $(t,p)$ and stacks of the shape $(\pi,q)$, where $p$ and $q$ are forcing conditions:
 $$ (t,p) \pole (\pi,q) \Leftrightarrow (t\star\pi,p\land q) \in \pole$$
 (The meet of forcing conditions is indeed a refinement containing somewhat the ``union'' of information contained in each, 
 just like the union of two compatible stores.)
\end{remark}

We can now relate closed terms and contexts by orthogonality with respect to a given pole. 
This allows us to define for any formula $A$ 
the sets $\tvv{A},\tvV{A},\tvp{A}$ (resp. $\fvF{A}$,$\fvE{A}$, $\fve{A}$) 
of realizers (or reducibility candidates) at level $v$, $V$, $p$ (resp $F$, $E$, $e$) for the formula $A$. 
It is to be observed that realizers are here closed {\cps}.

\begin{definition}[Realizers]
\label{def:realizers}
 Given a fixed pole $\pole$, we set:
 $$\begin{array}{ccl}
     \fve{A} 	 & = & \{\tis{e}{\tau} : \forall t \tau', \compat{\tau}{\tau'}\land \tis{p}{\tau'}\in\tvp{A} \Rightarrow \tis{p}{\tau'}\orth \tis{e}{\tau}\}\\
     \tvp{A} 	 & = & \{\tis{p}{\tau} : \forall E \tau', \compat{\tau}{\tau'}\land \tis{E}{\tau'}\in\fvE{A} \Rightarrow \tis{p}{\tau} \orth \tis{E}{\tau'}\}\\
     \fvE{A} 	 & = & \{\tis{E}{\tau} : \forall V \tau', \compat{\tau}{\tau'}\land \tis{V}{\tau'}\in\tvV{A} \Rightarrow \tis{V}{\tau'}\orth \tis{E}{\tau}\}\\
     \tvV{A} 	 & = & \{\tis{V}{\tau} : \forall F \tau', \compat{\tau}{\tau'}\land \tis{F}{\tau'}\in\fvF{A} \Rightarrow \tis{V}{\tau} \orth \tis{F}{\tau'}\}\\     
     \fvF{A} 	 & = & \{\tis{F}{\tau} : \forall v \tau', \compat{\tau}{\tau'}\land \tis{v}{\tau'}\in\tvv{A} \Rightarrow \tis{v}{\tau'}\orth \tis{F}{\tau}\}\\
     \tvv{A\imp B} & = & \{\tis{\lambda x .t}{\tau} : \forall u \tau', \compat{\tau}{\tau'}\land \tis{u}{\tau'}\in\tvp{A} \Rightarrow \tis{p}{\overline{\tau\tau'}[a:=u]}\in\tvp{B}\}\\
     \tvv{X} 	 & = & \{\tis{\cbold}{\tau} : \quad\vdash \cbold:{X}\}\\     
   \end{array}$$
\end{definition}
\begin{remark}
We draw the reader attention to the fact that we should actually write $\tvv{A}^\pole,\fvF{A}^\pole$, etc...
and $\tau\real_{\!\!\pole}\!\Gamma$, because the corresponding definitions 
are parameterized by a pole $\pole$. 
As it is common in Krivine's classical realizability, we ease the notations by
removing the annotation $\pole$ whenever there is no ambiguity on the pole.
\end{remark}
If the definition of the different sets might seem complex at first sight, 
we insist on the fact that they naturally follow from the abstract machine in context-free form
where the term and the context (in a command) behave independently of each other.
Intuitively, a realizer at a given level is precisely a term which is going to behave well (be in the pole)
in front of any opponent chosen in the previous level (in the hierarchy $v,F,V$,etc...). 
The definition of the different sets $\tvv{A},\fvF{A},\tvV{A}$, etc... directly stems from this intuition.

In comparison with the usual definition of Krivine's classical realizability, we only considered 
orthogonal sets restricted to some syntactical subcategories.
However, the definition still satisfies the usual monotonicity properties of bi-orthogonal sets:
\begin{proposition}
\label{prop:monotonicity}
For any type $A$ and any given pole $\pole$, we have the following inclusions:
  \begin{enumerate}
   \item $\tvv{A}\subseteq \tvV{A} \subseteq \tvp{A}$;
   \item $\fvF{A}\subseteq \fvE{A} \subseteq \fve{A}$.
  \end{enumerate}
\end{proposition}

\begin{proof}
 See \cite{MiqHer18}.
\end{proof}

We now extend the notion of realizers to stores, by stating that a store $\tau$ realizes a context $\Gamma$
if it binds all the variables $a$ and $\alpha$ in $\Gamma$ to a realizer of the corresponding formula. 

\begin{definition}Given a closed store $\tau$ and a fixed pole $\pole$, 
we say that $\tau$ \emph{realizes} $\Gamma$, which we write\footnote{Once again, 
we should formally write $\tau\real_{\!\!\pole}\!\Gamma$ but we will omit the annotation by $\pole$ as often as possible.} $\tau \Vdash \Gamma$, if:
\begin{enumerate}
 \item for any $(a:A) \in\Gamma$, $\tau\equiv \tau_0[a:=p]\tau_1$ and $\tis{p}{\tau_0} \in \tvp{A}$
 \item for any $(\alpha:A^\negt) \in\Gamma$, $\tau\equiv \tau_0[\alpha:=E]\tau_1$ and $\tis{E}{\tau_0} \in \fvE{A}$\
 \end{enumerate}
 \label{def:store_real}
 
\end{definition}

We are now equipped to prove the adequacy of the type system for the $\lbvtstar$-calculus with respect to the realizability interpretation.
%

\begin{theorem}[Adequacy]\label{lm:adequacy}
The typing rules of Figure~\ref{fig:typing-rules} for the $\lbvtstar$-calculus without co-constants
are adequate with any pole.
In other words, if $\Gamma$ is a typing context, $\pole$ a pole and $\tau$ a store such that ${\tau\real \Gamma}$,
then the following holds:
\begin{enumerate}
 \item If $v$ is a strong value 	such that $\Gamma\vdash_v v:A$, 	then $\tis{v}{\tau} \in\tvv{A}$.
 \item If $F$ is a forcing context 	such that $\Gamma\vdash_F F:A^\negt$, 	then $\tis{F}{\tau} \in\fvF{A}$.
 \item If $V$ is a weak value   	such that $\Gamma\vdash_V V:A$, 	then $\tis{V}{\tau} \in\tvV{A}$.
 \item If $E$ is a catchable context 	such that $\Gamma\vdash_E E:A^\negt$, 	then $\tis{E}{\tau} \in\fvF{A}$.
 \item If $p$ is a term    		such that $\Gamma\vdash_p p:A$, 	then $\tis{p}{\tau} \in\tvp{A}$.
 \item If $e$ is a context 		such that $\Gamma\vdash_e e:A^\negt$, 	then $\tis{e}{\tau} \in\fve{A}$.
 \item If $c$ is a command		such that $\Gamma\vdash_c c$, 		then $c\tau \in \pole$. 
 \item If $\tau'$ is a store	such that $\Gamma\vdash_\tau \tau':\Gamma'$, 	then $\tau\tau' \real \Gamma,\Gamma'$. 
 \item If $c\tau'$ is a closure		such that $\Gamma\vdash_l c\tau'$, 	then $c\tau\tau' \in \pole$. 
\end{enumerate}
\end{theorem}
\begin{proof}
 By induction on typing rules. Most of the case are similar to the proof of adequacy for the call-by-name \lmmt-calculus, hence
 we only give a few key cases. See~\cite[Chapter 6]{these} for a complete proof.
 \prfcase{\implrule}
Assume that $$\infer[\implrule]{\Gamma\vdash_F q\cdot E:(A\imp B)^\negt}{\Gamma\vdash_p q:A & \Gamma \vdash_E E:B^\negt}$$ 
and let $\pole$ be a pole and $\tau$ a store such that $\tau \real \Gamma$.
Let $\tis{\lambda a.p}{\tau'}$ be a closed term in the set $\tvv{A\to B}$ such that $\compat{\tau}{\tau'}$, then we have:
$$\cut{\lambda a.p}{q\cdot E}\overline{\tau\tau'}
~~\rightarrow ~~\cut{q}{\tmu a.\cut{p}{E}}\overline{\tau\tau'}
~~\rightarrow ~~\cut{p}{E}\overline{\tau\tau'}[a:=q]$$
By definition of $\tvv{A\to B}$, this closure is in the pole, and we can conclude by anti-reduction.

\prfcase{\arule} 
Assume that 
$$\infer[\arule]{\Gamma\vdash_V a:A}{(a:A)\in\Gamma}$$ 
and let $\pole$ be a pole and $\tau$ a store such that $\tau \real \Gamma$.
As $(a:A)\in\Gamma$, we know that $\tau$ is of the form $\tau_0[a:=p]\tau_1$
with $\tis{p}{\tau_0}\in\tvp{A}$.
Let $\tis{F}{\tau'}$ be in $\fvF{A}$, with  $\compat{\tau}{\tau'}$. By Lemma~\ref{lm:st_union},
we know that $\overline{\tau\tau'}$ is of the form $\overline{\tau_0}[a:=p]\overline{\tau_1}$.
Hence we have:
$$\cut{a}{F} \overline{\tau_0}[a:=p]\overline{\tau_1} ~~\rightarrow~~ \cut{p}{\tmu[a].\cut{a}{F}\overline{\tau_1}}\overline{\tau_0}$$
and it suffices by anti-reduction to show that the last closure is in the pole $\pole$. 
By induction hypothesis, we know that $\tis{p}{\tau_0}\in\tvt{A}$ thus we only need to show 
that it is in front of a catchable context in $\fvE{A}$.
This corresponds exactly to the next case that we shall prove now.

\prfcase{\eagerrule} 
Assume that
$$\infer[\eagerrule]{\Gamma\vdash_E \tmu[a].\cut{a}{F}\tau':A}{\Gamma,a:A,\Gamma'\vdash_F F:A & \Gamma,a:A\vdash \tau':\Gamma'}$$ 
and let $\pole$ be a pole and $\tau$ a store such that $\tau \real \Gamma$.
Let $\tis{V}{\tau_0}$ be a closed term in $\tvV{A}$ such that $\compat{\tau_0}{\tau}$.
We have that :
$$\cut{V}{\tmu[a].\cut{a}{F}\overline{\tau'}}\overline{\tau_0\tau}~~\rightarrow~~ \cut{V}{F} \overline{\tau_0\tau}[a:=V]\tau'$$
By induction hypothesis, we obtain $\tau[x:=V]\tau'\real \Gamma,x:A,\Gamma'$. 
Up to $\alpha$-conversion in $F$ and $\tau'$, so that the variables in $\tau'$ are disjoint from those in $\tau_0$.
Therefore, we have that $\overline{\tau_0\tau}\real\Gamma$ and then $\tau''\defeq\overline{\tau_0\tau}[a:=V]\tau'\real \Gamma,a:A,\Gamma'$.
By induction hypothesis again, we obtain that $\tis{F}{\tau''}\in\fvF{A}$ 
(this was an assumption in the previous case) 
and as $\tis{V}{\tau_0}\in\tvV{A}$, we finally get that $\tis{V}{\tau_0}\orth\tis{F}{\tau''}$ 
and conclude again by anti-reduction.
\end{proof}

In particular, we can now prove the normalization of typed closures. 
As we already saw in Proposition~\ref{prop:norm_pole}, the set $\pole_{\Downarrow}$ of normalizing closure is a valid pole,
so that it only remains to prove that any typing rule for co-constants is adequate with $\pole_{\Downarrow}$.

\begin{lemma}
 Any typing rule for co-constants is adequate with the pole $\pole_{\Downarrow}$, \emph{i.e.} if $\Gamma$ is a typing context,
 and $\tau$ is a store such that $\tau\real\Gamma$,
 if $\alphabold$ is a co-constant such that $\Gamma\vdash_F \alphabold:A^\negt$, then $\tis{\alphabold}{\tau}\in\fvF{A}$.
\end{lemma}
 \begin{proof}
This lemma directly stems from the observation that for any store $\tau$ and any closed strong value $\tis{v}{\tau'}\in\tvv{A}$, 
 $\cut{v}{\alphabold}\overline{\tau\tau'}$ does not reduce and thus belongs to the pole $\pole_{\Downarrow}$.
 \end{proof}

As a consequence, we get:
\begin{theorem}
\label{thm:normalization}
 If $c\tau$ is a closure of the $\lbvtstar$-calculus such that $\vdash_l c\tau$ is derivable, then $c\tau$ normalizes.
\end{theorem}

This concludes our study of classical call-by-need, and we shall use the same methodology to derive a proof of normalization for {\dlpaw}
in \Cref{s:dlpaw}. We shall now turn to our second preliminary problem, namely the definition of a sequent calculus with dependent types.

\section{A classical sequent calculus with dependent types}
\label{s:dl}
We shall now turn to the second problem prior to the definition of {\dlpaw}:
the definition of classical sequent calculus with dependent types.
In addition of being a necessary step in our process, this question was also
of great interest in itself. 
Indeed, not only are we interested in the definition of such calculus, but also we
are considering the possibility of using it in order to define a dependently typed
continuation-passing style translation. 
The latter was actually regarded as impossible\footnote{  
To quote their paper, they indeed say:
\emph{``We investigate CPS translatability of typed $\lambda$-calculi with
inductive and coinductive types. [...]
These translations also work in the
presence of control operators [...]
No translation is possible along the
same lines for small $\Sigma$-types and sum types with dependent
case.''}} since an article by Barthe and Uustalu 
on the subject~\cite{BarUus02}. 
Nonetheless, such a translation would provide us with a way of soundly compiling a calculus
with dependent types and a form of classical logic into a usual type theory.
Regarding the translation as a syntactic model, it would thus enable to give a way
of soundly extending a type theory (for instance, those underlying the foundations of proof assistants, \emph{e.g.}
CIC for Coq) with a form of computational classical logic.

Yet, in addition to the problem of safely combining control operators and dependent types~\cite{Herbelin05},
the presentation of a dependently typed language under the form of a sequent calculus is a challenge in itself. 
In~\cite{Miquey19}, we introduced such a system, called \dltp, which is a call-by-value sequent calculus with classical control  
and dependent types. 
In comparison with usual type systems, we decorate typing derivations with a list of dependencies
to ensure subject reduction. 
We managed to prove the soundness of the calculus 
by means of a CPS translation taking the dependencies into account.
The very definition of the translation constrained us to use delimited
continuations in the calculus when reducing dependently typed terms. 
At the same time, this unveiled the need for the syntactic restriction of dependencies 
to the \emph{negative-elimination-free} fragment as in {\dpaw}~\cite{Herbelin12}.
Additionally, we showed how to relate our calculus to a similar system by Lepigre~\cite{Lepigre16}, 
whose consistency is proved by means of a realizability interpretation.
In \Cref{s:dlpaw}, we will use the same techniques, namely a list of dependencies and delimited continuations,
to ensure the soundness of {\dlpaw}, and we will follow Lepigre's interpretation of dependent types for the definition
of our realizability model.
Let us then recall here the main intuitions leading to the definition of {\dltp} in~\cite{Miquey19}.

\subsection{Herbelin's paradox} 
\label{s:Herbelin}
The first step in order to soundly mix
dependent types and control operators is to 
understand the difficulty in doing so,
and in particular Herbelin's paradox~\cite{Herbelin05}.
Let us briefly recap his argument here.
Consider a minimal logic of $\Sigma$-types and equality, whose formulas,
terms (only representing natural number) and proofs are defined as follows:
$$\begin{array}{l@{\qquad}lll}
\text{\bf Formulas }&A,B  & ::= & t = u \mid \exists x^\N.A\\
  \text{\bf Terms } & t,u & ::= & n \mid \wit p \\
  \text{\bf Proofs} & p,q & ::= & \refl \mid \subst{p}{q} \mid (t,p) \mid \prf p\qquad
  \end{array}
  \eqno(n\in\N)$$
Let us explain the different proof terms by presenting their typing rules. 
First, the pair $(t,p)$ is as expected a proof for an existential formula $\exists x^\N. A$ (or $\Sigma (x:\N).A$)
where $t$ is a witness for $x$ and $p$ is a certificate for $A[t/x]$. 
This implies that both formulas and proofs are dependent on terms, which is usual in mathematics.
What is less usual in mathematics is that, as in Martin-Löf type theory, dependent types
also allow for terms (and thus for formulas) to be dependent on proofs, 
by means of the constructions $\wit p$ and $\prf p$.
The corresponding typing rules are given by:
$$\renewcommand{\hsep}{\quad~~}
\infer[\!\!\exirule]{\Gamma \vdash (t,p):\exists x^\N A}{\Gamma \vdash p:A(t) & \Gamma \vdash t:\N}
\hsep
\infer[\!\!\prfrule]{\Gamma \vdash \prf p:A[\wit p/x]}{\Gamma \vdash (t,p):\exists x^\N.A}
\hsep
\infer[\!\!\witrule]{\Gamma \vdash \wit t:\N}{\Gamma \vdash t:\exists x^\N. A}
\hsep
\infer{\Gamma\vdash n:\N}{n\in\N}
$$

Then, $\refl$ is a proof term for equality, and $\subst{p}{q}$ allows
to use a proof of an equality $t=u$ to convert a formula $A(t)$ into $A(u)$:
$$
\infer[\reflrule]{\Gamma \vdash \refl : t=u}{t \to u}
\qquad\qquad\qquad
\infer[\substrule]{\Gamma \vdash \subst{p}{q} : B[u]}{\Gamma \vdash p: t=u  & \Gamma\vdash q:B[t]}
$$

The reduction rules for this language, which are safe with respect to typing, are then:
$$ 
\wit(t,p) \to t 
\qquad\qquad
\prf (t,p) \to p 
\qquad\qquad 
\subst{\refl}{p} \to p$$

Starting from this (sound) minimal language, Herbelin showed that its classical extension with the control operators 
$\texttt{call/cc}_k$ and $\throw\,k$ permits to derive a proof of $0=1$~\cite{Herbelin05}.
The $\texttt{call/cc}_k$ operator, which is a binder for the variable $k$, is intended to catch its surrounding evaluation context.
On the contrary, $\throw\,k$ (in which $k$ is bound) discards the current context and restores the context captured by $\texttt{call/cc}_k$.
The addition to the type system of the typing rules for these operators:
$$
\infer{\Gamma \vdash \mathtt{call/cc}_k\, p: A}{\Gamma,k:\neg A\vdash p:A} 
\qquad\qquad\qquad
\infer{\Gamma,k:\neg A \vdash \throw\,k\,p: B}{\Gamma,k:\neg A\vdash p:A} 
$$
allows the definition of the following proof:
$$p_0\defeq \texttt{call/cc}_k\,(0,\throw\, k \, (1,\refl)):\exists x^\N\!.x=1$$
Intuitively 
such a proof catches the context, give 0 as witness (which is incorrect), and a certificate 
that will backtrack and give 1 as witness (which is correct) with a proof of the equality.

 If besides, the following reduction rules\footnote{We do not want 
 to enter into the details of the reduction rules etc., but rather focus on the intuition
of the causes of the problem. For a detailed proof, please refer to~\cite[Section 2]{Herbelin05}.},
are added:
$$\begin{array}{rcl}
 \wit (\mathtt{call/cc}_k\, p) & \red & \mathtt{call/cc}_k (\wit (p[k(\wit\{\;\})/k])) \\
\mathtt{call/cc}_k\, t & \red & t \\
  \end{array}
  \eqno\begin{array}{r}\\(k\notin \FV(t))\end{array}$$
then we can formally derive a proof of $1=0$.
Indeed, the seek of a witness by the term $\wit p_0$ will reduce to $\mathtt{call/cc}_k\, 0$, which itself reduces to $0$.
The proof term $\refl$ is thus a proof of $\wit p_0 = 0$, and we obtain indeed a proof of $1=0$:
$$\infer[\substrule]{\vdash \subst{(\prf p_0)}{\refl} : 1=0}{
\infer[\prfrule]{\vdash \prf p_0:\wit p_0 = 1}{\vdash p_0:\exists x^\N.x=1}
&
\infer[\reflrule]{\vdash \refl:\wit p_0 = 0}{\wit p_0 \red 0}
}
$$

The bottom line of this example is that the same proof $p_0$ is behaving differently in different contexts thanks to control operators, 
causing inconsistencies between the witness and its certificate.
The easiest and usual approach to prevent this is to impose a restriction to values (which are already reduced) for proofs appearing inside dependent types and 
within the operators $\wit$ and $\prf$\!, together with a call-by-value discipline.  
In particular, in the present example this would prevent us from writing $\wit p_0$ and $\prf p_0$.

\subsection{A naive sequent calculus with dependent types}
Here again, rather than directly trying to define a continuation-passing style
we first pay attention to the possibility of
extending the \lmmt-calculus with a form of dependent types.
Let us momentarily assume that we naively add a dependent product $\dptprod{a:A}{B[a]}$
to the \lmmt-calculus type system while
restricting dependencies to values (in order to prevent Herbelin's paradox from occurring).
In other words, we authorize functions $\lambda a.p$ to inhabit the type $\dptprod{a:A}{B[a]}$
if $p$ is of typed $B[a]$ under the assumption that $a:A$ and, dually, 
stacks $q\cdot e$ if $q$ is a value of type $A$ and $e$ a context of type $B[q]$.
Assuming that we have such terms at hands, we can thus get the following derivation:
  $$
  \infer[\cutrule]{{\cut{\lambda a.p}{q\cdot e}}:(\Gamma\vdash\Delta)}{
\infer[\imprrule]{\Gamma \vdash \lambda a.p : \dptprod{a:A}{B[a]}\mid \Delta}{
  \infer{\Gamma,a:A \vdash p : B[a] \mid \Delta}{\Pi_p}
  }
  &
\infer[\implrule]{\Gamma \mid q\cdot e : \dptprod{a:A}{B[a]}\vdash \Delta}{
  \infer{\Gamma \vdash q : A \mid \Delta}{\Pi_q}
  & 
  \infer{\Gamma \mid e : B[q] \vdash \Delta}{\Pi_e}
  &
  q \in V
}
}
$$

The first difficulties arise when trying to prove subject reduction. Indeed, the previous command
should reduce (following a call-by-value discipline as we eventually 
would like to have for \dlpaw) as follows:
$$\cut{\lambda a.p}{q\cdot e} \quad\to\quad \cut{q}{\mut a .\cut{p}{e}}$$
On the right-hand side, we see that $p$, whose type is $B[a]$, is now cut with 
$e$ whose type is $B[q]$. 
Consequently, we are not able to derive a typing 
judgment\footnote{Observe that the problem here arises independently of the value
restriction or not (that is whether we consider that $q$ is a value or not),
and is peculiar to the sequent calculus presentation).} for this command anymore:
$$
\infer[\cutrule]{{\cut{q}{\mut a .\cut{p}{e}}} :(\Gamma\vdash\Delta)} {
  \infer{\Gamma\vdash q:A\mid \Delta}  {\Pi_q}
  &\hspace{-8pt}
  \infer[\mutrule]{\Gamma\mid \mut a.\cut{p}{e}:A \vdash\Delta} {
   \infer=[\scriptsize Mismatch]{\cut{p}{e} :( \Gamma,a:A\vdash\Delta)} {
      {\Gamma,a:A \vdash p:{{\cancel{B[a]}}}\mid\Delta}
      &
      {\Gamma,a:A\mid e:{{\cancel{B[q]}}}\vdash\Delta}
   }
  }
}
$$
Our intuition is that in the full command, $a$ has been linked to $q$ at a previous
level of the typing judgment. 
In particular,  the command is still safe, since the
head-reduction imposes that the command $\cmd{p}{e}$ will not be executed before
the substitution of $a$ by $q$\footnote{Note that even if we were not 
restricting ourselves to values, this would still hold: 
if at some point the command $\cut{p}{e}$ is executed, it is necessarily after that
$q$ has produced a value to substitute for $a$.} is performed and by then 
the problem would have been solved. Roughly speaking, this phenomenon can be seen as a 
desynchronization of the typing process with respect to computation.
The synchronization can be re-established by making explicit a \emph{list of dependencies}
in the typing rules, which links $\tmu$ variables (here $a$) to the associate proof term on the 
left-hand side of the command (here $q$):
$$
\infer[\cutrule]{{\cut{q}{\mut a .\cut{p}{e}}} :(\Gamma\vdash\Delta){;\vide}} {
  \infer*{\Gamma\vdash q:A\mid \Delta}  {\Pi_q}
  &\hspace{-8pt}
  \infer[\mutrule]{\Gamma\mid \mut a.\cut{p}{e}:A \vdash\Delta{;\dpt{.|q}}} {
   \infer[\cutrule]{\cut{p}{e} : \Gamma,a:A\vdash\Delta{;\dpt{a|q}}} {
      {\Gamma,a:A \vdash p:{B[a]}\mid\Delta}
      &
      {\Gamma,a:A\mid e:{B[q]}\vdash\Delta{;\rdpt{p}\dpt{a|q}}}
	}
  }
}
$$

We spare the reader from the formal definition of the corresponding type system\footnote{See \cite{Miquey19} for further details.},
the main idea lying in the use of list of dependencies when typing commands. 
In particular, this is enough to extend the \lmmt-calculus with dependent types (restricted to values) and prove
that the resulting calculus satisfies subject reduction, is normalizing and consistent as a logic.

\subsection{A dependently typed continuation-passing style}
\label{s:nef}
\label{s:cps}

Nonetheless, in addition to the fact that the value restriction is unsatisfactory (it is too restrictive),
a problem still subsists: such a calculus remains incompatible with the definition of a continuation-passing style translation.
Indeed, considering the naive translation of the very same command, assuming that
a type $A$ is translated into $\neg\neg A$\footnote{Recall the translation
given for the call-by-name calculus in \Cref{s:lmmt:cbn_cps}.}, we get:
$$\tr{\cut{q}{\mut a.{\cut{p}{e}}}}\, = \tr{q}\,\tr{\mut a.{\cut{p}{e}}}\, = 
\underbrace{\tr{q}}_{\neg\neg A}\, (\lambda a.\underbrace{{\tr{p}}}_{\text{ $\neg\neg B[a]$}}~~\underbrace{{\tr{e}}}_{\neg B[q]})$$
We are thus facing the same problem when trying to type the translation of the command $\cut{p}{e}$. 
This does not come as a surprise, insofar as we observed a desynchronization between the computation and the typing process
which we only have compensated within the type system.
In order to obtain a well-typed continuation-passing style translation, 
we thus need to understand how to tackle 
 we need to tackle the problem through the operational semantics.

To this end, 
we follow the idea that the correctness is guaranteed by the head-reduction 
strategy, preventing $\cmd{p}{e}$ from reducing before the substitution of 
$a$ was made.
We would like to ensure the same thing happens in the target language (that will
also be equipped with a head-reduction strategy), namely that $\tr{p}$ 
cannot be applied to $\tr{e}$ before $\tr{q}$ has furnished a value to 
substitute for $a$. This would correspond informally to the term: 
$$(\tr{q}  (\lambda a.\tr{p})) \tr{e}$$
Assuming that $q$ eventually produces a value $V$, the previous term would indeed
reduce as follows:
$$(\tr{q}  (\lambda a.\tr{p})) \tr{e} \rightarrow ((\lambda a.\tr{p})\,\tr{V})\,\tr{e} \rightarrow \tr{p} [\tr{V}/a]\,\tr{e}$$
Since $\tr{p} [\tr{V}/a]$ now has a type convertible to $\neg\neg B[q]$,
the term that is produced in the end is well-typed\footnote{ 
Readers should now be familiar with realizability may observe that 
intuitively such a term defines an appropriate realizer, since it eventually
terminates on a correct term $\tr{p[q/a]}\,\tr{e}$.}.
We are now facing three questions about the term $(\tr{q}  (\lambda a.\tr{p})) \tr{e}$:
(1) Would any term $q$ be compatible with such a reduction?
(2) Is this term typable? (3) Does it matches the translation of a term in the source calculus?

Regarding the first question, we observe that if $q$, instead of producing a value,
was a classical proof throwing the current continuation away (for instance $\mu\alpha.c$ where $\alpha\notin FV(c)$), 
this would lead to the following unsafe reduction:
$$(\lambda\alpha.\tr{c}  (\lambda a.\tr{p})) \tr{e} \rightarrow \tr{c} \,\tr{e}.$$
Indeed, through such a translation, $\mu\alpha$ would only be able to catch the local continuation, 
and the term ends in $\tr{c}\tr{e}$ instead of $\tr{c}$.
We thus need to restrict ourselves at least to proof terms that could not throw the
current continuation. 
Syntactically, these are proof terms that can be 
expressed (up to $\alpha$-conversion) with only one continuation variable $\star$
and without applicative context of the shape $q\cdot e$. 
In other words, this corresponds exactly to Herbelin's restriction to the fragment of 
\emph{negative-elimanation-free} (\nef) proofs (see \Cref{s:dpaw}).

The second question concerns the typability of the term $(\tr{q}  (\lambda a.\tr{p})) \tr{e}$,
which is clearly not possible using the naive translation $\neg\neg A$ for the $\tr{q}$.
Nonetheless, the whole term can be typed by turning the type $A\imp \bot$ 
of the continuation that $\tr{q}$ is waiting for into a (dependent) type 
$\dptprod{a:A}{R[a]}$ parameterized by $R$. 
This way, we can have $\tr{q}:\forall R. (\dptprod{a:A}{R[a]}\imp R[q])$ instead of 
$\tr{q}:((A\imp \bot) \imp \bot)$, and for 
$R[a]:=(B(a)\imp \bot)\imp\bot$, the whole term is well-typed\footnote{This is precisely where
lies the difference with Barthe and Uustalu's work, 
who implicitly assume the return type to be fixed to $\bot$~\cite{BarUus02}.}.
It only remains to wonder if any translated term $\tr{q}$ can be given such a type,
and, fortunately, this is the case for all {\nef} terms.
Indeed, \emph{\nef} proofs are precisely terms which, through the translation, 
can only use once the continuation. They hence satisfy some kind of parametricity
equation $\tr{q}\,k = k\,(\tr{q}\,\lambda x.x)$ \cite{Wadler89} 
and can be given a parametric (and dependent) return type (generalizing Friedman $A$-translation~\cite{Friedman78}).
We prefer not enter into the details of the complete continuation-passing style translation
and refer the interested reader to~\cite{Miquey19}.

Last, the term $(\tr{q}  (\lambda a.\tr{p})) \tr{e}$ suggests the use of delimited 
continuations\footnote{We stick here to the presentations of delimited continuations 
in~\cite{HerGhi08,AriHerSab09}, where $\reset$ is used to denote the top-level delimiter.} 
to temporarily encapsulate the evaluation of $q$ when reducing such a command:
$$\cmd{\lambda a.p}{q\cdot e}~\reduce~\cmd{\shift\cmd{q}{\mut a.\cmd{p}{\reset}}}{e}.$$
Intuitively, a term $\shift c$ momentarily moves the top-level focus to $c$ and freezes
its own evaluation context until $c$ reduces to a command of the shape $\cut{p}{\reset}$:
$$ 
\cut{\shift 	{\cut{p}{\reset}}}{e}  \longrightarrow  \cut{p}{e}\qquad
\qquad \mid\qquad
\cut{\shift {c}}{e}  \longrightarrow  \cut{\shift {c'}}{e}
\eqno   (\text{if } c \rightarrow c')
   $$
Once more, the command $\cmd{\shift\cmd{q}{\mut a.\cmd{p}{\reset}}}{e}$ is 
safe under the guarantee that $q$ will not throw away the continuation 
$\mut a.\cmd{p}{\reset}$, and will mimic the aforedescribed reduction:
$$\cmd{\shift\cmd{q}{\mut a.\cmd{p}{\reset}}}{e}~\reduce~\cut{\shift\cut{V}{\mut a.\cmd{p}{\reset}}}{e}~\reduce~\cut{\shift\cut{p[V/a]}{\reset}}{e}~\reduce~\cut{p[V/a]}{e}.$$
It is now easy to define the translation of delimited continuations so that 
our modified reduction systems semantics matches the expected translation:
$$ 
\trp{\shift c}      \defeq  \lambda k.\trtp{c} k \qquad \qquad\qquad
\trtp{\cut{p}{\reset}} \defeq \trp{p}  
$$
In addition to providing an operational semantics that is now compatible with 
a continuation-passing style translation, delimited continuations
bring us a finer control on the use of  the list of dependencies in typing derivations.
Indeed, the places where a desynchronization needs to be compensated correspond exactly
to commands within delimited continuations. Even more, the only type that needs to
be adjusted with the list of dependencies is the one of $\reset$.
This allows us to distinguish between two typing mode, 
a regular mode with sequents without list of dependencies,
and a dependent one, which is activated when going beyond a $\mu\reset$ and whose
sequents we denote by $\Gamma \vdash_d p:A \mid \Delta; \sigma$ (where $\sigma$ is
a list of dependencies).
The main novelties of this extended syntax, which we summarize in \Cref{fig:extension},
are thus: (1) the fragment of \emph{\nef} proofs,
(2) the use of delimited continuations (3) a distinction between a regular and a dependent typing modes.
\begin{figure}[t]
\myfig{
\begin{tabular}{c}
\begin{tabular}{c|c}
 \begin{tabular}{lc@{~~}c@{~~}l}
    \textbf{Proofs}     &  $p$        & $::=$  & $\dots \mid \shift c_\reset$ \\[0.5em]

\textbf{Delimited}\hfill&  $c_\reset$ & $::=$  & $\cmd{p_N}{e_\reset}\mid \cmd{p}{\reset}$\\
\textbf{continuations}  &  $e_\reset$ & $::=$  & $\tmu a.c_\reset$\\[0.5em]

    \end{tabular}
&
\begin{tabular}{l@{\quad~~}c@{~~}c@{~~}l}
{\textbf{\scriptsize{NEF}}}      &  $p_N$      & $::=$  & $V \mid (t,p_N)\mid \mustar. c_N$\\
	       &	     & \multicolumn{2}{c}{$\mid\prf p_N\mid\subst{p_N}{q_N}$} \\
	       &  $c_N$      & $::=$  & $\cmd{p_N}{e_N}$\\
	       &  $e_N$      & $::=$  & $\star \mid \tmu a.c_N$
    \end{tabular}
\end{tabular}\\\\[-0.3em]    
(a) Language
\\[-0.3em]
\rule{0.9\textwidth}{0.4pt}\\[-0.5em]
{
$\begin{array}{c@{\qquad}c}
\\
\infer[\small \reset_I]{\Gamma \vdash \shift c : A \mid \Delta}{c: (\Gamma \vdash_d \Delta,\reset : A;\vide)}
\qquad
\infer[]{{\cmd{p}{e}}:\Gamma\vdash_d \Delta,\reset:B;\sigma}{\Gamma \vdash {p}:{A}  \mid \Delta & \Gamma \mid  {e}:{A} \vdash_d \Delta,\reset:B;\sigma\dpt{\cdot|p} }
\\\\
\infer[\small \reset_E]{\Gamma \mid  \reset : A \vdash_d \Delta,\reset:B;{\sigma\dpt{\cdot|p}}}{B\in A_\sigma}
\qquad
\infer[]{\Gamma \mid  {\mut a.c} : {A} \vdash_d \Delta,\reset:B;{\sigma\dpt{\cdot|p}}}{{c}:(\Gamma,{a}:{A} \vdash_d \Delta,\reset:B;{\sigma\dpt{a|p}})}
\\\\
\end{array}$}
\\ 
(b) Typing rules \\
\end{tabular} 
 }
 \caption{\dltp: extending {\dl} with delimited continuations (excerpt)}
 \label{fig:extension}
\end{figure}

As we shall explain in \Cref{s:small_step}, similar phenomena will occur
in {\dlpaw} with the dependent types $\forall x^T.A$ and $\exists x^T.A$ (remember that we
consider a stratified presentation with terms and proofs).
While the former can be solved with the exact same ideas, the latter will require
the introduction of the dual notion of co-delimited continuations, 
just as the general dependent sum $\dptsum{x:T}{A}$ would have.
Let us briefly go through the same process to highlight the corresponding intuitions.
Consider now a command formed by a pair $(t,p)$ of type $\dptsum{x:T}{A}$ and a context $e$:
$$
\infer{\Gamma \vdash \cut{(t,p)}{e} \mid \Delta}{
  \infer{\Gamma\vdash (t,p):\dptsum{x:T}{A}\mid \Delta}{
   \Gamma \vdash t:T\mid \Delta & \Gamma\vdash p:A[t]\mid \Delta
  }
  &
  \infer{\Gamma \mid {e}:\dptsum{x:T}{A} \vdash \Delta}{\Pi_e}
}$$
This situation is exactly dual to the case of a stack $q\cdot e$ above\footnote{In fact,
they could even be identified in a polarized version of the calculus inspired from Munch-Maccagnoni's system L~\cite{Munch13PhD},
observing that $(\dptprod{a:A}{B})^\negt = \dptsum{a:A}{B^\negt}$.
In other words, $(t,p)$ and $q\cdot e$ are both a pair whose second component depends on the first one.},
hence it does not come as a surprise that the command it naturally reduces to:
$$\cut{(t,p)}{e} \rightarrow \cut{t}{\mut x.\cut{p}{\mut a.\cut{{(x,a)}_{}}{e}}}$$
can not be typed without using a list of dependencies. 
Indeed, the pair $(x,a)$ that is now put in front of $e$ require to remember a link between $x$ and $t$
to be typed:
$$
\infer=[\scriptsize Mismatch]{\Gamma,x:T,a:A[t]\vdash (x,a):\dptsum{x:T}A\mid \Delta}{
       \Gamma,x:T,a:A[t]\vdash x:T\mid \Delta
       &
       \Gamma,x:T,a:{\cancel{A[t]}} \vdash a:{\cancel{A[x]}} \mid \Delta
}
$$
Similarly, if the former can be solved within the type system with the simple addition of a list of dependencies,
the naive continuation-passing style translation of the pair is again ill-typed:
{
$$\trp{(t,p)}\,k \defeq \trt{t}\,(\lambda x.\underbrace{\trp{p}}_{{\neg\neg A[t]}}\,(\underbrace{\lambda a.k\,(x,a)}_{{\neg A[x]}})$$
}
Following the intuition that $\tr{p}$ should not be applied to its continuation 
before $\tr{t}$ has reduced and provided a value to substitute for $x$,
we can twist this term into:
$$\trp{(t,p)}\,k \defeq  {\trp{p}} \,({\trt{t}}\, (\lambda xa.k\,(x,a)) $$
This translated term is now well-typed provided that $\tr{t}$ can be given the 
parametric dependent type $\forall {R}.{\dptprod{x:T}{{R[x]}}}\to {R[t]}$ (which is possible 
for any {\nef} term $t$, and in particular for any term $t$ in {\dpaw}---remember that terms 
are objects of the arithmetic in finite types and do not have access to control operators). 
Once again, such a translation can be obtained by extending the original operational semantics
with co-delimited continuations:
$$\cut{(t,p)}{e} \rightarrow  \cut{p}{{\coshift} \cut{t}{\mut x.\cut{{\coreset}}{\tmu a.\cut{(x,a)}{e}}}} 	$$
where $p$ is somehow frozen to put the focus on the command $\cut{t}{\mut x.\cut{{\coreset}}{\tmu a.\cut{(x,a)}{e}}}$.
It should now be clear to the reader why, when defining a small-step reduction system that
explicit the evaluation of {\dlpaw} terms in \Cref{s:small_step}, we will introduce this notion 
of co-delimited continuations.

\subsection{Realizability interpretation}\label{s:dl_real}
Last but not least, we shall say a few words about the realizability 
interpretation of {\dltp}.
This interpretation takes advantage of a recent paper in which Lepigre
presented a classical system  allowing the use of dependent types with 
a semantic value restriction~\cite{Lepigre16},.
In practice, the type system of his calculus does not contain a dependent product
$\dptprod{a:A}{B}$ strictly speaking, but it contains a predicate $a\in A$ 
allowing the decomposition of the dependent product into: 
$$\forall a.((a\in A) \imp B)$$
as it is usual in Krivine's classical realizability~\cite{Krivine09}.
In his system, the relativization $a\in A$ is restricted to values, so that 
we can only type $V:V\in A$:
$$\infer[\exists_i]{\Gamma \vdash_{val} V:V\in A}{\Gamma\vdash_{val} V:A}$$
However, typing judgments are defined up to 
observational equivalence, so that if $t$ is observationally equivalent to $V$,
one can derive the judgment $t:t\in A$.

Interestingly, as highlighted through the continuation-passing style 
translation\footnote{More precisely, this is exactly the counterpart of the usual
parametricity equation evoked earlier. See \cite[Lemma 4.1]{Miquey19}.},
any {\nef} proof $p:A$ is observationally equivalent to some value $p^+$,
so that we can derive $p:(p\in A)$ from $p^+:(p^+\in A)$.
The {\nef} fragment is thus compatible with the semantical value restriction.
The converse is obviously false, observational equivalence
allowing us to type realizers that would be untyped otherwise\footnote{In particular,
Lepigre's semantical restriction is so permissive that it is not decidable, while
it is easy to decide whether a proof term of {\dltp} is in \nef.}.
As a matter of fact, in~\cite{Miquey19} we defined an embedding from {\dltp}
into Lepigre's calculus that not only preserves typing, but also to transfer 
normalization and correctness properties along this translation. 
Additionally, this embedding has the benefits of providing us with 
an adequate realizability  interpretation for {\dltp} our calculus. 
We shall also mention that the translation from {\dltp} to Lepigre's calculus 
could not have been defined without the use of delimited continuations.
Actually we would have encountered a problem very similar to the one for the 
continuation-passing style translation.
Moreover, the translation of delimited continuations is informative in that
it somehow decompiles them in order to simulate the corresponding reductions in a natural deduction 
fashion. We refer to~\cite[Section 5]{Miquey19} the reader interested into
further details.

In the current paper, we will follow Lepigre's realizability interpretation 
regarding dependent types. We shall thus introduce a predicate $p\in A$
whose interpretation will be defined for proof terms $p$ observationally 
equivalent to a value.

\section{A sequent calculus with dependent types for classical arithmetic}
\label{s:dlpaw}
Drawing on the calculi we introduced in the last sections, we shall now present {\dlpaw}, 
our sequent calculus counterpart of Herbelin's {\dpaw}. 
This calculus provides us with dependent types restricted to the {\nef} fragment, 
for which {\dlpaw} is an extension of \dltp. 
In addition to the language of {\dltp}, {\dlpaw} has terms for classical arithmetic in finite types (PA$^\omega$).
More importantly, it includes a lazily evaluated co-fixpoint operator. 
To this end, the calculus uses a shared store, as in the $\lbvtstar$-calculus. 

We first present the language of {\dlpaw} with its type system and its reduction rules.
We prove that the calculus verifies the property of subject reduction and that it is as expressive as {\dpaw}.
In particular, the proof terms for {\acn} and {\dc} of {\dpaw} can be directly defined in {\dlpaw}. 
We then apply once again the methodology of Danvy's semantic artifacts to derive a small-step calculus (\Cref{s:small_step}),
from which we deduce a realizability interpretation (\Cref{s:realizability}),
which relies on a combination of the corresponding ones developed for the $\lbvtstar$-calculus and {\dltp}.
This interpretation will lead us to the main result of this paper: the normalization of \dlpaw.

Nonetheless, we should say before starting this section that we already have a
guardrail for the normalization.
Informally, we could argue that authorizing infinite stores in the $\lbvtstar$-calculus 
would not alter its normalization. 
Indeed, from the point of view of existing programs (which are finite and typed in finite contexts),
they are computing with a finite knowledge of the memory (and it is easy to see that 
through the realizability interpretation, terms are compatible with store extensions~\cite{MiqHer18}). 
Note that in the store, we could theoretically replace any 
co-fixpoint that produces a stream by the (fully developed) stream in
question.   Due to the presence of backtracks in co-fixpoints, the
store would contain all the possible streams (possibly an infinite
number of it) produced when reducing co-fixpoints.  In this setting,
if a term were to perform an infinite number of reductions
steps, it would necessarily have to explore an infinite number
of cells in the pre-computed memory, independently from its
production. This should not be possible.

This argument is actually quite close from Herbelin's original proof sketch, which this work precisely aims at replacing with a formal proof.
These unprecise explanations should be taken more as spoilers of the final result than as proof sketches.
We shall now present formally {\dlpaw} and prove its normalization, which will then not come as a surprise.
~\\[0.5em]

{ 
\emph{Most of the proofs in this section will resemble a lot to the corresponding ones in the previous sections. 
Yet, as {\dlpaw} gathers all the expressive power and features of the $\lbvtstar$-calculus and {\dltp},
the different proofs also combined all the tools and tricks used in each case. 
We will hence try to avoid repetitions and only highlight the most interesting parts as much as possible. 
}

\subsection{Syntax}

The language of {\dlpaw} is based on the syntax of {\dltp}~\cite{Miquey19}, 
extended with the expressive power of {\dpaw}~\cite{Herbelin12} and with explicit stores as in the $\lbvtstar$-calculus~\cite{AriEtAl12}.
We stick to the stratified presentation of dependent types, that is to say 
that we syntactically distinguish terms—that represent \emph{mathematical objects}—from proof terms—that represent \emph{mathematical proofs}.
In particular, types and formulas are separated as well, matching the syntax of \dpaw's formulas.
Types are defined as finite types with the set of natural numbers as the sole ground type,
while formulas are inductively built on atomic equalities of terms,
by means of conjunctions, disjunctions, first-order quantifications, dependent products and co-inductive formulas:
$$\begin{array}{>{}l@{~~\quad}r@{~~}c@{~~}l}
\text{\bf Types}        &T,U & ::= & \N \mid T\to U \\
\text{\bf Formulas    } &A,B  &::= &\top\mid \bot \mid t = u\mid A\land B \mid A\lor B \mid \exists x^T.A \mid \forall x^T. A \mid \dptprod{a:A}{B} \mid \nu^t_{x,f}A
\end{array}$$

The syntax of terms is identical to the one in {\dpaw}, including functions $\lambda x.t$ 
and applications $t u$, as well as a recursion operator $\rec{t}{xy}{t_0}{t_S}$, 
so that terms represent objects in arithmetic of finite types.
As for proof terms (and contexts, commands), they are now defined with all the expressiveness of {\dpaw}.
Each constructor in the syntax of formulas is reflected by a constructor in the syntax of proofs 
and by the dual co-proof (\emph{i.e.} destructor) in the syntax of evaluation contexts.
Amongst other things, the syntax includes 
pairs $(t,p)$ where $t$ is a term and $p$ a proof, which inhabit the dependent sum type $\exists x^T\!.A$;
dual co-pairs $\tmu(x,a).c$ which bind the (term and proof) variables $x$ and $a$ in the command $c$;
functions $\lambda x.p$ inhabiting the type $\forall x^T\!.A$ together with their dual,
stacks $t\cdot e$ where $e$ is a context whose type might be dependent in $t$;
functions $\lambda a.p$ which inhabit the dependent product type $\dptprod{a:A}{B}$, 
and, dually, stacks $q\cdot e$, where $e$ is a context whose type might be dependent in $q$;
a proof term $\refl$ which is the proof of atomic equalities $t=t$ and
a destructor $\muteq.c$ which allows us to type the command $c$ modulo an equality of terms;
operators $\ind{t}{p_0}{ax}{p_S}$ and $\cofix{t}{bx}{p}$, as in \dpaw, for inductive and coinductive reasoning;
delimited continuations through proofs $\shift c_\tp$ and the context $\reset$;
a distinguished context $[]$ of type $\bot$, which allows us to reason ex-falso.

\begin{figure*}[t]
  \framebox{\vbox{
 $
\begin{array}{l@{\qquad}r@{~~}c@{~~}l}
 \textbf{Closures} & l	& ::= & c\tau \\
\textbf{Commands}  & c	& ::= & \cmd{p}{e} \\[0.8em]

  \textbf{Proof terms } & p,q 	& ::= & a \mid \injec{i}{p} \mid (p,q) \mid (t,p) \mid \lambda x.p \mid \lambda a.p\mid \refl \\
&	& \mid&  \ind{t}{p_0}{ax}{p_S} \mid \cofix{t}{bx}{p}\mid \mu \alpha.c \mid \shift {c_\reset}\\
  \textbf{Proof values} & V	& ::= & a  \mid \injec{i}{V} \mid (V,V) \mid (V_t,V) \mid \lambda x.p \mid \lambda a.p \mid \refl \\[0.8em]

\textbf{Terms} & t,u 	& ::= & x \mid 0 \mid S(t) \mid \rec{t}{xy}{t_0}{t_S}\mid \lambda x.t \mid t~u \mid \wit p \\
\textbf{Terms values}  & V_t 	& ::= & x \mid S^n(0) \mid  \lambda x.t\\\\

\textbf{Stores} & \tau 	& ::= & \varepsilon \mid \tau [a := p_\tau] \mid \tau[\alpha := e]\\ 
\textbf{Storables} & p_\tau 	& ::= & V \mid \ind{V_t}{p_0}{ax}{p_S} \mid \cofix{V_t}{bx}{p}\\[0.8em]
	    
\textbf{Contexts} & e	& ::= & f \mid  \alpha \mid \mut a.c\tau \\[0.5em] 
\textbf{\begin{tabular}{>{\!\!}l}
Forcing\\contexts
\end{tabular}}    & f	& ::= & [] \mid \mut[a_1.c_1\mid a_2.c_2]  \mid \mut(a_1,a_2).c  \mid \mut(x,a).c \mid t\cdot e \mid p\cdot e \mid \muteq.c\\[0.8em]

\textbf{Delimited} &  c_\reset& ::= & \cmd{p_N}{e_\reset}\mid \cmd{p}{\reset} \\
\textbf{continuations} & e_\reset& ::= &  \mut a.c_\reset\tau\mid \mut[a_1.c_\reset\mid a_2.c'_\reset]  \mid  \mut(a_1,a_2).c_\reset \mid \mut(x,a).c_\reset\\[0.8em]

& c_N 	& ::= & \cmd{p_N}{e_N} \\
\textbf{\footnotesize NEF} & p_N,q_N & ::= & a \mid \injec{i}{p_N} \mid (p_N,q_N) \mid   (t,p_N) \mid\lambda x.p \mid \lambda a.p \mid \refl\\
& 	& \mid&  \ind{t}{p_N}{ax}{q_N} \mid \cofix{t}{bx}{p_N}  \mid \mustar.c_N \mid \shift {c_\reset}\\
& e_N 	& ::= & \star \mid \mut[a_1.c_N\mid a_2.c'_N] \mid \mut a.c_N\tau \mid \mut(a_1,a_2).c_N \mid \mut(x,a).c_N  \\
\end{array}$

}}
  \caption{The language of \dlpaw}
  \label{fig:language}
\end{figure*}

As in {\dltp}, the syntax of {\nef} proofs, contexts and commands is defined 
as a restriction of the previous syntax. Technically, they are defined (modulo $\alpha$-conversion)
with only one distinguished context variable $\star$ (and consequently only one binder $\mustar.c$), 
and without stacks of the shape $t\cdot e$ or $q\cdot e$ to avoid applications 
(recall that one can understand {\nef} proofs as the proofs that cannot drop their continuation). 
The commands $c_\reset$ within delimited continuations are defined as commands of the shape $\cut{p}{\reset}$ or formed by a {\nef} proof
and a context of the shape $\mut a.c_\reset\tau$, $\mut[a_1.c_\reset| a_2.c'_\reset]$\vspace{-1mm},
$\mut(a_1,a_2).c_\reset$ or $\mut(x,a).c_\reset$.

We adopt a call-by-value evaluation strategy except for fixpoint operators\footnote{To highlight the duality 
between inductive and coinductive fixpoints, we evaluate both in a lazy way.
Even though this is not indispensable for inductive fixpoints, we find this approach more natural in that
we can treat both in a similar way in the small-step reduction system and thus through the realizability interpretation.}, 
which are evaluated in a lazy way. 
To this purpose, we use \emph{stores} in the spirit of the $\lbvtstar$-calculus, 
which are again defined as lists of bindings of the shape $[a:=p]$ where $p$ is 
now a value or a (co-)fixpoint,
and of bindings of the shape $[\alpha:=e]$ where $e$ is any context.
We assume that each variable occurs at most once in a store $\tau$,
we thus reason up to $\alpha$-reduction and we assume the capability of generating fresh names.
Apart from evaluation contexts of the shape $\tmu a.c$ and co-variables $\alpha$,
all the contexts 
are \emph{forcing contexts} which eagerly require a value to be reduced and trigger the evaluation of lazily stored terms.
The resulting language is given in Figure\,\ref{fig:language}.

\begin{figure}[p]
 \framebox{\vbox{\input{figures/dlpaw/bigstep}}}
 \caption{Reduction rules of \dlpaw}
 \label{fig:bigstep}
\end{figure}

\subsection{Reduction rules}
The reduction system of {\dlpaw} is given in Figure\,\ref{fig:bigstep}.
The basic rules are those of the call-by-value \lmmt-calculus and of {\dltp}. 
The rules for delimited continuations are exactly the same as in {\dltp},
except that we have to prevent $\reset$ from being caught and stored by a proof $\mu\alpha.c$.
We thus distinguish two rules for commands of the shape $\cut{\mu\alpha.c}{e}$, depending
on whether $e$ is of the shape $e_\reset$ or not. 
In the former case, we perform the substitution $[e_\reset/\alpha]$, 
which is linear since $\mu\alpha.c$ is necessarily {\nef}.
We should also mention in passing that we abuse the syntax in every other rules, 
since $e$ should actually refer to $e$ or $e_\tp$ 
(or the reduction of delimited continuations would be stuck).
Elimination rules correspond to commands where the proof is a constructor (say of pairs) applied to values,
and where the context is the matching destructor. 
Call-by-value rules correspond to ($\varsigma$) rule of Wadler's sequent calculus~\cite{Wadler03}.
The next rules express the fact that (co-)fixpoints are lazily stored, and reduced only if their value is eagerly demanded 
by a forcing context. Observe that in that case, the unfolding of (co-)fixpoints is entirely done within the store.
Last, terms are reduced according to the usual $\beta$-reduction, 
with the operator $\texttt{rec}$ computing with the usual recursion rules. 
It is worth noting that the stratified presentation allows to define the reduction of terms as external: 
within proofs and contexts, terms are reduced in place. Consequently, as in {\dltp} the very same happen for 
{\nef} proofs embedded within terms. 
Computationally speaking, this corresponds indeed to the intuition that terms are reduced on an external device.


\subsection{Typing rules}
As often in Martin-L\"of's intensional type theory,
formulas are considered up to equational theory on terms.
We denote by $A\equiv B$ the reflexive-transitive-symmetric closure of the relation $\typered$ induced by the reduction of terms and {\nef} 
proofs as follows:
\begin{center}
\begin{tabular}{lcl}
   $A[t]$ &$\typered$ &$A[t']\quad$ whenever $\quad t\rightarrow_\beta t'$ \\
   $A[p]$ &$\typered$& $A[q] \quad$ whenever $\quad \forall \alpha\,(\cmd{p}{\alpha}\rightarrow\cmd{q}{\alpha})$ \\
\end{tabular}
\end{center}
in addition to the reduction rules for equality and for coinductive formulas:
$$
\begin{array}{r@{~~}c@{~~}l}
    0 = S(t) & \typered & \bot \\
    S(t) = 0 & \typered & \bot \\
\end{array}
\qquad
\begin{array}{r@{~~}c@{~~}l}
    S(t) = S(u) & \typered & t = u \\
   \nu ^t_{fx} A &\typered& A[t/x][\nu^y_{fx} A/f(y)=0] 
\end{array}$$

We work with one-sided sequents where typing contexts are defined by:
$$
\begin{array}{rcl}\Gamma,\Gamma' & ::= & \eps \mid \Gamma,x:T \mid \Gamma,a:A \mid \Gamma,\alpha:A^\negt\mid \Gamma,\reset:A^\negt.
\end{array}
$$
using the notation $\alpha:A^\negt$ for an 
assumption of the refutation of $A$.
This allows us to mix hypotheses over terms, proofs and contexts while keeping 
track of the order in which they are added (which is necessary because of the dependencies).
We assume that a variable occurs at most once in a typing context.

We define nine syntactic kinds of typing judgments:
\begin{itemize}
 \item six in regular mode, that we write $\Gamma \sigdash J$:\nomidem
 \begin{multicols}{2}
\begin{enumerate}
 \item $\Gamma\sigdash t:T$       ~~~\;for  terms,
 \item $\Gamma\sigdash p:A$       ~~\;\,for  proofs,
 \item $\Gamma\sigdash e:A^\negt$ ~for  contexts, 
 \item $\Gamma\sigdash c$         ~~~for  commands, 
 \item $\Gamma\sigdash c\tau$     ~for  closures, 
 \item $\Gamma\sigdash \tau':(\Gamma';\sigma')$ for  stores;
\end{enumerate}
\end{multicols}\nomidem
\item three more for the dependent mode, that we
write $\Gamma\vdash_d J;\sigma$:\\[-0.7em]
\begin{enumerate}\setcounter{enumi}{6}
 \item $\Gamma\vdash_d e:A^\negt;\sigma$ ~for typing contexts, 
 \item $\Gamma\vdash_d c    ;\sigma$         ~~~for typing commands, 
 \item $\Gamma\vdash_d c\tau;\sigma$     ~for typing closures.
\end{enumerate}
\end{itemize}
In each case, $\sigma$ is a list of dependencies---we explain the presence of a list of dependencies in each case thereafter---, which are defined from the following grammar:
$$
\sigma ::= \varepsilon \mid \sigma\dpt{p|q}
$$
The substitution on formulas according to a list of dependencies $\sigma$ is defined by:\nomidem
$$
\eps(A) \defeq \{A\}\qquad\qquad\qquad
\sigma\dpt{p|q}(A) \defeq \begin{cases}
 \sigma (A[q/p]) & \text{if $q\in\nef$} \\
 \sigma(A) & \text{otherwise}
\end{cases}
$$
Because the language of proof terms include constructors for pairs, injections, etc, 
the notation $A[q/p]$ does not refer to usual substitutions properly speaking: $p$ can be a pattern (for instance $(a_1,a_2)$)
and not only a variable.

\begin{figure}[t]
\framebox{\vbox{\input{figures/dlpaw/types}}}
  \caption{Type system for {\dlpaw} - Regular mode}
 \label{fig:dlpaw_types}
\end{figure}
\begin{figure}[t]
\ContinuedFloat
\framebox{\vbox{\input{figures/dlpaw/types2}}}
  \caption{Type system for {\dlpaw} - Dependent mode and terms}
 \label{fig:dlpaw_types}
\end{figure}

We shall attract the reader's attention to the fact that 
all typing judgments include a list of dependencies.
Indeed, as in the $\lbvtstar$-calculus, when a proof or a context is caught by a binder, say $V$ and $\tmu a$, 
the substitution $[V/a]$ is not performed but rather put in the store: $\tau[a:=V]$.
Now, consider for instance the reduction of a dependent function $\lambda a.p$ (of type $\dptprod{a:A}{B}$)
applied to a stack $V\cdot e$:
\begin{align*}\cut{\lambda a .p}{V\cdot e}\tau  &\red \cut{\shift\cut{V}{\tmu a.\cut{p}{\reset}}}{e}\tau\\
&\red \cut{\shift\cut{p}{\reset}}{e}\tau[a:=V]
 \red \cut{p}{e}\tau[a:=V]\nomidem
\end{align*}
Since $p$ still contains the variable $a$, whence his type is still $B[a]$, whereas the type of $e$ is $B[V]$.
We thus need to compensate the missing substitution\footnote{On the contrary, 
the reduced command in {\dltp}  would have been $\cut{p[V/a]}{e}$, which is 
typable with the \cutrule~rule over the formula $B[V/a]$.}. 

We are mostly left with two choices.
Either we mimic the substitution in the type system, which would amount to the following typing rule:
$$\qquad\qquad
\infer{\sigmaopt\Gamma\vdash c\tau}{\Gamma,\Gamma'\vdash \tau(c) & \Gamma \vdash \tau:\Gamma'}
\qquad\quad
\raisebox{1em}{\mbox{$
\begin{array}{ll}
\begin{array}{l}
 \text{where:}\\
\end{array}
&
\begin{array}{r@{~}c@{~}l}
\tau[\alpha:=e](c) &\defeq& \tau(c) \\
\tau[a:=p](c)      &\defeq& \tau(c)\hfill(p\notin\nef) \\
 \tau[a:=p_N](c)    &\defeq& \tau(c[p_N/a]) \quad~~ (p\in\nef)\\ 
\end{array}
\end{array}
$}
}
$$
Or we type stores in the spirit of the $\lbvtstar$-calculus, 
and we carry along the derivations all the bindings liable to be used in types, which constitutes again a list of dependencies.

The former solution has the advantage of solving the problem before typing the command, 
but it has the flaw of performing computations which would not occur in the reduction system.
For instance, the substitution $\tau(c)$ could duplicate co-fixpoints (and their typing derivations), which would never happen in the calculus.
That is the reason why we favor the other solution, which is closer to the calculus in our opinion. 
Yet, it has the drawback that it forces us to carry a list of dependencies even in regular mode. 
Since this list is fixed (it does not evolve in the derivation except when stores occur), 
we differentiate the denotation of regular typing judgments, written $\Gamma \sigdash J$, from the one of judgments in dependent mode, 
which we write $\Gamma \vdash_d J;\sigma$ to highlight that $\sigma$ grows along derivations.
The type system we obtain is given in \Cref{fig:dlpaw_types}.

\subsection{Subject reduction}
\label{s:sub_red}



We shall now prove that typing is preserved along reduction. As for the $\lbvtstar$-calculus,
the proof is simplified by the fact that substitutions are not performed (except for terms), which keeps us from proving 
the safety of the corresponding substitutions.
Yet, we first need to prove some technical lemmas about dependencies. 
To this aim, we define a relation $\sigma\dptimp \sigma'$
between lists of dependencies, which expresses the fact that any typing derivation obtained with $\sigma$ could be obtained as well as with $\sigma'$:
 $$\sigma \dptimp \sigma' ~\defeq~\sigma(A) =\sigma(B) \limp \sigma'(A) =\sigma'(B) \eqno(\text{for any}~ A,B)$$
We first show that the cases which we encounter in the proof of subject reduction satisfy this relation:
\begin{lemma}[Dependencies implication]The following holds for any $\sigma,\sigma',\sigma''$:
\begin{enumerate}
 \item $\sigma\sigma'' \dptimp \sigma\sigma'\sigma'$
 \item $\sigma\dpt{(a_1,a_2)|(V_1,V_2)} \dptimp \sigma\dpt{a_1|V_1}\dpt{a_2|V_2}$
 \item $\sigma\dpt{\injec i a|\injec i V} \dptimp \sigma\dpt{a|V}$
 \item $\sigma\dpt{(x,a)|(t,V)} \dptimp \sigma\dpt{a|V}\dpt{x|t}$
 \item $\sigma\rdpt{(p_1,p_2)} \dptimp \sigma\dpt{a_1|p_1}\dpt{a_2|p_2}\rdpt{(a_1,a_2)}$
 \item $\sigma\rdpt{\injec i p} \dptimp \sigma\dpt{a|p}\rdpt{\injec i a}$
 \item $\sigma\rdpt{(t,p)} \dptimp \sigma\dpt{a|p}\rdpt{(t,a)}$
\end{enumerate}
\noindent where the fourth item abuse the definition of list of dependencies to include a substitution of terms.
\label{lm:dlpaw:dpt_imp}
\end{lemma}
\begin{proof}
 All the properties are trivial from the definition of the substitution $\sigma(A)$.
\end{proof}

We can now prove that the relation $\dptimp$ indeed matches the expected intuition:
\begin{proposition}[Dependencies weakening]\label{prop:dlpaw:dpt_weak} 
 If $\sigma,\sigma'$ are two lists of dependencies such that $\sigma\dptimp\sigma'$, then any derivation using $\sigma$ can be done using $\sigma'$ instead. 
 In other words, the following rules are admissible:
 $$
 \infer[\weakrule]{\Gamma \vdash^{\sigma'} J}{\Gamma \vdash^{\sigma} J}
 \qquad\qquad\qquad
 \infer[\weakdrule]{\Gamma \vdash_d J;\sigma'}{\Gamma \vdash_d J;\sigma}
 $$
 for any judgment $J$ among $p:A$,$c$,$e:A^\negt,t:T$.
\end{proposition}
\begin{proof}
 Simple induction on the typing derivations. The rules {\resetrule} and {\cutrule} where the list of dependencies is used 
 exactly match the definition of $\dptimp$.
 Every other case is direct using the first item of \Cref{lm:dlpaw:dpt_imp}.
\end{proof}

To simplify the proof of subject reduction, we prove that the concatenation of stores is typable:
\begin{lemma}\label{lm:store_cat}
 The following rule is admissible:
 $$
 \infer[\taucatrule]{\Gamma\sigdash \tau_0\tau_1:(\Gamma_0,\Gamma_1;\sigma_0,\sigma_1)}{
    \Gamma\sigdash \tau_0:(\Gamma_0;\sigma_0)
    &
    \Gamma,\Gamma_0\vdash^{\sigma\sigma_0}\tau_1:(\Gamma_1;\sigma_1)
  }
  $$
\end{lemma}
\begin{proof}
 By induction on the structure of $\tau_1$.
\end{proof}

As explained, we only need to prove that term substitution is safe:
\begin{lemma}[Safe term substitution]\label{lm:safet}
 If $~\Gamma\sigdash t:T $ then for any conclusion $J$ for typing proofs, contexts, terms, etc; the following holds:
 \begin{enumerate}
  \item If ~$ \Gamma,x:T,\Gamma'\sigdash J       $ \quad~	then~~{$        \Gamma,\Gamma'[t/x]\vdash^{\sigma[t/x]} J[t/x]$.}
  \item If ~$ \Gamma,x:T,\Gamma'\vdash_d J;\sigma$ ~		then~~{$        \Gamma,\Gamma'[t/x]\vdash_d J[t/x];\sigma[t/x] $.}
 \end{enumerate}
\end{lemma}
\begin{proof}
 By induction on typing rules.
\end{proof}

We can prove the safety of reduction with respect to typing:
\begin{theorem}[Subject reduction]
\label{thm:subject_reduction}
For any context $\Gamma$ and any closures $c\tau$ and $c'\tau'$ such that $c\tau \rightarrow c'\tau'$, we have:\nomidem
\begin{multicols}{2}
 \begin{enumerate}
 \item If ~$\Gamma\vdash c\tau$~then~$\Gamma\vdash c'\tau'$.
 \item If ~$\Gamma\vdash_d c\tau;\varepsilon$~then~$\Gamma\vdash_d c'\tau';\varepsilon$.
 \end{enumerate}
 \end{multicols}
\end{theorem}
\begin{proof}
The proof follows the usual proof of subject reduction, by induction on the typing derivation and the reduction $c\tau \rightarrow c'\tau'$.
Since there is no substitution but for terms (proof terms and contexts being stored), there is no need for auxiliary lemmas 
about the safety of substitution. 
We sketch it by examining all the rules from \Cref{fig:dlpaw_types} from top to bottom.
~\\[0.2em]{\textbullet\quad} The cases for reductions of $\lambda$ are identical to the cases proven in the previous chapter for $\dltp$.
~\\[0.2em]{\textbullet\quad} The rules for reducing $\mu$ and $\tmu$ are almost the same
except that elements are stored, which makes it even easier.
For instance in the case of $\tmu$, the reduction rule is:
$$ \cmd{V}{\mut a.c\tau_1}\tau_0 				 \red c\tau_0[a:=V]\tau_1$$
A typing derivation in regular mode for the command on the left-hand side is of the shape:
{
$$
\infer[\lrule]{\Gamma\sigdash\cmd{V}{\mut a.c\tau_1}\tau_0}{
  \infer[\cutrule]{\Gamma,\Gamma_0\vdash^{\sigma\sigma_0}\cmd{V}{\mut a.c\tau_1}}{
    \infer{\Gamma,\Gamma_0\vdash^{\sigma\sigma_0}V:A}{\Pi_V}
    &
    \infer[\mutrule]{\Gamma,\Gamma_0\vdash^{\sigma\sigma_0}\mut a.c\tau_1:A^\negt}{
       \infer[\lrule]{\Gamma,\Gamma_0,a:A\vdash^{\sigma\sigma_0}c\tau_1}{
	  \infer{\Gamma,\Gamma_0,a:A,\Gamma_1\vdash^{\sigma\sigma_0\sigma_1}c}{\Pi_c}
	  &
	  \infer{\Gamma,\Gamma_0,a:A\vdash^{\sigma\sigma_0}\tau_1:(\Gamma_1;\sigma_1)}{\Pi_{\tau_1}}  
       }
    }
  }
  &\hspace{-1cm}
  \infer{\Gamma\vdash^{\sigma}\tau_0:(\Gamma_0;\sigma_0)}{\Pi_{\tau_0}}
}
$$
}
Thus we can type the command on the right-hand side:
{
$$
\scalebox{0.95}{
\infer[\!\lrule]{\Gamma\sigdash c\tau_0[a:=V]\tau_1}{
  \infer[\!\!\weakrule]{\Gamma_\tau\vdash^{\sigma\sigma_0\dpt{a|V}\sigma_1}c}{
    \infer{\Gamma_\tau\vdash^{\sigma\sigma_0\sigma_1}c}{\Pi_c} 
  }
  &\!\!
  \infer[\!\!\taucatrule]{\Gamma\sigdash \tau_0[a:=V]\tau_1:(\Gamma_0,a:A,\Gamma_1;\sigma_0\dpt{a|V}\sigma_1)}{
    \infer[\!\!\tauprule]{\Gamma\sigdash \tau_0[a:=V]:(\Gamma_0,a:A;\sigma_0,\dpt{a|V})}{
      \infer{\Gamma\vdash^{\sigma}\tau_0:(\Gamma_0;\sigma_0)}{\Pi_{\tau_0}}
      &
      \infer{\Gamma,\Gamma_0\vdash^{\sigma\sigma_0}V:A}{\Pi_V}
    }
    &
    \infer{\Gamma,\Gamma_0,a:A\vdash^{\sigma\sigma_0}\tau_1:(\Gamma_1;\sigma_1)}{\Pi_{\tau_1}}
  }
}
}
$$
}
\noindent \!\! where $\Gamma_\tau\defeq \Gamma,\Gamma_0,a:A,\Gamma_1$.
As for the dependent mode, the binding $\dpt{a|p}$ within the list of dependencies is compensated when typing the store as shown in the last derivation. 

~\\{\textbullet\quad} Similarly, elimination rules for contexts $\tmu[a_1.c_1|a_2.c_2]$, $\tmu(a_1,a_2).c$, $\tmu(x,a).c$ or $\muteq.c$ 
are easy to check, using \Cref{lm:dlpaw:dpt_imp} and the rule {\tauprule} in dependent mode to prove the safety with respect to dependencies.

~\\{\textbullet\quad} The cases for delimited continuations are identical to the corresponding cases for $\dltp$.

~\\{\textbullet\quad} The cases for the so-called ``call-by-value'' rules opening constructors are straightforward, 
using again \Cref{lm:dlpaw:dpt_imp} in dependent mode to prove the consistency with respect to the list of dependencies.

~\\{\textbullet\quad} The cases for the lazy rules are trivial.

~\\{\textbullet\quad} The first case in the ``lookup'' section is trivial. 
The three lefts correspond to the usual unfolding of inductive and co-inductive fixpoints. We only sketch the latter in regular mode.
The reduction rule is:
$$\cut{a}{f}\tau_0[a:=\cofix{t}{bx}{p}]\tau_1 ~\red~ \cmd{p[t/x][b'/b]}{\mut a.\cut{a}{f}\tau_1}\tau_0[b':=\lambda y.\cofix{y}{bx}{p}] $$
The crucial part of the derivation for the left-hand side command is the derivation for the cofix in the store:
{
$$
\infer[\tauprule]{\Gamma\sigdash \tau_0[a:=\cofix{t}{bx}{p}]:(\Gamma_0,a:\nu_{fx}^t A;\sigma_0)}{
  \infer{\Gamma\sigdash \tau_0:(\Gamma_0;\sigma_0)}{\Pi_{\tau_0}}
  &
  \infer[\cofixrule]{\Gamma,\Gamma_0\vdash^{\sigma\sigma_0} \cofix{t}{bx}{p}:\nu_{fx}^t A}{
    \infer{\Gamma\vdash^{\sigma\sigma_0} t:T}{\Pi_t}
    & 
    \infer{\Gamma,\Gamma_0, f:T\imp \N,x:T,b:\forall y^T. f(y)=0\vdash^{\sigma\sigma_0} p:A}{\Pi_p}
  }
}
$$
}
Then, using this derivation, we can type the store of the right-hand side command:
$$
\infer[\tauprule]{\Gamma\sigdash \tau_0[b':=\lambda y.\cofix{y}{bx}{p}]:\Gamma_0,b':-\forall y.\nu_{fx}^y A}{
 \infer{\Gamma\sigdash \tau_0:(\Gamma_0;\sigma_0)}{\Pi_{\tau_0}}
  &\hspace{-1cm}
  \infer[\farrule]{\Gamma,\Gamma_0\vdash^{\sigma\sigma_0} \lambda y.\cofix{y}{bx}{p}:\forall y.\nu_{fx}^t A}{
    \infer[\cofixrule]{\Gamma,\Gamma_0,y:T\vdash^{\sigma\sigma_0} \cofix{y}{bx}{p}:\nu_{fx}^y A}{
      \infer{\Gamma,\Gamma_0,y:T\vdash^{\sigma\sigma_0} y:T}{}
      & 
      \infer{\Gamma,\Gamma_0, f:T\imp \N,x:T,b:\forall y^T. f(y)=0\vdash^{\sigma\sigma_0} p:A}{\Pi_p}
    }
  }
}
$$
It only remains to type (we avoid the rest of the derivation, which is less interesting) the proof $p[t/x]$
with this new store to ensure us that the reduction is safe (since the variable $a$ will still be of  type $\nu^t_{fx} A$
when typing the rest of the command):
$$
  \infer[\convrrule]{\Gamma,\Gamma_0,b:\forall y.\nu_{fx}^y A\sigdash p[t/x]:\nu_{fx}^t A}{
    \infer{\Gamma,\Gamma_0,b:\forall y.\nu_{fx}^y A\sigdash p[t/x]:A[t/x][\nu_{fx}^y A/f(y)=0]}{\Pi_p}
    &
    \nu_{fx}^t A \equiv A[t/x][\nu_{fx}^y A/f(y)=0]
  }
$$

~\\{\textbullet\quad} The cases for reductions of terms are easy.
Since terms are reduced in place within proofs, the only things to check is that the reduction 
of $\wit$ preserves types (which is trivial) and that the $\beta$-reduction 
verifies the subject reduction (which is a well-known fact).  \qedhere

 \end{proof}

\subsection{Natural deduction as macros}
\label{s:macros}

\begin{figure*}[t]
 \frame{\vbox{
 \input{figures/dlpaw/ded_nat}
 }}
 \caption{Typing rules of $\dpaw$}
 \label{dpaw_types}
\end{figure*}

We can recover the usual proof terms for elimination rules in natural deduction systems
by defining them as macros in our language. 
In particular, this provides us with an embedding of {\dpaw} proof terms into {\dlpaw} syntax.
The definitions are straightforward and follow the same intuition than for defining 
the application in the \lmmt-calculus (see \Cref{rmk:application}): 
we consider the expected reduction rule in an abstract machine (for instance 
$\cmd{\throw{e}p}{e'} \to \cmd{p}{e}$) and define the macro by solving the induced equation
 (in that case, $\throw{e}p \defeq \mu\beta.\cmd{p}{e}$).
Observe that we use delimited continuations
to define $\letop \dots\inop $ and the constructors over {\nef} proofs which
 might be dependently typed:
$$
\begin{array}{l}
 \begin{array}{@{}r@{~~\defeq~~}l}

  {\letin{a}{p}q} 		& \mu\alpha_p.\cmd{{p}}{\mut a.\cmd{{q}}{\alpha_p}} \\
  {\split{p}{a_1,a_2}q} 	& \mu\alpha_p.\cmd{{p}}{\mut (a_1,a_2).\cmd{{q}}{\alpha_p}} \\
  {\case{p}{a_1.p_1}{a_2.p_2}}  & \mu\alpha_p.\cmd{{p}}{\mut[a_1.\cmd{{p_1}}{\alpha_p}|a_2.\cmd{{p_2}}{\alpha_p}]}\\
  {\dest{p}{a,x}q}  		& \mu\alpha_p.\cmd{{p}}{\mut (x,a).\cmd{{q}}{\alpha_p}}  \\
  \prf p 			&  \shift\cmd{p}{\tmu (x,a).\cmd{a}{\reset}}\\
  
\end{array}\\
\\[-0.5em]
\begin{array}{c@{~~}|@{~~}c}
\begin{array}{r@{~\defeq~}l@{}}
  \subst p q			& \mu \alpha.\cmd{p}{\muteq.\cmd{q}{\alpha}} \\
  {\exf{p}}  			& \mu\alpha.\cmd{{p}}{[]} 
\end{array}&
\begin{array}{r@{~\defeq~}l@{}}
  {\catch{\alpha}p}  		& \mu\alpha.\cmd{{p}}{\alpha} \\
  {\throw{\alpha}~p}  		& \mu\_.\cmd{{p}}{\alpha}\\
\end{array}
\end{array}
\end{array}
$$
where $\alpha_p = \reset$ if $p$ is {\nef} and $\alpha_p = \alpha$ otherwise.
It is then easy to check that the macros match the expected typing rules from \dpaw's type system~\cite{Herbelin12}.

\begin{proposition}[Natural deduction]
 The typing rules from $\dpaw$, given in Figure~\ref{dpaw_types}, are admissible.
\end{proposition}
\begin{proof}
Straightforward derivations, as an example we give here the two cases for the macros $\subst p q$ and $\prf p$ 
 (in natural deduction) that are admissible in {\dlpaw}.
 Recall that we have the following typing rules in {\dlpaw}:
 $$
 \infer[\!\!\exlrule]{\sigmaopt\Gamma \sigdash \mut (x,a).c: (\exists x^T. A)^\negt\optsigma }{\sigmaopt\Gamma, x:T, a:A\sigdash c\optsigma}
  \qquad
  \infer[\eqrule]{\sigmaopt\Gamma \sigdash \muteq.\cut{p}{e} : (t=u)^\negt\optsigma }{\sigmaopt\Gamma \sigdash p:A\optsigma & \sigmaopt\Gamma \sigdash e : A[u/t]\optsigma}	
  $$
  and that we defined $\prf p$ and $\subst p q$ as syntactic sugar: 
  $$\prf p \defeq \shift\cmd{p}{\tmu (x,a).\cmd{a}{\reset}}\qquad\qquad
  \subst p q\defeq \mu \alpha.\cmd{p}{\muteq.\cmd{q}{\alpha}}.$$
  Observe that $\prf p$ is now only definable if $p$ is a {\nef} proof term.
  For any $p\in\nef$ and any variables $a,\alpha$, we can prove the admissibility of the $\prfrule$-rule:
{

  $$
  \scalebox{0.96}{
  \infer{\Gamma\sigdash \shift\cmd{p}{\tmu (x,a).\cmd{a}{\reset}}:A(\wit p)\mid \Delta}{
	\infer[\cutrule]{\cmd{p}{\tmu (x,a).\cmd{a}{\alpha}}:\Gamma\vdash_d \Delta,\reset:A(\wit p);\sigma\rdpt{p}}{
	  \Gamma\sigdash p:\exists x^{T}\!.A\mid \Delta
	  \hspace{-2cm}
	  & 
	  \infer{\Gamma\mid \tmu (x,a).\cmd{a}{\reset}:\exists x^{T}\!. A\vdash_d \Delta,\reset:A(\wit p);\sigma\rdpt{p}}{
	  \infer[\cutrule]{\cmd{a}{\alpha}:\Gamma,x:T,a:A(x)\vdash_d \Delta,\reset:A(\wit p);\sigma\dpt{(x,a)|p}}{
	    \infer[\!\!\convrrule]{a:A(x)\sigdash a:A(\wit(x,a))}{
	      \infer{a:A(x)\sigdash a:A(x)}{}
	    }
	    &
	    \infer[\!\!\resetrule]{\Gamma\mid \reset:A(\wit (x,a)) \vdash_d\reset:A(\wit p)\mid\Delta;\sigma}{
	      \sigma\dpt{(x,a)|p}(A(\wit p))=\sigma\dpt{(x,a)|p}(A(\wit (x,a)))}
	    }
	    }
	 }
      }
      }
      $$
}  
Similarly, we can prove that the ($\subst{\!}{\!}$)-rule is admissible:
{  $$\scalebox{0.96}{
      \infer[\murule]{\Gamma\sigdash \mu \alpha.\cmd{p}{\muteq.\cmd{q}{\alpha}}:B[u]\mid \Delta}{
	\infer[\cutrule]{\cmd{p}{\muteq.\cmd{q}{\alpha}}:\Gamma\sigdash \Delta,\alpha:B[u]}{
	  \Gamma\sigdash p:t=u\mid \Delta
	  & \infer[\eqrule]{\Gamma\mid \muteq.\cmd{q}{\alpha}:t=u\sigdash \Delta,\alpha:B[u] }{	
	      \Gamma\sigdash q:B[t]\mid\Delta;\sigma
	      &
	      \infer[\axlrule]{\Gamma\mid \alpha:B[u] \sigdash\alpha:B[u]\mid\Delta}{}
	    }
	 }
      }
      }\eqno\qedhere
      $$
}  
\end{proof}

 \begin{figure}[t]
\framebox{\vbox{
\input{figures/dlpaw/ac_n}
}
}
\caption{Proof of the axiom of countable choice in \dlpaw}
\label{fig:ac_n}
\end{figure}

One can even check that the reduction rules in {\dlpaw} for these proofs almost mimic the ones of \dpaw. 
To be more precise, the rules of {\dlpaw} do not allow to simulate each rule of {\dpaw}, 
due to the head-reduction strategy amongst other things. 
Nonetheless, up to a few details the reduction of a command in {\dlpaw} follows one particular reduction path of 
the corresponding proof in {\dpaw}, or in other words, one reduction strategy.


Noteworthily, through the embedding of {\dpaw} the same proof terms are suitable for
the axioms of countable and dependent choices~\cite{Herbelin12}.
We do not state it here, but following the approach of~\cite{Herbelin12}, 
we could also extend {\dlpaw} to obtain a proof for the axiom of bar induction.
\begin{theorem}[Countable choice~\cite{Herbelin12}]\label{thm:acn}
 We have:
$$\begin{array}{@{~~}rc@{~~}l@{}l}
AC_\N		&:=& \lambda H.&\letop a=\cofix{0}{bn}{(H n, b(S(n))}\\
&&&\,\inop\, (\lambda n.\wit(\nthn n a),\lambda n.\prf(\nthn n a)\\
		&: & \multicolumn{2}{l}{\forall x^\N \exists y^T P(x,y) \imp \exists f^{\N\imp T} \forall x^\N P(x,f(x))}\\
\end{array}$$
where $\nthn n a:= \pi_1(\ind{n}{a}{x,c}{\pi_2(c)})$.
\end{theorem}
\begin{proof}
The complete typing derivation of the proof term for $AC_\N$ from Herbelin's paper~\cite{Herbelin12}
is given in Figure \ref{fig:ac_n}.
\end{proof}


\begin{theorem}[Dependent choice~\cite{Herbelin12}]\label{thm:dc}
 We have:
$$\begin{array}{r@{~~}c@{~~}l@{}l}
DC &:=& \lambda H.\lambda x_0.\letop~ a=(x_0,\cofix{0}{bn}{d_n})fsix~\\ 
   & & \inop~ (\lambda n.\wit(\nthn n a),(\refl,\lambda n.\pi_1(\prf(\prf(\nthn n a)))))\\
   &:& \forall x^T\!. \exists y^T\!. P(x,y) \imp \\ 
&& \qquad  \forall x_0^T\! .\exists  f\in T^\N.( f(0) = x_0 
 \land \forall n^\N\!. P( f (n), f (S (n))))\\
\end{array}$$
where $d_n:=\dest{H n}{y,c} (y,(c,b\,y)))$\\
and ~~
$\nthn n a:= \ind{n}{a}{x,d}{(\wit(\prf d),\pi_2(\prf(\prf(d))))}$.
\end{theorem}
\begin{proof}
Left to the reader.\note{DONNER CELLE DE DC, même si c'est chiant !!!}
\end{proof}

\section{Normalization and consistency of {\dlpaw}}
We shall now present our realizability interpretation for \dlpaw, which will eventually
allow us to prove its normalization and consistency.
Once more, we will follow Danvy's methodology of semantic artifacts~\cite{DanEtAl10,AriEtAl12}
to first define a small-step calculus (or context-free abstract machine).
The resulting realizability interpretation will be similar in its structure to the one
presented in \Cref{s:lbvtstar} for the \lbvtstar-calculus.

\subsection{Small-step calculus}
\label{s:small_step}
We start by decomposing the reduction system of {\dlpaw} into small-step reduction rules, that we denote by $\reds$. 
This requires a refinement and an extension of the syntax, that we shall now present.
To keep us from boring the reader stiff with new (huge) tables for the syntax, typing rules and so forth,
we will introduce them step by step. 
We hope it will help the reader to convince herself of the necessity and of the somewhat naturality of these extensions.

\subsubsection{Values}
First of all, we need to refine the syntax to distinguish between strong and weak values
in the syntax of proof terms.
As in the $\lbvtstar$-calculus, this refinement is induced by the computational behavior of the calculus:
weak values are the ones which are stored by $\tmu$ binders, but which are not
values enough to be eliminated in front of a forcing context, that is to say variables.
Indeed, if we observe the reduction system, we see that in front of a forcing context $f$, 
a variable leads a search through the store for a ``stronger'' value, which could incidentally provoke the
evaluation of some fixpoints. 
On the other hand, strong values are the ones which can be
reduced in front of the matching forcing context, that is to say functions, $\refl$, pairs of values, 
injections or dependent pairs:
$$
\begin{array}{>{\!\!}lc@{~}c@{~}l}
\textbf{Weak values}   & V	& ::= & a  \mid v \\
\textbf{Strong values} & v	& ::= & \injec{i}{V} \mid (V,V) \mid (V_t,V)  \mid  \lambda x.p \mid \lambda a.p\mid \refl 
\end{array}
$$
This allows us to distinguish commands of the shape $\cut{v}{f}\tau$, where the forcing context (and next the strong
value) are examined to determine whether the command reduces or not; from commands of the shape $\cut{a}{f}\tau$
where the focus is put on the variable $a$, 
which leads to a lookup for the associated proof in the store.

\subsubsection{Terms}
Next, we need to explicit the reduction of terms. To this purpose,
we include a machinery to evaluate terms in a way which resemble the evaluation of proofs. 
In particular, we define new commands which we write $\cut{t}{\pi}$ where 
$t$ is a term and $\pi$ is a context for terms (or co-term).
Co-terms are either of the shape $\tmu x.c$ or stacks of the shape $u\cdot \pi$. 
These constructions are the usual ones of the \lmmt-calculus (which are also the ones for proofs).
We also extend the definitions of commands with delimited continuations to include the corresponding commands for terms:
$$
\begin{array}{c@{~}|@{~}c}
\begin{array}{@{}l}
  \textbf{Commands}\\
  \textbf{Co-terms} \\
\end{array}\quad
\begin{array}{c@{~}c@{~}l}
c	& ::= & \cmd{p}{e} \mid \cut{t}{\pi}\\
\pi 	& ::= & t\cdot \pi \mid \mut x.c \\ 
\end{array}
&
\begin{array}{c@{~}c@{~}l}
c_\reset& ::= & \cdots\mid \cmd{t}{\pi_\reset}\\
\pi_\reset & ::= & t\cdot \pi_\reset \mid \mut x.c_\reset
\end{array}
\end{array}
$$
We give typing rules for these new constructions, which are the usual rules for typing contexts in the \lmmt-calculus:
$$
\infer[\implrule]{\Gamma \vdash t \cdot \pi:(T\to U)^\negt}
      {\Gamma \vdash t:T  & \Gamma\vdash \pi:U^\negt}
\qquad	\qquad
\infer[\autorule{\tmu_x}]{\Gamma\vdash \tmu x.c : T^\negt }
      {c : (\Gamma, x:T)}
\qquad
\infer[\autorule{\textsc{cut}_t}]{\Gamma\sigdash \cmd{t}{\pi}}{\Gamma\sigdash {t}:{T}  & \Gamma\sigdash \pi :T^\negt }
$$
It is worth noting that the syntax as well as the typing and reduction rules for terms now match exactly
the ones for proofs\footnote{Except for substitutions of terms, which we could store as well.}. In other words, with these definitions, 
we could abandon the stratified presentation without any trouble, since reduction rules for
terms will naturally collapse to the ones for proofs.

\begin{figure}[t]
\framebox{\vbox{
\input{figures/dlpaw/smallstep1}
}
}
\caption{Small-step reduction rules (1/2)}
\label{fig:smallstep}
\end{figure}

\begin{figure}[t]
\ContinuedFloat
\framebox{\vbox{
\input{figures/dlpaw/smallstep2}
}
}
\caption{Small-step reduction rules (2/2)}
\label{fig:smallstep}
\end{figure}

\subsubsection{Co-delimited continuations}
\label{s:codelimited}
Finally, in order to maintain typability when reducing dependent pairs of the strong existential type,
 we need to add what we call \emph{co-delimited continuations}. 
As observed in \Cref{s:cps}, the \cps~translation of pairs $(t,p)$ in {\dltp} is not the expected one,
reflecting the need for a special reduction rule (remember that the usual reduction rule for opening pairs,
following the global call-by-value evaluation strategy, was causing subject reduction to fail). 
%
To remediate this, we instead use a rule where $t$ is reduced within 
a context where $a$ is not linked to $p$ but to a co-reset $\coreset$ (dually to reset $\reset$), 
whose type can be changed from $A[x]$ to $A[t]$ thanks to a list of dependencies:
$$\cmdp{(t,p)}{e}\tau    \reds  \cmdp{p}{\coshift \cmd{t}{\mut x.\cmd{\coreset}{\tmu a.\cut{(x,a)}{e}}}}\tau 	$$
We thus equip the language with new contexts $\coshift c_\coreset$, which we call \emph{co-shifts} and where $c_\coreset$ is a command
whose last cut is of the shape $\cut{\coreset}{e}$.
This corresponds formally to the following syntactic sets, which are dual to the ones introduced for delimited continuations:
$$
\begin{array}{ccl}
e	& ::= & \cdots \mid  \coshift {c_\coreset} \\[0.8em]  
c_\coreset& ::= & \cmd{p_N}{e_\coreset}\mid \cmd{t}{\pi_\coreset}\mid \cut{\coreset}{e}\\
e_\coreset& ::= &  \mut a.c_\coreset\mid \mut[a_1.c_\coreset\mid a_2.c'_\coreset]  \mid   \mut(a_1,a_2).c_\coreset \mid \mut(x,a).c_\coreset \\
\pi_\coreset & ::= & t\cdot \pi_\coreset \mid \mut x.c_\coreset \\[0.8em]   
e_N 	& ::= & \cdots \mid \coshift c_\coreset  \\
\end{array}\leqno 
\begin{array}{@{}l}
  \textbf{Contexts}\\[0.8em]
  \textbf{Co-delimited}\\ 
  \textbf{continuations}\\\\[0.8em] 
  \textbf{\scriptsize NEF}\\ 
  \end{array}$$
This might seem to be a heavy addition to the language, 
but we insist on the fact that these artifacts are merely 
the dual constructions of delimited continuations
introduced in \dltp, with a very similar intuition. 
In particular, it might be helpful for the reader to think of the fact that
we introduced delimited continuations for type safety 
of the evaluation of dependent products in $\dptprod{a:A}{B}$ 
(which naturally extends to the case $\forall x^T\!.A$).
Therefore, to maintain type safety of dependent sums in $\exists x^T\!.A$,
we need to introduce the dual constructions of co-delimited continuations.
We also give typing rules to these constructions, which are dual to the typing rules for delimited-continuations:
$$
{\infer[\!\!\autorule{\tmu\coreset}]{\Gamma\sigdash \coshift c_\coreset : A^\negt}{
\Gamma,\coreset:A\vdash_d c_\coreset;\sigma
  }
\qquad\qquad
\infer[\!\!\autorule{\coreset}]{\Gamma,\coreset:B,\Gamma' \vdash_d \cmd{\coreset}{e};\sigma}{\Gamma,\Gamma'\sigdash e:A^\negt & \sigma(A)=\sigma(B)}
}
$$
Note that we also need to extend the definition of list of dependencies to include bindings of the shape $\dpt{x|t}$ for terms, 
and that we have to give the corresponding typing rules to type commands of terms in dependent mode:
$$
{
\infer[\!\!\dxmutrule]{\Gamma\vdash_d \tmu x.c : T^\negt;\sigma\rdpt{t}\! }{c : (\Gamma, x:T;\sigma\dpt{x|t})}
\qquad\qquad
\infer[\!\!\dtcutrule]{\Gamma,\coreset:B,\Gamma' \vdash_d \cmd{t}{\pi};\sigma}{\Pi_t\!\! & \Gamma,\coreset:B,\Gamma'\vdash_d \pi:A^\negt;\sigma\dpt{\cdot|t}\! }
}
$$
where $\Pi_t \defeq \Gamma,\Gamma'\sigdash {t}:{T}$.

The small-step reduction system we obtained is given in \Cref{fig:smallstep}.
We write $c_\iota\tau \reds c'_o\tau'$ for the reduction rules,
where the annotation $\iota,p$ on commands 
are indices (\emph{i.e.} $c,p,e,V,f,t,\pi,V_t$) indicating which
part of the command is in control. 
As in the $\lbvtstar$-calculus, we observe an alternation of steps 
descending from $p$ to $f$ for proofs and from $t$ to $V_t$ for terms.
The descent for proofs can be divided in two main phases.
During the first phase, from $p$ to $e$ we observe the call-by-value 
process, which extracts values from proofs, opening recursively the 
constructors and computing values. 
In the second phase, the core computation takes place from $V$ to $f$, 
with the destruction of constructors and the application of function to
their arguments.
The laziness corresponds precisely to skipping the first phase, waiting
to possibly reach the second phase before actually going through the first one.

\begin{proposition}[Subject reduction]\label{p:small_sub_red}
 The small-step reduction $\reds$ satisfies subject reduction.
\end{proposition}
\begin{proof}
 The proof is again a tedious induction on the reduction $\reds$.
 There is almost nothing new in comparison with the cases for the big-step reduction rules: 
 the cases for reduction of terms are straightforward, as well as the administrative reductions changing the focus on a command.
 We only give the case for the reduction of pairs $(t,p)$.
 The reduction rule is:
 $$\cmdp{(t,p)}{e}\tau    \reds  \cmdp{p}{\coshift \cmd{t}{\mut x.\cmd{\coreset}{\tmu a.\cut{(x,a)}{e}}}}\tau 	$$
 Consider a typing derivation for the command on the left-hand side, which is of the shape (we omit the rule \lrule~and the store for conciseness):
 $$
   \infer[\cutrule]{\Gamma\sigdash \cmd{(t,p)}{e}}{
      \infer[\exrrule]{\Gamma\sigdash (t,p):\exists x^T\!.A}{
	\infer{\Gamma \sigdash t:T}{\Pi_t}
	&
	\infer{\Gamma \sigdash p:A[t/x]}{\Pi_p}
      }
      &
      \infer{\Gamma\sigdash e:(\exists x^T\!.A)^\negt}{\Pi_e}
   }
 $$
 Then we can type the command on the right-hand side with the following derivation:
 $$
 \infer[\!\!\cutrule]{\Gamma\sigdash \cmdp{p}{\coshift \cmd{t}{\mut x.\cmd{\coreset}{\tmu a.\cut{(x,a)}{e}}}}}{
    \Pi_p
    &\hspace{-0.5cm}
    \infer[\!\!\autorule{\tmu\coreset}]{\Gamma\sigdash \coshift \cmd{t}{\mut x.\cmd{\coreset}{\tmu a.\cut{(x,a)}{e}}}:A[t]^\negt}{
      \infer[\!\!\dcutrule]{\Gamma,\coreset:A[t]\vdash  \cmd{t}{\mut x.\cmd{\coreset}{\tmu a.\cut{(x,a)}{e}}};\sigma}{
	\Pi'_t
	&\hspace{-0.9cm}
	\infer[\!\!\autorule{\tmu_x}]{\Gamma,\coreset:A[t/x]\vdash_d \mut x.\cmd{\coreset}{\tmu a.\cut{(x,a)}{e}}:T;\sigma\rdpt{t}}{
	  \infer[\!\!\dcutrule]{\Gamma,\coreset:A[t],x:T\vdash_d \cmd{\coreset}{\tmu a.\cut{(x,a)}{e}};\sigma\dpt{x|t}}{
	    \infer[\!\!\mutrule]{\Gamma,x:T\sigdash \tmu a.\cut{(x,a)}{e}:A[x]^\negt}{
	      \infer[\!\!\cutrule]{\Gamma,x:T,a:A[x]\sigdash \cut{(x,a)}{e}:A[x]^\negt}{
		\infer{\Gamma\sigdash (x,a):\exists x^T\!.A}{\Pi_(x,a)}
		&
		\infer{\Gamma\sigdash e:(\exists x^T\!.A)^\negt}{\Pi_e}
	      }
	    }
	    &\hspace{-0.5cm}
	    A[t] = (\dpt{x|t})(A[x])
	  }
	}
      }
    }
}
$$
where $\Pi_{(x,a)}$ is as expected.
\end{proof}

It is also direct to check that the small-step reduction system simulates the big-step one, 
and in particular that it preserves the normalization :
\begin{proposition}\label{prop:reduction}
 If a closure $c\tau$ normalizes for the reduction $\reds$, then it also normalizes for the reduction $\red$.
\end{proposition}
\begin{proof}
 By contraposition, one proves that if a command $c\tau$  produces an infinite 
 number of steps for the reduction $\red$, then it does not normalize for $\reds$ either.
 This is proved by showing by induction on the reduction $\red$ that each step, except for the contextual reduction of terms,
 is reflected in  at least on for the reduction $\reds$. 
 The rules for term reductions require a separate treatment, which is really not interesting at this point.  
 We claim that the reduction of terms, which are usual simply-typed $\lambda$-terms, is known to be normalizing anyway and does not deserve
 that we spend another page proving it in this particular setting.
\end{proof}

\subsection{A realizability interpretation of {\textbf{\boldmath dLPA$^\omega$}}}
\label{s:realizability}
We shall now present the realizability interpretation of $\dlpaw$, which will finally give us a proof of its normalization.
Here again, the interpretation combines ideas of the interpretations for the $\lbvtstar$-calculus~\cite{MiqHer18}
and for {\dltp} through its embedding in Lepigre's calculus (see \Cref{s:dl_real}).
Namely, as for the $\lbvtstar$-calculus, formulas will be interpreted by sets of proofs-in-store 
of the shape $\tis{p}{\tau}$, and the orthogonality will be defined between proofs-in-store $\tis{p}{\tau}$ 
and contexts-in-store $\tis{e}{\tau'}$ such
that the stores $\tau$ and $\tau'$ are compatible. 
On the other hand, we will follow Lepigre's decomposition of dependent types into a quantification
bounded by a predicate $a\in A$ whose interpretation is defined modulo observational equivalence.

We recall the main definitions necessary to the realizability interpretation:
\begin{definition}[Proofs-in-store]
We call \emph{closed \cp} (resp. \emph{closed \ce}, \emph{closed \ct}, etc) 
the combination of a proof $p$ (resp. context $e$, term $t$, etc) with a closed store $\tau$ such that
$FV(p)\subseteq \dom(\tau)$. 
We use the notation $\tis{p}{\tau}$ to denote such a pair. 
In addition, we denote by $\Lambda_p$ (resp. $\Lambda_e$, etc.) the set of all proofs 
and by $\Lambda_p^\tau$ (resp. $\Lambda_e^\tau$, etc.) the set of all proofs-in-store.
We denote the sets of closed closures by $\C_0$, and we identify $\tis{c}{\tau}$ with the closure $c\tau$ when $c$ is closed in $\tau$.
\end{definition}

We recall the definition of compatible stores~(\Cref{def:compat}) in this context, which allows us to define
an orthogonality relation between proofs- and contexts-in-store.
\begin{definition}[Compatible stores and union]  
Let $\tau$ and $\tau'$ be stores, we say that:
\begin{itemize}
 \item they are \emph{independent} and note $\indpt{\tau}{\tau'}$ if
 ${\dom(\tau)\cap\dom(\tau')=\emptyset}$.
 \item they are \emph{compatible} and note $\compat{\tau}{\tau'}$ 
 if for all variables $a$ (resp. co-variables $\alpha$) present in both stores: ${a\in \dom(\tau)\cap\dom(\tau')}$;
 the corresponding proofs (resp. contexts) in $\tau$ and $\tau'$ 
 coincide. 
 \item $\tau'$ is an \emph{extension} of $\tau$ and we write $\tau\stext \tau'$ whenever 
 $\compat{\tau}{\tau'}$ and $\dom(\tau)\subseteq\dom(\tau')$.

 \item $\overline{\tau\tau'}$ is \emph{the compatible union } of compatible closed stores $\tau$ and $\tau'$.
 It is defined as $\overline{\tau\tau'}\defeq \stjoin{\tau}{\tau'}$, which itself given by:\vspace{-0.4em}
 $$\begin{array}{r@{~~\defeq~~}l}
  \stjoin{\tau_0[a:=p]\tau_1}{\tau'_0[a:=p]\tau'_1} & \tau_0\tau'_0[a:=p]\stjoin{\tau_1}{\tau'_1} \\
  \stjoin{\tau_0[\alpha:=e]\tau_1}{\tau'_0[\alpha:=e]\tau'_1} & \tau_0\tau'_0[\alpha:=e]\stjoin{\tau_1}{\tau'_1} \\
  \stjoin{\tau_0}{\tau_0'} & \tau_0\tau_0'                                                             \\
 \end{array}\nomidem
 $$
 where $\indpt{\tau_0}{\tau_0'}$.
\end{itemize}
 \end{definition}
The next lemma (which follows from the previous definition) states the main property  
we will use about union of compatible stores.
\begin{lemma}
\label{lm:st_union}
If $\tau$ and $\tau'$ are two compatible stores, then $\tau\stext\overline{\tau\tau'}$ and $\tau'\stext\overline{\tau\tau'}$.
Besides, if $\tau$ is of the form $\tau_0[x:=t]\tau_1$,
then $\overline{\tau\tau'}$ is of the form $\overline{\tau_0}[x:=t]\overline{\tau_1}$ with $\tau_0 \stext \overline{\tau_0}$
and $\tau_1\stext\overline{\tau_1}$.
\end{lemma}

We can now define the notion of pole, which has to satisfy an extra condition due to the presence of delimited continuations
\begin{definition}[Pole]
 A subset $\pole\in \C_0$ is said to be \emph{saturated} or \emph{closed by anti-reduction} 
 whenever for all $\tis{c}{\tau},\tis{c'}{\tau'}\in\C_0$, we have:
 $$(c'\tau' \in \pole) ~ \land ~(c\tau\rightarrow c'\tau')~\Rightarrow~ (c\tau\in\pole)$$
 It is said to be \emph{closed by store extension} if whenever $c\tau$ is in $\pole$, for any store $\tau'$ extending $\tau$,
 $c\tau'$ is also in $\pole$:
 $$(c\tau\in\pole) ~\land~ (\tau\stext\tau') ~\Rightarrow~ (c\tau'\in\pole)$$
 It is said to be \emph{closed under delimited continuations} if whenever $c[e/\reset]\tau$ (resp. $c[V/\coreset]\tau$) is in $\pole$, 
 then $\cut{\shift c}{e}\tau$ (resp .$\cut{V}{\coshift c}\tau$) belongs to $\pole$:\nomidem
 $$(c[e/\reset]\tau\in\pole) ~\Rightarrow~ (\cut{\shift c}{e}\tau\in\pole)\nomidem$$
 $$ (c[V/\coreset]\tau\in\pole) ~\Rightarrow ~(\cut{V}{\coshift c}\tau\in\pole)$$
 A \emph{pole} is defined as any subset of $\C_0$ that is closed by anti-reduction, by store extension and under delimited continuations.
\end{definition}

\begin{proposition}
\label{prop:dlpaw:norm_pole}
 The set $\pole_{\Downarrow}=\{c\tau\in\C_0:~c\tau\text{ normalizes }\}$ is a pole.
\end{proposition}
\begin{proof}
 The first two conditions are already verified for the $\lbvtstar$-calculus~\cite{MiqHer18}. 
 The third one is straightforward, since if a closure $\cut{\shift c}{e}\tau$ is not normalizing, 
 it is easy to verify that $c[e/\reset]$ is not normalizing either. 
 Roughly, there is only two possible reduction steps for a command $\cut{\shift c}{e}\tau$:
 either it reduces to $\cut{\shift c'}{e}\tau'$, in which case $c[e/\reset]\tau$ also reduces
 to a closure which is almost $(c'\tau')[e/\reset]$;
 or $c$ is of the shape $\cut{p}{\reset}$ and it reduces to $c[e/\reset]\tau$. 
 In both cases, if $\cut{\shift c}{e}\tau$ can reduce, so can $c[e/\reset]\tau$. 
 The same reasoning allows us to show that if $c[V/\coreset]\tau$ normalizes,
 then so does $\cut{V}{\coshift c}\tau$ for any value $V$.
\end{proof}

We finally recall the definition of the orthogonality relation w.r.t. a pole,
which is identical to the one for the $\lbvtstar$-calculus:
\begin{definition}[Orthogonality]
Given a pole $\pole$, we say that a {\cp} $\tis{p}{\tau}$ is {\em orthogonal} to a {\ce} $\tis{e}{\tau'}$
and write $\tis{p}{\tau}\orth\tis{e}{\tau'}$
if $\tau$ and $\tau'$ are compatible and $\cut{p}{e}\overline{\tau\tau'}\in\pole$.
The orthogonality between terms and co-terms is defined identically.
\end{definition}

We are now equipped to define the realizability interpretation of $\dlpaw$.
Firstly, in order to simplify the treatment of coinductive formulas, 
we extend the language of formulas with second-order variables $X,Y,\dots$ 
and we replace ${\nu^t_{fx} A}$ by $\nu^t_{Xx} A[X(y)/f(y)=0]$. 
The typing rule for co-fixpoint operators then becomes:
$$\infer[\cofixrule]{\sigmaopt\Gamma\sigdash \cofix{t}{bx}{p}: \nu^t_{Xx} A \optsigma}{
	\sigmaopt\Gamma\sigdash t:T  \optsigma
	& 
	\sigmaopt\Gamma,x:T,b:\forall y^T\!. X(y)\sigdash p:A  \optsigma
	& 
	X \notin\FV(\Gamma)
}$$
where $X$ has to be \text{ positive in } $A$.

Secondly, as in the interpretation of {\dltp} through Lepigre's calculus,
we introduce two new predicates, $p\in A$ for {\nef} proofs and $t\in T$ for terms.
This allows us to decompose the dependent products and sums into:
$$\hspace{-3.5mm}\begin{array}{c|c}
\begin{array}{c@{~}c@{~}l}
\forall x^T\!.A &\defeq& \forall x.(x\in T \imp A)\qquad \\
\exists x^T\!.A &\defeq& \exists x.(x\in T \imp A)\qquad \\
\end{array}
&
\begin{array}{c@{~}c@{~}l}
\dptprod{a:A}{B}&\defeq& A \imp B \hfill(a\notin \FV(B))\\
\dptprod{a:A}{B}&\defeq& \forall a.(a\in A \imp B) \quad~(\text{otherwise})\\
\end{array}
\end{array}
$$
This corresponds to the language of formulas and types defined by:
$$
\begin{array}{ccl}
T,U & ::= & \N \mid T \to U \mid t\in T \\
A,B  &::= &\top\mid \bot \mid X(t) \mid t = u\mid A\land B \mid A\lor B \\
&\mid& \forall x. A \mid \exists x. A \mid \forall a.A \mid \nu^t_{Xx}A \mid a\in A
\end{array}\leqno
\begin{array}{l}
\textbf{Types}\\
\textbf{Formulas}\\\\
\end{array}
$$
and to the following inference rules:
\setlength{\saut}{-0.5em}
$$
\begin{array}{c@{\qquad\quad}c}
\infer[\autorule{\forall^a_r}]{\Gamma \sigdash v:\forall a.A}{\Gamma\sigdash v:A & a\notin \FV(\Gamma)} &
\infer[\autorule{\forall^a_l}]{\Gamma \sigdash e:(\forall a.A)^\negt}{\Gamma\sigdash e:A[q/a] & q~\nef} 
\end{array}$$
$$\begin{array}{c@{\qquad\quad}c}
\infer[\autorule{\forall^x_r}]{\Gamma \sigdash v:\forall x.A}{\Gamma\sigdash v:A & x\notin \FV(\Gamma)} &
\infer[\autorule{\forall^x_l}]{\Gamma \sigdash e:(\forall x.A)^\negt}{\Gamma\sigdash e:A[t/x]}
\\\\[\saut]
\infer[\autorule{\exists^x_r}]{\Gamma \sigdash v:\exists x.A}{\Gamma\sigdash v:A[t/x]} &
\infer[\autorule{\exists^x_l}]{\Gamma \sigdash e:(\exists x.A)^\negt}{\Gamma\sigdash e: A & x\notin \FV(\Gamma)}
\\\\[\saut]
\infer[\autorule{\in^p_r}]{\Gamma \sigdash p:p\in A}{\Gamma \sigdash p:A & p~\nef}           &
\infer[\autorule{\in^p_l}]{\Gamma \sigdash e:(q\in A)^\negt}{\Gamma \sigdash e:A^\negt}     
\\\\[\saut]
\infer[\autorule{\in^t_r}]{\Gamma \sigdash t:t\in T}{\Gamma \sigdash t:T}                   &
\infer[\autorule{\in^t_l}]{\Gamma \sigdash \pi:(t\in T)^\negt}{\Gamma \sigdash \pi:T^\negt}
\end{array}
$$
These rules are exactly the same as in Lepigre's calculus~\cite{Lepigre16} up to our stratified presentation in
a sequent calculus fashion, and modulo our syntactic restriction to {\nef} proofs instead of his semantical restriction.
It is a straightforward verification to check that the typability is maintained through the decomposition of dependent products and sums.

Another similarity with Lepigre's realizability model is that truth/falsity values 
will be closed under observational equivalence of proofs and terms. 
To this purpose, for each store $\tau$ we introduce the relation $\equiv_\tau$, which we define as 
the reflexive-transitive-symmetric closure of the relation $\typered_\tau$:
\begin{center}
\begin{tabular}{lcl}
   $t$ &$\typered_\tau$ &$t'\quad$ whenever $\quad \exists \tau',\forall \pi, (\cmd{t}{\pi}\tau\rightarrow\cmd{t'}{\pi}\tau'$ \\
   $p$ &$\typered_\tau$& $q \quad$ whenever $\quad \exists \tau',\forall f\,(\cmd{p}{f}\tau\rightarrow\cmd{q}{f}\tau')$ \\
\end{tabular}
\end{center}

All this being settled, it only remains to determine how to interpret coinductive formulas. 
While it would be natural to try to interpret them by fixpoints in the semantics,
this poses difficulties for the proof of adequacy.
{We discuss this matter in \Cref{app:cofix}, but as for now,
we will give a simpler interpretation.}
We stick to the intuition that since \texttt{cofix} operators are lazily evaluated,
they actually are realizers of every finite approximation of the (possibly infinite) coinductive formula.
Consider for instance the case of a stream: 
$$\texttt{str}^{0}_\infty {p}\defeq \cofix{0}{bx}{(p x,b(S(x)))}$$ of type $\nu^0_{Xx} A(x)\land X(S(x))$.
Such stream will produce on demand any tuple 
$(p0,(p1,...(pn,\square)...))$
where $\square$ denotes the fact that it could be any term, in particular $\texttt{str}_\infty^{n+1} {p}$.
Therefore, $\texttt{str}_\infty^0 p$ should be a successful defender of the formula 
$$(A(0)\land (A(1) \land ...(A(n)\land \top)...)$$
Since $\texttt{cofix}$ operators only reduce when they are bound to a variable in front of a forcing context,
it suggests interpreting the coinductive formula $\nu^0_{Xx} A(x)\land X(S(x))$ at level $f$ 
as the union of all the opponents to a finite approximation.

To this end, given a coinductive formula $\nu^0_{Xx} A$ where $X$ is positive in $A$, we define its finite approximations by:
$$F^0_{A,t} \defeq \top \qquad\qquad\qquad F^{n+1}_{A,t} \defeq A[t/x][F^{n}_{A,y}/X(y)]$$
Since $X$ is positive in $A$, we have for any integer $n$ and any term $t$
that $\fvf{F^{n}_{A,t}}\subseteq \fvf{F^{n+1}_{A,t}}$.
We can finally define the interpretation of coinductive formulas by:
$$\fvf{\nu^t_{Xx}A}   \defeq  \bigcup_{n\in\N}\fvf{F^n_{A,t}}$$

\begin{figure}[t]
  \myfig{
  \input{figures/dlpaw/real.tex}
   }
   \caption{Realizability interpretation for \dlpaw}
   \label{fig:dlpaw_real}
   \end{figure}

The realizability interpretation of closed formulas and types is defined in \Cref{fig:dlpaw_real}
by induction on the structure of formulas at level $f$, and by orthogonality at levels $V,e,p$.
When $S$ is a subset of $\P(\Lambda_p^\tau)$ (resp. $\P(\Lambda_e^\tau),\P(\Lambda_t^\tau),\P(\Lambda_\pi^\tau)$), 
we use the notation $S^{\pole_f}$ (resp. $S^{\pole_V}$, etc.)
to denote its orthogonal set restricted to $\Lambda_f^\tau$:
$$S^{\pole_f}\defeq \{\tis{f}{\tau}\in\Lambda^\tau_f : \forall \tis{p}{\tau'}\in S, \compat{\tau}{\tau'} \Rightarrow \cut{p}{f}\overline{\tau\tau'} \in\pole\}$$

At level $f$, closed formulas are interpreted by sets of strong forcing contexts-in-store $\tis{f}{\tau}$.
As explained earlier, these sets are besides closed under the relation $\equiv_\tau$ along their component $\tau$,
we thus denote them by $\P(\Lambda_f^\tau)_{/{\equiv_\tau}}$. 
Second-order variables $X,Y,\dots$ are then interpreted by functions from the set of terms $\Lambda_t$ to $\P(\Lambda_f^\tau)_{/{\equiv_\tau}}$
and as is usual in Krivine realizability~\cite{Krivine09}, for each such function $F$ we add a predicate symbol $\dot F$ in the language.

\subsection{Adequacy of the interpretation}

We shall now prove the adequacy of the interpretation with respect to the type system.
To this end, we need to recall a few definitions and lemmas. 
Since stores only contain proof terms, we need to define valuations for term variables in order 
to close formulas\footnote{Alternatively, we could have modified the small-step reduction rules to include substitutions of terms.}.
These valuations are defined by the usual grammar:
$$\rho ::= \varepsilon \mid \rho[x\mapsto V_t]\mid \rho[X\mapsto \dot F]$$
We denote by $\tis{p}{\tau}_\rho$ (resp. $p_\rho$, $A_\rho$) the proof-in-store $\tis{p}{\tau}$ where
all the variables $x\in\dom(\rho)$ (resp. $X\in\dom(\rho)$) have been substituted by the corresponding term $\rho(x)$ 
(resp. falsity value $\rho(x)$).

\begin{definition}Given a closed store $\tau$, a valuation $\rho$ and a fixed pole $\pole$, 
we say that the pair $(\tau,\rho)$ \emph{realizes} $\Gamma$, which we write\footnote{Once again, 
we should formally write $(\tau,\rho)\real_{\!\!\pole}\!\Gamma$ but we will omit the annotation by $\pole$ as often as possible.} $(\tau,\rho) \Vdash \Gamma$, if:
\begin{enumerate}
 \item for any $(a:A) \in\Gamma$, $\tis{a}{\tau}_\rho \in \tvV{A_\rho}$,
 \item for any $(\alpha:A_\rho^\negt) \in\Gamma$, $\tis{\alpha}{\tau}_\rho \in \fve{A_\rho}$,
 \item for any $\dpt{a|p}\in\sigma$, $a\equiv_{\tau} p$,
 \item for any $(x:T) \in\Gamma$, $x\in\dom(\rho)$ and $\tis{\rho(x)}{\tau} \in \tvVt{T_\rho}$.
 \end{enumerate}
 \label{def:store_real}
\end{definition}

We recall two key properties of the interpretation, whose proofs are similar to the proofs
for the corresponding statements in the $\lbvtstar$-calculus~\cite{MiqHer18}:
\begin{lemma}[Store weakening]
\label{lm:dlpaw:st_weak}
 Let $\tau$ and $\tau'$ be two stores such that $\tau\stext\tau'$, let $\Gamma$ be a typing context,
 let $\pole$ be a pole and $\rho$ a valuation. The following statements hold:
 \begin{enumerate}
  \item $\overline{\tau\tau'} = \tau'$
  \item If ~$\tis{p}{\tau}_\rho \in \tvp{A_\rho}$~ for some closed proof-in-store $\tis{p}{\tau}_\rho$ and formula $A$, then ~$\tis{p}{\tau'}_\rho\in\tvp{A_\rho}$. 
  The same holds for each level $e,E,V,f,t,\pi,V_t$ of the interpretation.
  \item If ~$(\tau,\rho)\real \Gamma$~ then ~$(\tau',\rho) \real \Gamma$.
 \end{enumerate}
\end{lemma}

\begin{proposition}[Monotonicity]
\label{prop:dlpaw:monotonicity}
For any closed formula $A$, any type $T$ and any given pole $\pole$, we have the following inclusions:
$$\tvV{A} \subseteq \tvp{A} \qquad\qquad \fvf{A}\subseteq \fve{A} \qquad\qquad \tvVt{T}\subseteq \tvt{T}$$
\end{proposition}

We can check that the interpretation is indeed defined up to the relations $\equiv_\tau$:
\begin{proposition}\label{lm:dlpaw:equiv}
 For any store $\tau$ and any valuation $\rho$, the component along $\tau$ of the truth and falsity values defined in \Cref{fig:dlpaw_real}
 are closed under the relation $\equiv_\tau$:
 \begin{enumerate}
 \item if $\tis{f}{\tau}_\rho \in \fvf{A_\rho}$ and $A_\rho\equiv_\tau B_\rho$, then $\tis{f}{\tau}_\rho\in\fvf{B_\rho}$,
 \item if $\tis{V_t}{\tau}_\rho \in \tvVt{A_\rho}$ and $A_\rho\equiv_\tau B_\rho$, then $\tis{V_t}{\tau}_\rho\in\tvv{B_\rho}$.
\end{enumerate}
The same applies with $\tvp{A_\rho}$, $\fve{A_\rho}$, etc.
\end{proposition}
\begin{proof}
 By induction on the structure of $A_\rho$ and the different levels of interpretation. 
 The different base cases ($p\in A_\rho$, $t\in T$, $t=u$) are direct since their components along $\tau$ are defined modulo $\equiv_\tau$, 
 the other cases are trivial inductions.
\end{proof}

We can now prove the main property of our interpretation:

\begin{proposition}[Adequacy]\label{prop:dlpaw:adequacy}
  The typing rules are adequate with respect to the realizability interpretation.
 In other words, if~ $\Gamma$ is a typing context, $\pole$ a pole, $\rho$ a valuation and $\tau$ a store such that ${(\tau,\rho)\real \Gamma;\sigma}$,
then the following hold:
\begin{enumerate}
 \item If~ $v$ is a strong value 	s.t. $\Gamma\sigdash v:A$      ~or~~$\Gamma\vdash_d v:A      ;\sigma$, then $\tis{v}{\tau}_\rho   \in\tvV{A_\rho}$.
 \item If~ $f$ is a forcing context 	s.t. $\Gamma\sigdash\! f:A^\negt$~or~~$\Gamma\vdash_d f:A^\negt;\sigma$, then $\tis{f}{\tau}_\rho   \in\fvf{A_\rho}$.
 \item If~ $V$ is a weak value   	s.t. $\Gamma\sigdash V:A$      ~or~~$\Gamma\vdash_d V:A      ;\sigma$, then $\tis{V}{\tau}_\rho   \in\tvV{A_\rho}$.
 \item If~ $e$ is a context 		s.t. $\Gamma\sigdash e:A^\negt$~or~~$\Gamma\vdash_d e:A^\negt;\sigma$, then $\tis{e}{\tau}_\rho   \in\fve{A_\rho}$.
 \item If~ $p$ is a proof term  	s.t. $\Gamma\sigdash p:A$ 	    ~or~~$\Gamma\vdash_d p:A      ;\sigma$, then $\tis{p}{\tau}_\rho   \in\tvp{A_\rho}$.
 \item If~ $V_t$ is a term value   	s.t. $\Gamma\sigdash V_t:T$,	                                    then $\tis{V_t}{\tau}_\rho \in\tvVt{T_\rho}$.
 \item If~ $\pi$ is a term context   	s.t. $\Gamma\sigdash \pi:T$,	                                    then $\tis{\pi}{\tau}_\rho \in\fvpi{T_\rho}$.
 \item If~ $t$ is a term  		s.t. $\Gamma\sigdash t:T$, 	                                    then $\tis{t}{\tau}_\rho \in\tvt{T_\rho}$.
 \item If~ $\tau'$ is a store		s.t. $\Gamma\sigdash\tau':(\Gamma';)\sigma'$,                          then $(\tau\tau',\rho) \real (\Gamma,\Gamma';\sigma\sigma')$. 
 \item If~ $c$ is a command		s.t. $\Gamma\sigdash c$        ~or~~$\Gamma\vdash_d c        ;\sigma$, then $(c\tau)_\rho \in \pole$. 
  \item If~ $c\tau'$ is a closure	s.t. $\Gamma\sigdash c\tau'$   ~or~~$\Gamma\vdash_d c\tau'   ;\sigma$, then $(c\tau\tau')_\rho \in \pole$. 
\end{enumerate}
\end{proposition}
\begin{proof}
The proof is done by induction on the typing derivation such as given in the system
extended with the small-step reduction $\reds$. 
Most of the cases correspond to the proof of adequacy for the interpretation of the $\lbvtstar$-calculus,
so that we only give the most interesting cases. To lighten the notations, we omit the annotation by the valuation $\rho$ whenever it is possible.
\prfcase{\exrrule}
We recall the typing rule through the decomposition of dependent sums:
$$\infer{\sigmaopt\Gamma \sigdash (t,p) : (u\in T \land A[u]) \optsigma}{\sigmaopt\Gamma \sigdash t:u\in T \optsigma & \sigmaopt\Gamma \sigdash p: A[u/x] \optsigma }$$
By induction hypothesis, we obtain that $\tis{t}{\tau}\in\tvt{u\in T}$ and $\tis{p}{\tau}\in\tvp{A[u]}$.
Consider thus any context-in-store $\tis{e}{\tau'}\in\fve{u\in T \land A[u]}$ such that $\tau$ and $\tau'$ are compatible, and let us denote by $\tau_0$ the union $\overline{\tau\tau'}$.
We have:
$$\cmdp{(t,p)}{e}\tau_0 \reds  \cmdp{p}{\coshift \cmd{t}{\mut x.\cmd{\coreset}{\tmu a.\cut{(x,a)}{e}}}}\tau_0$$ 
so that by anti-reduction, we need to show that $\coshift \cmd{t}{\mut x.\cmd{\coreset}{\tmu a.\cut{(x,a)}{e}}}\in \fve{A[u]}$.
Let us then consider a value-in-store $\tis{V}{\tau_0'}\in\tvV{A[u]}$ such that $\tau_0$ and $\tau_0'$ are compatible, and let us denote by $\tau_1$ the union $\overline{\tau_0\tau_0'}$.
By closure under delimited continuations, to show that $\cmdp{V}{\coshift \cmd{t}{\mut x.\cmd{\coreset}{\tmu a.\cut{(x,a)}{e}}}}\tau_1$ is in the pole 
it is enough to show that the closure $\cmd{t}{\mut x.\cmd{V}{\tmu a.\cut{(x,a)}{e}}}\tau_1$ is in $\pole$,. Thus it suffices to show that 
the coterm-in-store $\tis{\mut x.\cmd{V}{\tmu a.\cut{(x,a)}{e}}}{\tau_1}$ is in $\fvpi{u\in T}$.

Consider a term value-in-store $\tis{V_t}{\tau_1'}\in\tvVt{u\in T}$, such that $\tau_1$ and $\tau_1'$ are compatible, and let us denote by $\tau_2$ the union $\overline{\tau_1\tau_1'}$.
We have:
$$\cut{V_t}{\mut x.\cmd{V}{\tmu a.\cut{(x,a)}{e}}}{\tau_2}\reds \cmd{V}{\tmu a.\cut{(V_t,a)}{e}}\tau_2\reds \cut{(V_t,a)}{e}{\tau_2[a:=V]}$$ 
It is now easy to check that $\tis{(V_t,a)}{\tau_2[a:=V]}\in\tvV{u\in T \land A[u]}$ and to conclude, using \Cref{lm:dlpaw:st_weak} to get $\tis{e}{\tau_2[a:=V]}\in\fve{u\in T\land A[u]}$,
that this closure is finally in the pole.

\prfcase{\convrrule,\convlrule}
These cases are direct consequences of \Cref{lm:dlpaw:equiv} since if $A,B$ are two formulas such that $A\equiv B$,
in particular $A\equiv_\tau B$ and thus $\tvv{A} = \tvv{B}$.

\prfcase{\reflrule,\eqrule}
The case for $\refl$ is trivial, while it is trivial to show that $\tis{\muteq.\cut{p}{e}}{\tau}$ is in $\fvf{t=u}$
if $\tis{p}{\tau}\in\tvp{A[t]}$ and $\tis{e}{\tau}\in\fve{A[u]}$.
Indeed, either $t \equiv_{\tau} u$ and thus $A[t] \equiv_{\tau} A[u]$ (\Cref{lm:dlpaw:equiv},
or $t \nequiv_{\tau} u$ and $\fvf{t=u}=\Lambda_f^\tau$.

\prfcase{\autorule{\forall_r^x}} This case is standard in a call-by-value language with value restriction.
We recall the typing rule:
$$\infer[\autorule{\forall^x_r}]{\Gamma \sigdash v:\forall x.A}{\Gamma\sigdash v:A & x\notin \FV(\Gamma)}\qquad$$
The induction hypothesis gives us that $\tis{v}{\tau}_\rho$ is in $\tvV{A_\rho}$ for any valuation $\rho[x\mapsto t]$.
Then for any $t$, we have $\tis{v}{\tau}_\rho\in \fvf{A_\rho[t/x]}^{\pole_v} $
so that $\tis{v}{\tau}_\rho\in (\bigcap_{t\in\Lambda_t} \fvf{A[t/x]}^{\pole_v})$. 
Therefore, if $\tis{f}{\tau'}_\rho$ belongs to  $\fvf{\forall x.A_\rho} = (\bigcap_{t\in\Lambda_t} \fvf{A[t/x]}^{\pole_v})^{\pole_f}$,
 we have by definition that $\tis{v}{\tau}_\rho\pole \tis{f}{\tau'}_\rho$.

\prfcase{\indrule}
We recall the typing rule:
$$\infer[\indrule]{\sigmaopt\Gamma\sigdash \ind{t}{p_0}{ax}{p_S} : A[t/x]\optsigma}{
	  \sigmaopt\Gamma\sigdash t: \N  \optsigma
	  &
	  \sigmaopt\Gamma \sigdash p_0:A [0/x]  \optsigma
	  &
	  \sigmaopt\Gamma,x:T,a:A\sigdash p_S:A[S(x)/x]\optsigma
	}
$$
We want to show that $\tis{\ind{t}{p_0}{ax}{p_S}}{\tau}\in\tvp{A[t]}$, let us then consider $\tis{e}{\tau'}\in\fve{A[t]}$
 such that $\tau$ and $\tau'$ are compatible, and let us denote by $\tau_0$ the union $\overline{\tau\tau'}$.
By induction hypothesis, we have\footnote{Recall that any term $t$ of type $T$ can be given the type $t\in T$.}
$t\in \tvt{t\in \N}$ and we have:
$$\cmdp{\ind{t}{p_0}{bx}{p_S}}{e}\tau_0\reds  \cmdp{\shift\cmd{t}{\mut y.\cmd{a}{\reset}[a:=\ind{y}{p_0}{bx}{p_S}]}}{e}\tau_0$$
so that by anti-reduction and closure under delimited continuations, it is enough to show that  the coterm-in-store
$\tis{\mut y.\cmd{a}{e}[a:=\ind{y}{p_0}{bx}{p_S}]}{\tau_0}$ is in $\fvpi{t\in \N}$.
Let us then consider $\tis{V_t}{\tau_0'}\in\tvVt{t\in \N}$
such that $\tau_0$ and $\tau_0'$ are compatible, and let us denote by $\tau_1$ the union $\overline{\tau_0\tau_0'}$.
By definition, $V_t=S^n(0)$ for some $n\in\N$ and $t\equiv_{\tau_1} S^n(0)$, and 
we have: 
$$\cmd{S^n(0)}{\mut y.\cmd{a}{e}[a:=\ind{y}{p_0}{bx}{p_S}]}\tau_1 \reds \cmd{a}{e}\tau_1[a:=\ind{S^n(0)}{p_0}{bx}{p_S}] $$
We conclude by showing by induction on the natural numbers that for any $n\in N$,
the value-in-store $\tis{a}{\tau_1[a:=\ind{S^n(0)}{p_0}{bx}{p_S}]}$ is in $\tvV{A[S^n(0)]}$.
Let us consider $\tis{f}{\tau_1'}\in\fvf{A[S^n(0)]}$ such that  the store
$\tau_1[a:=\ind{S^n(0)}{p_0}{bx}{p_S}]$ and $\tau_1'$ are compatible, 
and let us denote by $\tau_2[a:=\ind{S^n(0)}{p_0}{bx}{p_S}]\tau_2'$ their union.
\begin{itemize}
\item 
If $n=0$, we have:
$$
\cmd{a}{f}\tau_2[a:=\ind{0}{p_0}{bx}{p_S}]\tau_2' \reds \cmd{p_0}{\tmu a.\cut{a}{f}\tau'_2}\tau_2
$$
We conclude by anti-reduction and the induction hypothesis for $p_0$, since it is easy to show that $\tis{\tmu a.\cut{a}{f}\tau'_2}{\tau_2}\in\fve{A[0]}$.

\item 
If $n=S(m)$, we have:
$$
\cmd{a}{f}\tau_2[a:=\ind{S(S^m(0))}{p_0}{bx}{p_S}]\tau_2' \reds\cmdp{p_S[S^m(0)/x][b'/b]}{\mut a.\cmd{a}{f}\tau_2'}\tau_2[b':=\ind{S^m(0)}{p_0}{bx}{p_S}]
$$
Since we have by induction that $\tis{b'}{\tau_2[b':=\ind{S^m(0)}{p_0}{bx}{p_S}]}$ is in $\tvV{A[S^m(0)]}$,
we can conclude by anti-reduction, using the induction hypothesis for $p_S$
and the fact that $\tis{\tmu a.\cut{a}{f}\tau'_2}{\tau_2}$ belongs to $\fve{A[S(S^m(0))]}$.
\end{itemize}\noem

\prfcase{\cofixrule}
We recall the typing rule:
$$\infer[\cofixrule]{\sigmaopt\Gamma\sigdash \cofix{t}{bx}{p}: \nu^t_{Xx} A \optsigma}{
	\sigmaopt\Gamma\sigdash t:T  \optsigma
	& 
	\sigmaopt\Gamma,x:T,b:\forall y^T\!. X(y)\sigdash p:A  \optsigma
	& 
	X\text{ positive in } A & X\notin \FV(\Gamma)
}$$
We want to show that $\tis{\cofix{t}{bx}{p}}{\tau}\in\tvp{\nu^t_{Xx} A}$, let us then consider $\tis{e}{\tau'}\in\fve{\nu^t_{Xx} A}$
such that $\tau$ and $\tau'$ are compatible, and let us denote by $\tau_0$ the union $\overline{\tau\tau'}$.
By induction hypothesis, we have $t\in \tvt{t\in T}$ and we have:
$$\cmdp{\cofix{t}{bx}{p}}{e}\tau_0\reds  \cmdp{\shift\cmd{t}{\mut y.\cmd{a}{\reset}[a:=\cofix{y}{bx}{p}]}}{e}\tau_0$$
so that by anti-reduction and closure under delimited continuations, it is enough to show that the coterm-in-store
$\tis{\mut y.\cmd{a}{e}[a:=\cofix{y}{bx}{p}]}{\tau_0}$ is in $\fvpi{t\in \N}$.
Let us then consider $\tis{V_t}{\tau_0'}\in\tvVt{t\in T}$
such that $\tau_0$ and $\tau_0'$ are compatible, and let us denote by $\tau_2$ the union $\overline{\tau_0\tau_0'}$.
We have:
$$\cmd{V_t}{\mut y.\cmd{a}{e}[a:=\cofix{y}{bx}{p}]}\tau_1 \reds \cmd{a}{e}\tau_1[a:=\cofix{V_t}{bx}{p}] $$
It suffices to show now that the value-in store $\tis{a}{\tau_1[a:=\cofix{V_t}{bx}{p}]}$ is in $\tvV{\nu^{V_t}_{Xx}A}$.
By definition, we have:
$$\tvV{\nu^{V_t}_{Xx}A} = (\bigcup_{n\in\N}\fvf{F^n_{A,{V_t}}})^{\pole_V}  
= \bigcap_{n\in\N}\fvf{F^n_{A,{V_t}}}^{\pole_V} 
= \bigcap_{n\in\N}\tvV{F^n_{A,{V_t}}}$$
We conclude by showing by induction on the natural numbers that for any $n\in N$ and any $V_t$,
the value-in-store $\tis{a}{\tau_1[a:=\cofix{V_t}{bx}{p}]}$ is in $\tvV{F^n_{A,V_t}}$.

The case $n=0$ is trivial since $\tvV{F^0_{A,V_t}}=\tvV{\top}=\Lambda^\tau_V$.
Let then $n$ be an integer and any $V_t$ be a term value.
Let us consider $\tis{f}{\tau_1'}\in\fvf{F^{n+1}_{A,V_t}A}$ 
such that $\tau_1[a:=\cofix{V_t}{bx}{p}]$ and $\tau_1'$ are compatible, and let us denote by $\tau_2[a:=\cofix{V_t}{bx}{p}]\tau_2'$
their union.
By definition, we have: 
$$
\cmd{a}{f}\tau_2[a:=\cofix{V_t}{bx}{p}]\tau_2' \reds \cmd{p[V_t/x][b'/b]}{\tmu a.\cut{a}{f}\tau'_2}\tau_2[b':=\lambda y.\cofix{y}{bx}{p}]
$$
It is straightforward to check, using the induction hypothesis for $n$, that $\tis{b'}{\tau_2[b':=\lambda y.\cofix{y}{bx}{p}]}$ is in 
$\tvV{\forall y.y\in T\imp F^n_{A,y}}$.
Thus we deduce by induction hypothesis for $p$, denoting by $S$ the function $t\mapsto \fvf{F^n_{A,t}}$, that: 
$$\tis{p[V_t/x][b'/b]}{\tau_2[b':=\lambda y.\cofix{y}{bx}{p}]} \in \tvp{A[V_t/x][\dot S/X]} =\tvp{A[V_t/x][F^n_{A,y}/X(y)]} = \tvp{F^{n+1}_{A,V_t}}$$
It only remains to show that $\tis{\tmu a.\cut{a}{f}\tau'_2}{\tau_2}\in\fve{F^{n+1}_{A,V_t}}$, which is trivial from the hypothesis for $f$.
\end{proof}

We can finally deduce that
$\dlpaw$ is normalizing and sound.
\begin{theorem}[Normalization]	
If~ $\Gamma \sigdash c$, then $c$ is normalizable. 
\end{theorem}
\begin{proof}
Direct consequence of \Cref{prop:dlpaw:norm_pole,prop:dlpaw:adequacy}.
\end{proof}


\begin{theorem}[Consistency]\label{thm:consistency} 
$\nvdash_{\text{\dlpaw}}p: \bot$
\end{theorem}
\begin{proof}
 Assume there is such a proof $p$, by adequacy $\tis{p}{\varepsilon}$ is in $\tvp{\bot}$ for any pole.
 Yet, the set $\pole \defeq \emptyset$ is a valid pole, and with this pole, $\tvp{\bot}=\emptyset$, which is absurd.
\end{proof}

\section{About the interpretation of coinductive formulas}
\label{app:cofix}
While our realizability interpretation give us a proof of normalization and soundness for {\dlpaw},
it has two aspects that we should discuss. 
First, regarding the small-step reduction system, one could have expected the lowest level of interpretation to be $v$ instead of $f$.
Moreover, if we observe our definition, we notice that most of the cases of $\fvf{\cdot}$ are in fact defined by orthogonality to
a subset of strong values. 
Indeed, except for coinductive formulas, we could indeed have defined instead an interpretation  $\tvv{\cdot}$ of formulas at level $v$
and then the interpretation $\fvf{\cdot}$ by orthogonality:
$$\begin{array}{ccl}
     \tvv{\bot}	    & \defeq & \emptyset\\ 
     \tvv{t=u}	    & \defeq & \begin{cases}\refl & \text{if~} t\equiv u\\ \emptyset & \text{otherwise}\end{cases}\\
     \tvv{p\in A}   & \defeq & \{\tis{v}{\tau}\in\tvv{A}: v \eqtau p \}\\
     \tvv{T\imp B}  & \defeq & \{\tis{\lambda x .p}{\tau}  : \forall V_t \tau', \compat{\tau}{\tau'}\land \tis{V_t}{\tau'}\in\tvV{T} \Rightarrow \tis{p[V_t/x]}{\overline{\tau\tau'}}\in\tvp{B}\}\\
     \tvv{A\imp B}  & \defeq & \{\tis{\lambda a .p}{\tau}  : \forall V \tau', \compat{\tau}{\tau'}\land \tis{V}{\tau'}\in\tvV{A} \Rightarrow \tis{p}{\overline{\tau\tau'}[a:=V]}\in\tvp{B}\}\\
     \tvv{T\land A} & \defeq & \{\tis{(V_t,V)}{\tau}    : \tis{V_t}{\tau}\in\tvVt{T}  \land \tis{V}{\tau}   \in\tvV{A_2}\}\\
 \tvv{A_1\land A_2} & \defeq & \{\tis{(V_1,V_2)}{\tau}  : \tis{V_1}{\tau}\in\tvV{A_1} \land \tis{V_2}{\tau} \in\tvV{A_2}\}\\
 \tvv{A_1\lor A_2}  & \defeq & \{\tis{\injec i V}{\tau} : \tis{V}{\tau}\in\tvV{A_i}\}\\
 \tvv{\exists x.A}  & \defeq & \bigcup_{t\in\Lambda_t} \tvv{A[t/x]}\\
 \tvv{\forall x.A}  & \defeq & \bigcap_{t\in\Lambda_t} \tvv{A[t/x]}\\
 \tvv{\forall a.A}  & \defeq & \bigcap_{p\in\Lambda_p} \tvv{A[p/x]}\\
     \fvf{A} 	    & \defeq & \{\tis{f}{\tau} : \forall v \tau', \compat{\tau}{\tau'}\land \tis{v}{\tau'}\in\tvv{A} \Rightarrow \tis{v}{\tau'}\orth \tis{F}{\tau}\}\\
   \end{array}$$

If this definition is somewhat more natural, it poses a problem for the definition of  coinductive formulas. 
Indeed, there is a priori no strong value in the orthogonal of $\fvf{\nu^t_{fv} A}$, which is:
$$(\fvf{\nu^t_{fv} A})^{\pole_v} = (\bigcup_{n\in\N}\fvf{F^n_{A,t}})^{\pole_v} = \bigcap_{n\in\N}(\fvf{F^n_{A,t}})^{\pole_v})$$
For instance, consider again the case of a stream of type $\nu^0_{fx} A(x)\land f(S(x))=0$, a strong value in the intersection
should be in every $\tvv{A(0)\land (A(1)\land \dots (A(n)\land \top)\dots)}$, 
which is not possible due to the finiteness of terms\footnote{Yet, it might possible to consider interpretation with infinite proof terms, 
the proof of adequacy for proofs and contexts (which are finite) will still work exactly the same. 
However, another problem will arise for the adequacy of the \texttt{cofix} operator. Indeed, with the interpretation above,
we would obtain the inclusion:
\begin{center}$\bigcup_{n\in\N}(\fvf{F^n_{A,t}})\subset (\bigcap_{n\in\N}\tvt{F^n_{A,t}})^{\pole_f} = \fvf{\nu^t_{fx} A}$\end{center}
which is strict in general. By orthogonality, this gives us that
$\tvV{\nu^t_{fx} A}\subseteq {\bigcup_{n\in\N}(\fvf{F^n_{A,t}}))^{\pole_V}} $, 
while the proof of adequacy only proves that $\tis{a}{\tau[a:=\cofix{t}{b}{x}{p}]}$ belongs to the latter set.}.
Thus, the definition $\tvv{\nu^t_{fv} A}\defeq \bigcap_{n\in\N}\tvv{F^n_{A,t}}$ would give $\tvv{\nu^t_{fx}A} = \emptyset =\tvv{\bot}$.

Interestingly, and this is the second aspect that we shall discuss here, we could have defined instead 
the truth value of coinductive formulas directly by :
$$\tvv{\nu^t_{fx}A}\defeq \tvv{A[t/x][\nu^y_{fx} A/f(y)=0]}$$
Let us sketch the proof that such a definition is well-founded. 
We consider the language of formulas without coinductive formulas and extended with formulas of the shape $X(t)$ where $X,Y,...$ are parameters.
At level $v$, closed formulas are interpreted by sets of strong values-in-store $\tis{v}{\tau}$,
and as we already observed, these sets are besides closed under the relation $\equiv_\tau$ along their component $\tau$.
If $A(x)$ is a formula whose only free variable is $x$, the function which associates to each term $t$ the set $\tvv{A(t)}$ is thus
a function from $\Lambda_t$ to $\P(\Lambda_v^\tau)/_{\equiv_\tau}$, let us denote the set of these functions by $\L$.
\begin{proposition}
 The set $\L$ is a complete lattice with respect to the order $\leq_\L$ defined by:\vspace{-0.5em}
 $$F\leq_\L G 	\defeq \forall t\in\Lambda_t. F(t)\subseteq G(t)$$
\end{proposition}
\begin{proof}
 Trivial since the order on functions is defined pointwise and the co-domain $\P(\Lambda^\tau_v)$ is itself a complete lattice.
\end{proof}

We define valuations, which we write $\rho$, as functions mapping each parameter $X$ to a function $\rho(X)\in\L$.
We then define the interpretations $\tvv{A}^\rho,\fvf{A}^\rho,...$ of formulas with parameters exactly as above with the additional
rule\footnote{Observe that this rule is exactly the same as in the previous section (see \Cref{fig:dlpaw_real}).}:
$$\tvv{X(t)}^\rho\defeq \{\tis{v}{\tau}\in\rho(X)(t)\}$$

Let us fix a formula $A$ which has one free variable $x$ and a parameter $X$ such that sub-formulas of the shape $X\,t$ only occur in positive positions in $A$.
\begin{lemma}\label{lm:compatibility}
Let $B(x)$ is a formula without parameters whose only free variable is $x$,
and let $\rho$ be a valuation which maps $X$ to the function $t\mapsto \tvv{B(t)}$.
Then $\tvv{A}^{\rho} = \tvv{A[B(t)/X(t)]}$
\end{lemma}
\begin{proof}
 By induction on the structure of $A$, all cases are trivial, and this is true for the basic case $A \equiv X(t)$:
 $$\tvv{X(t)}^\rho = \rho(X)(t) = \tvv{B(t)}\noem$$
\end{proof}

Let us now define $\varphi_A$ as the following function:
$$
\varphi_A:\left\{
  \begin{array}{ccc}
  \L& \to & \L \\
  F &\mapsto&t\mapsto \tvv{A[t/x]}^{[X\mapsto F]}
  \end{array}
\right.
$$

\begin{proposition}
 The function $\varphi_A$ is monotone.
\end{proposition}
\begin{proof}
 By induction on the structure of $A$, where $X$ can only occur in positive positions. 
 The case $\tvv{X(t)}$ is trivial, and it is easy to check that truth values are monotonic
 with respect to the interpretation of formulas in positive positions,
 while falsity values are anti-monotonic.
\end{proof}

We can thus apply Knaster-Tarski theorem to $\varphi_A$, and we denote by $\texttt{gfp}(\varphi_A)$ its greatest fixpoint.
We can now define:
$$\tvv{\nu^t_{Xx} A} \defeq \texttt{gfp}(\varphi_{ A})(t)$$
This definition satisfies the expected equality:
\begin{proposition}\label{prop:dlpaw:coind_def}
 We have:
 $$\tvv{\nu^t_{Xx} A} = \tvv{A[t/x][\nu^y_{Xx} A / X(y)]} $$
\end{proposition}
\begin{proof}
Observe first that by definition, the formula $B(z)=\tvv{\nu^z_{Xx} A}$ satisfies the hypotheses 
of \Cref{lm:compatibility} and that $\texttt{gfp}(\varphi_{ A}) = t\mapsto B(t)$.
Then we can deduce :
$$\begin{array}{c}
   \tvv{\nu^t_{Xx} A} = \texttt{gfp}(\varphi_{ A})(t)
		      = \varphi_{ A}(\texttt{gfp}(\varphi_A))(t)
		      = \tvv{ A[t/x]}^{[X\mapsto \texttt{gfp}(\varphi_A)]}\\
		      =  \tvv{A[t/x][\nu^y_{Xx} A/X(y)]}
\end{array}		      
$$
\end{proof}
Back to the original language, 
it only remains to define $\tvv{\nu^t_{fx} A}$ as the set $\tvv{\nu^t_{Xx} A[X(y)/f(y)=0]}$ that we just defined.
This concludes our proof that the interpretation of coinductive formulas
through the equation in \Cref{prop:dlpaw:coind_def} is well-founded.

We could also have done the same reasoning with the interpretation from the previous section, 
by defining $\L$ as the set of functions from $\Lambda_t$ to $\P(\Lambda_f^\tau)_{\equiv_\tau}$.
The function $\varphi_A$, which is again monotonic, is then:
$$
\varphi_A:\left\{
  \begin{array}{ccc}
  \L& \to & \L \\
  F &\mapsto&t\mapsto \tvv{A[t/x]}^{[X\mapsto F]}
  \end{array}
\right.
$$
We recognize here the definition of the formula $F^n_{A,t}$.
Defining $f^0$ as the function $t\mapsto\fvf{\top}$ and $f^{n+1} \defeq \varphi_A(f^n)$
we have:
$$\forall n\in\N, \fvf{F^n_{A,t}} = f^n(t) =\varphi_A^n(f^0)(t)$$

However, in both cases (defining primitively the interpretation at level $v$ or $f$),
this definition does not allow us to prove\footnote{To be honest, we should rather say that 
we could not manage to find a proof, and that we would welcome any suggestion from insightful readers.}
the adequacy of the {\cofixrule} rule.
In the case of an interpretation defined at level $f$, the best that we can do is to show 
that for any $n\in\N$, $f^n$ is a post-fixpoint since for any term $t$, we have:
$$f^{n}(t) = \fvf{F^{n}_{A,t}} \subseteq  \fvf{F^{n+1}_{A,t}} = f^{n+1}(t)=\varphi_A(f^n)(t) $$
With $\fvf{\nu^t_{fx} A}$ defined as the greatest fixpoint of $\varphi_A$, for any term $t$ and any $n\in\N$
we have the inclusion
$f^n(t) \subseteq \texttt{gfp}(\varphi_A)(t) = \fvf{\nu^t_{fx} A}$ and thus:
$$ \bigcup_{n\in\N} \fvf{F^{n}_{A,t}}  = \bigcup_{n\in\N} f^n(t)  \subseteq \fvf{\nu^t_{fx} A}$$
By orthogonality, we get: 
$$ \tvV{\nu^t_{fx} A}\subseteq \bigcap_{n\in\N} \tvV{F^{n}_{A,t}} $$
and thus our proof of adequacy from the last section is not enough to conclude that 
$\cofix{t}{bx}p \in \tvp{\nu^t_{fx} A}$.
For this, we would need to prove that the inclusion is an equality.
An alternative to this would be to show that the function $t\mapsto \bigcup_{n\in\N} \fvf{F^{n}_{A,t}}$
is a fixpoint for $\varphi_A$. In that case, we could stick to this definition and happily conclude that it
satisfies the equation:
$$\fvf{\nu^t_{Xx} A} = \fvf{A[t/x][\nu^y_{Xx} A / X(y)]} $$
This would be the case if the function $\varphi_A$ was Scott-continuous on $\L$ (which is a dcpo), 
since we could then apply Kleene fixed-point theorem\footnote{In fact, Cousot and Cousot proved a constructive version
of Kleene fixed-point theorem which states that without any continuity requirement, 
the transfinite sequence $(\varphi_A^\alpha(f^0))_{\alpha\in O_n}$ is stationary~\cite{Cousot79}. 
Yet, we doubt that the gain of the desired equality is worth a transfinite definition of the realizability interpretation.}
to prove that 
$t\mapsto \bigcup_{n\in\N} \fvf{F^{n}_{A,t}}$ is the stationary limit of $\varphi_A^n(f_0)$.
However, $\varphi_A$ is not Scott-continuous\footnote{In fact, this is nonetheless a good news about our interpretation. 
Indeed, it is well-know that the more ``regular'' a model is, the less interesting it is. 
For instance, Streicher showed that the realizability model induced by Scott domains (using it as a realizability structure) 
was not only a forcing model by also equivalent to the ground model. 
}
(the definition of falsity values involves double-orthogonal sets which 
do not preserve supremums), and this does not apply.

\section{Conclusion and perspectives}
\subsection{Conclusion}
At the end of the day, we met our main objective, namely proving the soundness and the normalization 
of a language which includes proof terms for dependent and countable choice in a classical setting.
This language, which we called {\dlpaw}, provides us with the same
computational features as {\dpaw} but in a sequent calculus fashion.
These computational features allow {\dlpaw} to internalize the realizability approach of~\cite{BerBezCoq98,EscOli14} 
as a direct proofs-as-programs interpretation: both proof terms for countable and dependent choices 
furnish a lazy witness for the ideal choice function which is evaluated on demand. 
This interpretation is in line with the slogan that 
with new programing principles---here the lazy evaluation and the co-inductive objects---come
new reasoning principles---here the axioms $AC_\N$ and $DC$.

Interestingly, in our search for a proof of normalization for {\dlpaw}, we developed novel tools to study 
these side effects and dependent types in presence of classical logic.
On the one hand, we set out in \cite{Miquey19} the difficulties related to the definition of a sequent calculus with dependent types. 
On the other hand, building on ~\cite{MiqHer18}, we developed a variant of Krivine realizability adapted to
a lazy calculus where delayed substitutions are stored in an explicit environment.
The sound combination of both frameworks led us to the definition of {\dlpaw} together with its realizability interpretation.

\subsection{
Krivine's interpretations of dependent choice}\label{s:dlpaw_quote}
The computational content we give to the axiom of dependent choice is pretty different of 
Krivine's usual realizer of the same~\cite{Krivine03}. 
Indeed, our proof uses dependent types to get witnesses of existential formulas, 
and we represent choice functions through the lazily evaluated stream of their values. A similar
idea can also be found in NuPrl BITT type theory, where choice sequences are used in place of functions~\cite{CohEtAl18}. 
In turn, Krivine realizes a statement which is logically equivalent to the axiom of dependent choice
thanks to the instruction \automath{\Quote}, which injectively associates a natural number 
to each closed $\lambda_c$-term.
In a more recent work~\cite{Krivine16}, Krivine proposes a realizability model which has a bar-recursor
and where the axiom of dependent choice is realized using the bar-recursion. 
This realizability model satisfies the continuum hypothesis and many more properties, 
in particular the real numbers have the same properties as in the ground model. 
However, the very structure of this model, where ${\bm \Lambda}$ is of cardinal $\aleph_1$ (in particular infinite streams of integer are terms),
makes it incompatible with $\Quote$.

It is clear that the three approaches are different in terms of programming languages. 
Nonetheless, it could be interesting to compare them from the point of view of the realizability models they give rise to.
In particular, our analysis of the interpretation of co-inductive formulas may suggest that
the interest of lazy co-fixpoints is precisely to approximate the limit situation where $\bm \Lambda$ has infinite objects.

\subsection{Reduction of the consistency of classical arithmetic in 
  finite types with dependent choice to the consistency of
  second-order arithmetic}
The standard approach to the computational content of classical
dependent choice in the classical arithmetic in finite types is via
realizability as initiated by Spector~\cite{Spector62} in the context
of Gödel's functional interpretation, and later adapted to the context
of modified realizability by Berardi {\em et al}~\cite{BerBezCoq98}. 
The aforementioned works of Krivine~\cite{Krivine03,Krivine16} in the different settings of PA2 and ZF$_\varepsilon$ 
also give realizers of dependent choice.
In all these approaches, the correctness of the realizer, which
implies consistency of the system, is itself justified by a use at the
meta-level of a principle classically equivalent to dependent choice
(dependent choice itself in~\cite{Krivine03}, bar induction or update
induction~\cite{Berger04} in the case of \cite{Spector62,BerBezCoq98}.).

Our approach is here different, since we directly interpret proofs
of dependent choice in classical arithmetic computationally.
Besides, the structure of our realizability interpretation for {\dlpaw} suggests
the definition of a typed CPS to an extension of system $F$\vlong{\footnote{See \cite[Chapitre 8]{these} for further details.}},
but it is not clear whether its consistency is itself conservative or not over system $F$.
Ultimately, we would be interested in a computational reduction 
of the consistency of {\dpaw} or {\dlpaw} to the one of PA2, that is 
to the consistency of second-order arithmetic.
While it is well-known that $DC$ is conservative over second-order arithmetic 
with full comprehension  (see \cite[Theorem VII.6.20]{Simpson09}),
it would nevertheless be very interesting to have such a direct computational reduction.
The converse direction has been recently studied by Valentin Blot, who presented in~\cite{Blot17} 
a translation of System F into a simply-typed total language with a variant of bar recursion.

\newcommand{\ac}{\automath{\text{AC}}}
\subsection{The axiom of choice}
The computational content of the axiom of countable and dependent choices
being understood, it is natural to wonder whether our approach
extends to the full axiom of choice. 
To that end, the main difficulty  seems to be related to 
the memoization technique we use: in both proof terms for {\acn} and \dc, 
the choice function ranges over the domain of natural numbers, hence it is easy
to know whether the computation for a given $n\in\N$ has already been performed.
In turn, when considering an arbitrary domain $A$ as is the case for the 
full axiom of choice \ac, if the equality on $A$ is not decidable, 
we might not be able to know whether the computation for $a\in A$ has already been 
done or not. At first sight, our approach seems to extend to any type $A$ with decidable
equality, but handling arbitrary types with undecidable equality would require 
a more careful treatment of equality during the memoization process.

On a similar line of work, there exists a wide literature on
different form of choices (\emph{e.g.}, the axiom of dependent choice, König's Lemma),
as well as on bar induction principles or the Fan theorem, which can be seen as 
their contrapositives. Recently, Herbelin and Brede undertook a systematical
classification of these principles, including also the Ultrafilter Theorem in the picture,
aiming at highlighting dualities and studying logical equivalences in different frameworks 
(classical, intuitionistic and linear logic)~\cite{HerBre19}. 
In that regard, having a precise understanding of the computational content of each of these
principles as well as on the translations existing between them is an interesting perspective, for which
this work or Blot's work on bar induction~\cite{Blot17} might be the first steps. 

\vspace{-0.3em}
\section*{Acknowledgment}
\noindent
 The author warmly thanks Hugo Herbelin for numerous discussions 
 and attentive reading of this work during his PhD years, 
 and Amina Doumane for her helpful comments on coinductive fixpoints 
 which led to the \Cref{app:cofix} of this paper.

\vspace{-0.3em}
\bibliographystyle{abbrvurl}
\vlong{\bibliography{biblio}}


\end{document}